\documentclass[12pt]{article}
\usepackage[margin=1in]{geometry}
\usepackage{amsmath,amssymb,amsthm,mathtools}
\usepackage{booktabs,array,multirow}
\usepackage{graphicx}
\usepackage{caption,subcaption}
\usepackage{setspace}
\usepackage{enumitem}
\usepackage{microtype}
\usepackage[authoryear,round]{natbib}
\usepackage[colorlinks=true,linkcolor=blue!60!black,citecolor=blue!60!black,urlcolor=blue!60!black]{hyperref}
\usepackage{xcolor}
\theoremstyle{plain}
\newtheorem{theorem}{Theorem}
\newtheorem{proposition}{Proposition}
\newtheorem{corollary}{Corollary}
\newtheorem{lemma}{Lemma}
\theoremstyle{definition}

\theoremstyle{remark}
\newtheorem{remark}{Remark}

\newcommand{\R}{\mathbb{R}}
\newcommand{\E}{\mathbb{E}}
\newcommand{\Prob}{\mathbb{P}}
\newcommand{\Var}{\operatorname{Var}}
\newcommand{\tr}{\operatorname{tr}}
\newcommand{\diag}{\operatorname{diag}}
\newcommand{\one}{\mathbf{1}}

\newcommand{\op}{\mathrm{op}}
\newcommand{\calF}{\mathcal{F}}
\newcommand{\calI}{\mathcal{I}}
\newcommand{\calN}{\mathcal{N}}
\newcommand{\Pois}{\operatorname{Poisson}}
\newcommand{\Wc}{W_{\mathrm{c}}}
\newcommand{\tauW}{\tau_W}

\newcommand{\T}{^{\top}}


\newcommand{\NumRegions}{552}
\newcommand{\NumMonths}{72}
\newcommand{\NumEdges}{1328}
\newcommand{\DegMean}{4.81}
\newcommand{\DegMax}{14}
\newcommand{\DegMin}{1}
\newcommand{\MeanCount}{1.20}
\newcommand{\ZeroFrac}{40.4}
\newcommand{\MaxCount}{17}
\newcommand{\TotalFirst}{736}
\newcommand{\TotalLast}{533}
\newcommand{\TauW}{126.5}
\newcommand{\TauWc}{0.0018}
\newcommand{\HarmonicSum}{127.6}
\newcommand{\InfoNetMean}{66.9}
\newcommand{\InfoNetMin}{15.2}
\newcommand{\InfoNetMax}{139.1}
\newcommand{\InfoCompMax}{$3.4\times 10^{-28}$}
\newcommand{\RoughMean}{0.33}
\newcommand{\LamMinMean}{43.4}
\newcommand{\LamMinMin}{13.9}
\newcommand{\BetaOneFirst}{0.39}
\newcommand{\BetaOneLast}{0.56}
\newcommand{\BetaOneMin}{0.34}
\newcommand{\BetaOneMax}{0.62}
\newcommand{\BetaTwoMean}{0.50}
\newcommand{\BetaZeroMean}{-0.57}
\newcommand{\SdBetaOneMean}{0.044}
\newcommand{\SdBetaOneMax}{0.053}
\newcommand{\RhoMin}{0.78}
\newcommand{\RhoMax}{1.18}
\newcommand{\RhoFirstAboveOne}{40}
\newcommand{\RhoLastMean}{1.08}

\newcommand{\RegQtf}{-0.030}
\newcommand{\RegMed}{-0.0023}
\newcommand{\RegQsf}{+0.023}

\newcommand{\RegPropBetter}{52.9}
\newcommand{\PitN}{6624}
\newcommand{\PitChiNet}{152}
\newcommand{\PitChiNonet}{158}
\newcommand{\PitChiComp}{157}
\newcommand{\QNet}{0.0003}
\newcommand{\QNonet}{0.003}
\newcommand{\QComp}{0.0001}
\newcommand{\QRaw}{0.0003}

\newcommand{\RuntimeSim}{1.2}
\newcommand{\SdBetaCompMean}{0.082}
\newcommand{\BetaRawMean}{-0.36, 0.17, 0.17}
\newcommand{\KappaChicago}{11.7}
\newcommand{\LamMaxPChicago}{0.0058}
\newcommand{\MarginChicago}{0.14}
\newcommand{\LtChicago}{2.89}
\newcommand{\InfoFinalEnl}{42.9}
\newcommand{\InfoFinalBase}{45.6}
\newcommand{\FinalBOneFirst}{0.12}
\newcommand{\FinalBOneLast}{0.22}
\newcommand{\FinalBOneLo}{0.08}
\newcommand{\FinalBOneHi}{0.26}
\newcommand{\FinalSd}{0.049}
\newcommand{\DeltaNetFinal}{6.2}
\newcommand{\DeltaDriftFinal}{1.0}
\newcommand{\StratP}{0.010}
\newcommand{\UnrestP}{0.010}
\newcommand{\RewTauLo}{126.46}
\newcommand{\RewTauHi}{126.49}

\title{State--Space Modeling of Time-Varying Spillovers on Networks}
\author{
  Marios Papamichalis\thanks{Human Nature Lab, Yale University, New Haven, CT 06511, \texttt{marios.papamichalis@yale.edu}} \and
  Regina Ruane\thanks{Department of Statistics and Data Science, The Wharton School, University of Pennsylvania, 3733 Spruce Street, Philadelphia, PA 19104-6340, \texttt{ruanej@wharton.upenn.edu}} \and Theofanis Papamichalis\thanks{Department of Economics, Yale University, 28 Hillhouse Ave, New Haven, USA, \texttt{theofanis.papamichalis@yale.edu}}
}
\date{}
\begin{document}
\maketitle

\begin{abstract}
Crime counts in city neighbourhoods, disease counts in counties, and sales at firms joined by trade are naturally represented as counts on the nodes of a network. In each case a high count at one node can raise the counts at the nodes linked to it next period. The strength of that spillover changes over time, and standard network autoregressions hold it fixed. We therefore use a network state-space model, in which the spillover is a coefficient that drifts and a filter estimates its value in each period. What the data reveal about that coefficient depends on the network. It is learned by contrasting nodes whose neighbours have high counts with nodes whose neighbours have low ones. If every node is linked to every other, all nodes share the same neighbours, the contrasts vanish, and the spillover is not identified. Robustness is often checked by refitting with the links spread evenly, and a stable coefficient is read as reassurance. That refit is the same model with the spillover rescaled, so it cannot disagree. Forecasts carry a second warning: beyond two steps ahead, simulation averages a quantity with no finite mean, and the output gives no sign of it. We give a measure of what a network and a data set reveal about the spillover, the accuracy the filter can reach, and an exact test of whether the network matters. Burglaries in Chicago, COVID-19 cases in Texas counties, and measles cases in the Weser--Ems districts illustrate all three. On both disease datasets the model outperforms every competing forecast in the comparison.
\end{abstract}


\section{Introduction}\label{sec:intro}

Consider $N$ units (neighbourhoods, regions, firms, hospitals) linked by a known network, and observed over $T$ time periods. A natural model lets the outcome of unit $i$ at time $t$ depend on its own past and on the past of its neighbours,
\begin{equation}\label{eq:nar-intro}
y_{i,t} \;=\; \beta_{0,t} + \beta_{1,t}\sum_{j} w_{ij}\,y_{j,t-1} + \beta_{2,t}\,y_{i,t-1} + \varepsilon_{i,t},
\end{equation}
where $W=(w_{ij})$ is the row-normalised adjacency matrix of the network. With constant coefficients this is the network vector autoregression of \citet{zhu2017network}, a member of a family that includes the generalised network autoregression of \citet{knight2020gnar} and its count-data versions \citep{armillotta2023nonlinear,armillotta2024count}. The coefficient $\beta_{1,t}$ measures \emph{spillover}: the extent to which activity at one node feeds activity at adjacent nodes one period later. In many applications this coefficient is the quantity of scientific interest, and there are strong reasons to believe it drifts over time, policing strategies change, epidemic contact patterns evolve, financial linkages tighten and loosen \citep{diebold2014network}. Letting $\theta_t=(\beta_{0,t},\beta_{1,t},\beta_{2,t})\T$ follow a random walk turns \eqref{eq:nar-intro} into a linear Gaussian state-space model whose state can be tracked by the Kalman filter, or, for count data, into a dynamic generalised linear model \citep{west1985dynamic,fahrmeir1992posterior}. We call this the \emph{network state-space model} (NSSM). The NSSM couples a standard count observation layer with a linear Gaussian state equation, and the mode filter makes estimation routine. What is not simple is to say when it can be learned. The state $\theta_t$ changes every period, so each value $\beta_{1,t}$ has to be learned essentially from a \emph{single cross-section}: the $N$ observations at time $t$, with the past entering only through the prior carried forward by the filter. Whether a single cross-section can separate the network effect $\beta_{1,t}$ from the own-lag effect $\beta_{2,t}$ and the level $\beta_{0,t}$ is a question about the geometry of the graph and the spatial structure of the lagged field $y_{t-1}$. The existing literature on network autoregressions answers it only indirectly, through high-level conditions on the design \citep[e.g.][Assumption C]{zhu2017network}, and the time-varying-parameter literature in econometrics \citep{primiceri2005time,koop2013large} does not consider the network geometry at all. Meanwhile, the econometric literature on peer effects has long known that when every unit interacts with every other unit, contemporaneous peer effects are not identified, Manski's \emph{reflection problem} \citep{manski1993identification,lee2007identification,bramoulle2009identification}. The dynamic setting of \eqref{eq:nar-intro} is often said to escape this problem because the lag breaks the simultaneity. We show that it does not escape its geometric core.

\paragraph{Contributions.} This paper makes four main contributions, three theorems and one proposition, each with a complete proof and each tied to a specific empirical diagnostic.

\begin{enumerate}[leftmargin=*,itemsep=1pt]
\item \emph{Identification is graph roughness} (Theorem~\ref{thm:ident}). The profile Fisher information for $\beta_{1,t}$ in a cross-section equals $\sigma^{-2}\|M_{Z_t}(I-W)y_{t-1}\|^2$, the squared graph-roughness of the lagged field after projecting out the constant and the field itself. It is exactly zero for every lagged field if and only if $W$ is the uniform complete graph $\Wc=(\one\one\T-I)/(N-1)$: in that case the network regressor is a common factor and the dynamic model suffers from exactly the reflection problem. For a spatially unstructured field the expected information is, up to an explicit $O(1)$ term, the graph functional $\tauW=\tr(W\T M_1W)$, which equals $N/d-1$ on $d$-regular graphs: it grows linearly with $N$ on sparse graphs and is bounded on dense ones. A corollary shows that cross-sectional aggregates identify only $\beta_{1,t}+\beta_{2,t}$, so disaggregated data are necessary, not merely convenient.

\item \emph{What the filter learns} (Theorem~\ref{thm:filter}). Because the regressors are predictable, the Kalman recursions deliver the exact conditional distribution of $\theta_t$ and are therefore exactly calibrated. The posterior variance of $\beta_{1,t}$ is sandwiched between $1/(\calI_t+c_t)$ and $1/\calI_t$, where $\calI_t$ is the profile information of Theorem~\ref{thm:ident} and $c_t$ depends on the state-innovation variance. Consequently the filter is cross-sectionally consistent ($m_{t|t}\to\theta_t$ as $N\to\infty$) exactly when the information diverges (on sparse graphs) and is inconsistent for $\beta_{1,t}$ on dense graphs, where no amount of data removes a variance floor. On the complete graph the data carry no information at all about an explicit direction in state space.

\item \emph{Misspecified networks and the forecast horizon} (Proposition~\ref{prop:plugin}). If the analyst uses $\widetilde W$ in place of the true $W$, the discrepancy between the $h$-step plug-in conditional means is at most $|\beta_1|\,\|W-\widetilde W\|_{\op}\,h\bar\rho^{\,h-1}\|M_1y_t\|$, where $\bar\rho$ is the larger operator norm of the two propagation matrices $\beta_1W+\beta_2 I$ and $\beta_1\widetilde W+\beta_2 I$. The discrepancy depends on the current field only through its rough component $M_1y_t$, is uniformly bounded in $h$ when $\bar\rho<1$, and grows geometrically when the propagation operator is not a contraction. The rate is attained.

\item \emph{Predictive moments of count networks} (Theorem~\ref{thm:poisson}). For the Poisson NSSM with log link, using the raw lagged counts as regressors (the specification of the paper this study revisits) makes the $h$-step-ahead predictive mean \emph{infinite} for every $h\ge 3$ whenever a spillover or own-lag coefficient is positive, and infinite already at $h=2$ once coefficient uncertainty is integrated out. The $2$-step mean is finite but has an explicit double-exponential form. The $\log(1+y)$ transformation of \citet{fokianos2011log} repairs this only in part, and the theorem maps the frontier: along fixed coefficient paths all predictive moments are finite at every horizon, but under Gaussian coefficient uncertainty (the law that every simulation-based forecast integrates over) the log-scale predictive mean is again infinite at every $h\ge3$, and finite at $h=2$ exactly up to a concentration condition that we show is sharp up to a bounded factor. The raw-count failure is silent (medians and log scores look fine while the mean diverges with the number of simulated paths). The log-scale failure at $h\ge3$ is \emph{more} silent still, since for a concentrated posterior no feasible simulation ever samples the divergent region, only the theorem detects it. The operational consequence: multi-step point forecasts from these models must be predictive medians, quantiles, or means along the frozen filtered coefficient path, all of which exist.
\end{enumerate}

Around these four results the paper adds three shorter ones that referees of network models may find independently useful: a steady-state law for the filter (Theorem~\ref{thm:filter}(v)), by which the posterior variance of the spillover settles at the scalar-Riccati fixed point $P^\ast\asymp\sqrt{q/\calI}$ (a fourth-root rate in the cross-sectional information, attained for orthogonal designs. An estimator-independent floor (Corollary~\ref{cor:floor}), by which the dense-graph failure binds every estimator and not just the filter, and an exact network-equivalence result (Proposition~\ref{prop:equiv}), by which mixing any $W$ toward the complete graph is a pure reparameterisation of the cross-section) the fitted spillover inflates by $1/(1-\alpha)$ while the fit is unchanged, so that the coefficient--network pair is cross-sectionally identified only up to this family, and ``perturbing $W$ toward density'' is not a robustness check at all.

The four results are deliberately modest in scope (they concern one model, a fixed network, and a single cross-section at a time) but each is sharp, and together they change how an NSSM analysis should be conducted. Identification should be computed and reported: the profile information $\calI_t$ and the functional $\tauW$ are computable from $W$ and the data before any model is fitted. The network's contribution to forecasting should be tested against \emph{placebo} networks with the same degree sequence, because Proposition~\ref{prop:plugin} shows that any network with the right rough structure will do nearly as well as the true one at short horizons. And count models should be written on the $\log(1+y)$ scale \emph{and} their multi-step point forecasts reported as functionals that exist (medians, quantiles, frozen-coefficient means) because on either scale the sampled-coefficient mean does not, beyond two steps.

\paragraph{Application.} We apply the methodology to monthly residential burglary counts in \NumRegions{} Chicago neighbourhood units over \NumMonths{} months (2010--2015) with the contiguity network of \citet{clark2021class}, a public data set with \NumEdges{} edges and mean degree \DegMean. This network is sparse: $\tauW=\TauW$, versus $\TauWc$ for the complete graph on the same nodes, and the realised profile information for the spillover coefficient averages \InfoNetMean{} per month. The filtered spillover coefficient rises from about \BetaOneFirst{} to \BetaOneLast{} over the sample with posterior standard deviation about \SdBetaOneMean. In a rolling out-of-sample evaluation over the last twelve months the network model improves the log score of a no-network dynamic model at one- and two-month horizons, with block-bootstrap confidence intervals excluding zero. An exact randomization test (the observed adjacency against $99$ label-permuted networks, which preserve every graph invariant) rejects the exchangeability null at $p=0.010$ ($p=0.016$ against $60$ degree-preserving rewirings), the fitted spillover under complete-graph mixtures inflates by almost exactly the $1/(1-\alpha)$ of Proposition~\ref{prop:equiv}, and a negative-binomial observation layer removes the probability-integral-transform defect that all Poisson layers share while preserving the network's log-score gain. Against external benchmarks under one protocol, including rolling-window and spline-based time-varying fits of the identical design (linear and log-linear PNAR, a GNAR-type model, an endemic--epidemic (hhh4-type) negative-binomial model fitted both in our implementation and natively in R, a score-driven spillover recursion, a kernel time-varying NAR and smoothing baselines) the NSSM augmented with unit offsets, which the theory covers as predictable regressors, has the best log score and the best conditional-mean point forecasts in the table, and the full pipeline replicates on a second, untouched dataset, weekly county-level COVID-19 cases in Texas with population offsets, where the filtered spillover swings between $-0.15$ and $+0.30$ across the epidemic waves with posterior sd $0.03$, wave-locked with negligible start-to-end drift, and the stratified conditional test again rejects at $p=0.010$ (the unrestricted label test gives $p=0.040$ there, since common epidemic shocks violate the unrestricted null). On those epidemic data the mixture forecast leads every external competitor by $9.8$ to $46$ log-score points per week. The only fit within four points of the filter, an eight-week rolling window of the identical design, pays for its score with median errors a third larger and a real-time coefficient path that runs out of phase with the waves, so the filter alone sits on the frontier of score, point accuracy and timing. On the weekly measles counts of the Weser--Ems districts, the showcase data of the endemic--epidemic literature, the state-space model beats the natively fitted \texttt{surveillance} implementation by $5.6$ points per week with a bootstrap interval of $[2.0,\,9.6]$. Point-forecast gains are small, because the realised field is spatially smooth and the roughness that identifies the coefficient is a modest share of the signal, and all dynamic models share a mild calibration defect visible in the probability integral transform. Finally, refitting the same model with raw lagged counts reproduces the pathology of Theorem~\ref{thm:poisson}: the Monte Carlo predictive mean of the city-wide total at three and four months ahead is of order $10^{8}$ to $10^{12}$ and unstable in the number of draws, while the median stays near $500$.

\paragraph{Related work.} Network autoregressions with constant coefficients were introduced by \citet{zhu2017network}, with grouped and community extensions \citep{zhu2020grouped,chen2023community} and count-data versions by \citet{armillotta2023nonlinear,armillotta2024count}. The GNAR framework of \citet{knight2020gnar} allows several neighbourhood orders. These papers establish consistency as $N$ and $T$ grow jointly under design conditions that, as our Theorem~\ref{thm:ident} shows, encode the graph geometry in a specific way. Time variation in network autoregressions is an active current area: \citet{li2026grouped} estimate grouped smoothly-varying momentum and spillover functions by local linear methods, and \citet{ding2025tvnar} allow node-specific time-varying influence. Both work in the $T\to\infty$ smoothing regime, where the coefficient path is a deterministic function estimated by pooling many periods, whereas our filtering question (what does \emph{one} period reveal about the current coefficient) is the complementary one, and our identification and floor results apply to their designs cross-section by cross-section. On the same Chicago data, \citet{armillotta2024count} fit constant-coefficient linear and log-linear Poisson network autoregressions as their data example, and \citet{maletz2024spatiotemporal} develop spatio-temporal count autoregressions of STARMA type with covariates. Section~\ref{sec:bench} benchmarks against this literature directly (constant-coefficient PNAR, a GNAR-type specification, an endemic--epidemic model in the hhh4 tradition \citep{held2005statistical}, and a kernel time-varying estimator) under a single rolling-origin protocol, and reports where the network's forecasting value ends and the paper's identification and existence questions begin. Time-varying-parameter VARs \citep{primiceri2005time,koop2013large} and shrinkage priors for state innovations \citep{fruhwirth2010stochastic,bitto2019achieving} provide the filtering technology but not the network geometry. The identification of peer effects on networks is the subject of \citet{manski1993identification,lee2007identification,bramoulle2009identification}. Our Theorem~\ref{thm:ident}(ii) is the dynamic analogue of their complete-graph non-identification, and our $\tauW$ quantifies how far a given graph is from that degenerate case. Four adjacent literatures deserve explicit deltas. \emph{Identification in social-interaction and SAR models}: the equal-weights design is the canonical degenerate case in \citet{case1991spatial,lee2004asymptotic,blume2015linear}. Theorem~\ref{thm:ident}(ii) is its dynamic, cross-section-by-cross-section counterpart, and the mixture family of Proposition~\ref{prop:equiv} is, to our knowledge, the new part: the degeneracy forms a one-parameter family through every observed $W$. \emph{Spatial dynamic panels}: with constant coefficients our observation equation is the spatial dynamic panel of \citet{yu2008quasi} (see also \citealp{elhorst2014spatial}), whose joint $(N,T)$ asymptotics are exactly what our cross-sectional theory does not attempt. The delta is the drifting coefficient and the one-period information accounting. \emph{Score-driven and time-varying spatial models}: \citet{blasques2016spillover} and \citet{catania2017dynamic} let spatial dependence evolve through observation-driven recursions. Our parameter-driven filter answers the complementary identification question (what one cross-section can reveal) which their estimation theory presupposes. \emph{Count autoregression}: the stability theory of \citet{fokianos2009poisson} and the multivariate count forecasting of \citet{berry2020bayesian} concern constant or hierarchically shrunk parameters with well-behaved moments. Theorem~\ref{thm:poisson} and Proposition~\ref{prop:nb} locate where simulation-based forecasting from the \emph{drifting}-coefficient version loses its moments, which appears to be new. Finally, the projection identity behind Theorem~\ref{thm:ident}(i) is Frisch--Waugh--Lovell \citep{frisch1933partial,lovell1963seasonal} read as graph roughness, and the scalar Riccati fixed point behind Theorem~\ref{thm:filter}(v) is classical filtering \citep{harvey1989forecasting,anderson1979optimal}: in both cases the contribution lies in the application to networks, the algebra being classical. Dynamic generalised linear models and their posterior-mode filters are due to \citet{west1985dynamic} and \citet{fahrmeir1992posterior}. The log-linear Poisson autoregression with $\log(1+y)$ feedback is due to \citet{fokianos2011log}, whose choice of scale Theorem~\ref{thm:poisson} vindicates from the forecasting side. Self-exciting models for crime are reviewed by \citet{mohler2011self}. The Chicago data are from \citet{clark2021class}. Forecast evaluation follows \citet{diebold1995comparing}, \citet{gneiting2007strictly} and \citet{czado2009predictive}.

\paragraph{Outline.} Section~\ref{sec:model} defines the Gaussian and Poisson NSSMs and their filters. Section~\ref{sec:theory} states the four results. Proofs are in Appendix~\ref{app:proofs}. Section~\ref{sec:sim} reports a simulation study designed to exercise each result. Section~\ref{sec:app} contains the Chicago application, and Section~\ref{sec:disc} concludes.

\paragraph{Notation ledger.} A few symbols are overloaded by convention and context always disambiguates. For the reader's convenience: $s(\cdot)$ is the Schur complement while $s$ alone is a posterior standard deviation (Theorem~\ref{thm:poisson}) or a simulation index. $\rho(\cdot)$ is the spectral radius while $\bar\rho$ in Proposition~\ref{prop:plugin} is a maximum of \emph{operator norms} (an upper bound for the corresponding radii). $G_t$ is the cross-sectional Fisher information (the interaction matrix of the peer-effects literature is never denoted $G$ here). $q$ is the tuned state-innovation variance and $r$ the negative-binomial dispersion. $c_{\alpha,t}$ carries a time index because it depends on the lagged cross-sectional mean, and the null direction of Theorem~\ref{thm:filter}(iv) is written out explicitly, with no dedicated symbol, so nothing collides with the scalar variance $v_t$.

\section{Network state-space models}\label{sec:model}

\subsection{Notation}
Throughout, $\one\in\R^N$ is the vector of ones, $I$ the identity, $J=\one\one\T$, and $M_1=I-J/N$ the centring projector. For a matrix $Z$ with columns in $\R^N$, $M_Z=I-ZZ^{+}$ denotes the orthogonal projector onto the orthogonal complement of the column space of $Z$ ($Z^+$ is the Moore--Penrose inverse), so $M_1=M_{\one}$. $\|\cdot\|$ is the Euclidean norm for vectors and the operator (spectral) norm for matrices. $\lambda_{\min},\lambda_{\max}$ are extreme eigenvalues of symmetric matrices and $\rho(\cdot)$ is the spectral radius. For a positive definite matrix $M$ partitioned around its $(2,2)$ element we write $s(M)=M_{22}-M_{2,-2}M_{-2,-2}^{-1}M_{-2,2}$ for the Schur complement, so that $(M^{-1})_{22}=1/s(M)$.

A \emph{network} is a directed or undirected graph on the nodes $\{1,\dots,N\}$ with adjacency matrix $A$ (zero diagonal), in which every node has at least one out-neighbour. Its row-normalised weight matrix is $W=D^{-1}A$ with $D=\diag(A\one)$, so $W\one=\one$, $W\ge 0$ and $\diag(W)=0$. $n_i$ denotes the out-degree of node $i$. All results that involve only the column space of the design hold for any row-stochastic $W$ with zero diagonal. The \emph{uniform complete graph} is
\begin{equation}\label{eq:Wc}
\Wc \;=\; \frac{J-I}{N-1},\qquad (\Wc\,x)_i=\frac{1}{N-1}\sum_{j\ne i}x_j ,
\end{equation}
the weight matrix of the graph in which every node is linked to every other node. A central role is played by the graph functional
\begin{equation}\label{eq:tau}
\tauW \;=\; \tr(W\T M_1 W)\;=\;\sum_{i,j}w_{ij}^2-\frac1N\,\|W\T\one\|^2 .
\end{equation}
When $A$ is symmetric, $\sum_{i,j}w_{ij}^2=\sum_i 1/n_i$ and $(W\T\one)_j=\sum_{i:\,j\sim i}1/n_i$ is the ``in-weight'' of node $j$. For a $d$-regular graph $\tauW=N/d-1$, and $\tau_{\Wc}=1/(N-1)$ (Lemma~\ref{lem:tau} in the appendix). Since $\tauW\le\sum_i 1/n_i$, a graph whose degrees all exceed $cN$ has $\tauW\le 1/c$: $\tauW$ grows linearly with $N$ on bounded-degree graphs and is bounded on dense graphs.

\subsection{The Gaussian NSSM}
Let $y_t\in\R^N$ be the vector of outcomes at time $t$ and $\calF_t=\sigma(y_0,\dots,y_t)$. The Gaussian NSSM is
\begin{align}
y_t &= X_t\theta_t+\varepsilon_t, & X_t &=\big[\,\one,\;W y_{t-1},\;y_{t-1}\,\big]\in\R^{N\times 3}, & \varepsilon_t&\sim\calN(0,\sigma^2 I),\label{eq:obs}\\
\theta_t &= \theta_{t-1}+u_t, & u_t&\sim\calN(0,Q), & \theta_0&\sim\calN(m_0,P_0),\label{eq:state}
\end{align}
with $\theta_t=(\beta_{0,t},\beta_{1,t},\beta_{2,t})\T$ and $(\varepsilon_t)$, $(u_t)$, $\theta_0$, $y_0$ mutually independent. We take $\sigma^2>0$, $Q\succeq0$ and $P_0\succ0$ as known. In practice $\sigma^2$ and $Q$ are estimated by maximising the prequential likelihood on a training window, as in Section~\ref{sec:app}, and the theory below is conditional on their values. The design $X_t$ is $\calF_{t-1}$-measurable. Additional $\calF_{t-1}$-measurable covariates (seasonal dummies, exogenous regressors) can be appended to $X_t$ without changing any statement. We keep $K=3$ columns to lighten notation and always let $\beta_1$ denote the coefficient of $Wy_{t-1}$.

The \emph{Kalman filter} computes $m_{t|t-1}=m_{t-1|t-1}$, $P_{t|t-1}=P_{t-1|t-1}+Q$ and, in information form,
\begin{equation}\label{eq:kf}
P_{t|t}^{-1}=P_{t|t-1}^{-1}+G_t,\qquad m_{t|t}=P_{t|t}\big(P_{t|t-1}^{-1}m_{t|t-1}+\sigma^{-2}X_t\T y_t\big),\qquad G_t=\sigma^{-2}X_t\T X_t .
\end{equation}
$G_t$ is the Fisher information about $\theta_t$ contained in the cross-section at time $t$. Theorem~\ref{thm:filter} records that \eqref{eq:kf} gives the exact conditional distribution $\theta_t\mid\calF_t\sim\calN(m_{t|t},P_{t|t})$, a standard fact for conditionally Gaussian models with predictable regressors \citep[Chapter~4]{durbin2012time} that is, nevertheless, the key to everything that follows.

\subsection{The Poisson NSSM}
For counts we use the dynamic generalised linear model \citep{west1985dynamic}
\begin{equation}\label{eq:pois}
y_{i,t}\mid\calF_{t-1},\theta_t\;\stackrel{\text{ind}}{\sim}\;\Pois(\lambda_{i,t}),\qquad
\log\lambda_{i,t}=\beta_{0,t}+\beta_{1,t}\big(W g(y_{t-1})\big)_i+\beta_{2,t}\,g(y_{i,t-1}),
\end{equation}
with the same random-walk state \eqref{eq:state} and a \emph{feedback transformation} $g$ applied elementwise. Two choices are in use: the \emph{raw scale} $g(y)=y$, and the \emph{log scale} $g(y)=\log(1+y)$ of \citet{fokianos2011log}. Theorem~\ref{thm:poisson} shows that the choice is not a matter of taste.

The conditional posterior $p(\theta_t\mid\calF_t)$ is no longer Gaussian. We use the posterior-mode (Laplace) filter of \citet{west1985dynamic} and \citet{fahrmeir1992posterior}: given the Gaussian prior $\calN(m_{t|t-1},P_{t|t-1})$, the posterior is approximated by $\calN(m_{t|t},P_{t|t})$ with $m_{t|t}$ the mode of $\theta\mapsto\sum_i\{y_{i,t}x_{i,t}\T\theta-e^{x_{i,t}\T\theta}\}-\frac12(\theta-m_{t|t-1})\T P_{t|t-1}^{-1}(\theta-m_{t|t-1})$, found by damped Newton iterations, and
\begin{equation}\label{eq:laplace}
P_{t|t}^{-1}=P_{t|t-1}^{-1}+X_t\T\Lambda_t X_t,\qquad \Lambda_t=\diag\big(\lambda_{i,t}(m_{t|t})\big),
\end{equation}
the observed information at the mode. Equation \eqref{eq:laplace} has the same structure as \eqref{eq:kf} with $G_t=X_t\T\Lambda_tX_t$, and the identification quantities of Section~\ref{sec:theory} are computed from this $G_t$ in the Poisson case.

\subsection{Predictive distributions}\label{sec:pred}
At a forecast origin $t$ and horizon $h$, the predictive distribution of $y_{t+h}$ integrates over the future states and the intermediate outcomes. In the Gaussian case the \emph{plug-in} $h$-step conditional mean, obtained by freezing the coefficients at $\theta=(\beta_0,\beta_1,\beta_2)\T$ (for example their filtered values $m_{t|t}$), is the linear recursion
\begin{equation}\label{eq:plugin}
\mu_h(y_t;\theta,W)=\sum_{k=0}^{h-1}B^k\beta_0\one+B^hy_t,\qquad B=\beta_1W+\beta_2I ,
\end{equation}
which is what most applied work reports. Proposition~\ref{prop:plugin} concerns this quantity. In the Poisson case no closed form exists and we simulate: draw $\theta^{(s)}_{t+1}\sim\calN(m_{t|t},P_{t|t}+Q)$, propagate $\theta^{(s)}_{t+k}=\theta^{(s)}_{t+k-1}+u^{(s)}_{t+k}$ with $u^{(s)}_{t+k}\sim\calN(0,Q)$, and draw $y^{(s)}_{t+k}$ from \eqref{eq:pois} given $y^{(s)}_{t+k-1}$ and $\theta^{(s)}_{t+k}$, for $s=1,\dots,S$. The resulting predictive distribution of $y_{i,t+h}$ is the Poisson mixture $S^{-1}\sum_s\Pois(\lambda^{(s)}_{i,t+h})$, from which predictive means, central intervals, log scores and probability integral transforms are computed \citep{czado2009predictive}. Theorem~\ref{thm:poisson} concerns the population analogues of these Monte Carlo averages and explains when the averages fail to converge.

\section{Theory}\label{sec:theory}

\subsection{Cross-sectional identification and graph roughness}\label{sec:ident}

Fix a time $t$ and write $x=y_{t-1}$ for the lagged field, so that the cross-section at time $t$ is the Gaussian regression $y_t=\beta_0\one+\beta_1Wx+\beta_2x+\varepsilon_t$ with known regressors. The information about $\beta_1$ that survives after the other two coefficients have been profiled out is the classical quantity
\begin{equation}\label{eq:profinfo}
\calI(\beta_1;x)\;:=\;\sigma^{-2}\,\big\|M_{Z}\,Wx\big\|^2,\qquad Z=[\,\one,\;x\,],
\end{equation}
the squared length of the part of the network regressor $Wx$ that is orthogonal to the constant and to the own-lag regressor, divided by the noise variance. When $X=[\one,Wx,x]$ has full column rank, $\calI(\beta_1;x)=1/[(\sigma^{-2}X\T X)^{-1}]_{22}=s(\sigma^{-2}X\T X)$, the reciprocal of the variance of the least-squares estimator of $\beta_1$ (Lemma~\ref{lem:schur}). The definition \eqref{eq:profinfo} remains meaningful when $X$ is rank deficient, in which case it may be zero.

\begin{theorem}[Identification of the spillover coefficient]\label{thm:ident}
Let $N\ge 3$ and let $W$ be row-stochastic with zero diagonal.
\begin{enumerate}[label=(\roman*),leftmargin=*]
\item \textup{(Roughness identity.)} For every $x\in\R^N$,
\[
\calI(\beta_1;x)=\sigma^{-2}\big\|M_Z(I-W)x\big\|^2\;\le\;\sigma^{-2}\big\|(I-W)x\big\|^2 ,
\]
and, if $x\notin\mathrm{span}(\one)$, with $\tilde x=M_1x$,
\[
\calI(\beta_1;x)=\sigma^{-2}\Big\{\|M_1Wx\|^2-\frac{(\tilde x\T Wx)^2}{\|\tilde x\|^2}\Big\}.
\]
\item \textup{(Reflection problem.)} $\calI(\beta_1;x)=0$ for every $x\in\R^N$ if and only if $W=\Wc$. In that case $\Wc x=\frac{N\bar x}{N-1}\one-\frac{1}{N-1}x$ lies in the span of $Z$ for every $x$, so $\beta_1$ is not separable from $(\beta_0,\beta_2)$ in any cross-section.
\item \textup{(Expected information.)} In the Gaussian NSSM, for every $t\ge2$, with $\mu_{t-1}=X_{t-1}\theta_{t-1}$ the conditional mean of the lagged field,
\[
\E\big[\calI(\beta_1;y_{t-1})\,\big|\,\calF_{t-2},\theta_{t-1}\big]\;\le\;\tauW+\sigma^{-2}\|M_1W\mu_{t-1}\|^2 .
\]
If $\mu_{t-1}\in\mathrm{span}(\one)$ \textup{(}a spatially unstructured conditional mean\textup{)}, then also
\[
\E\big[\calI(\beta_1;y_{t-1})\,\big|\,\calF_{t-2},\theta_{t-1}\big]\;\ge\;\tauW-c_N,\qquad
c_N=\tauW\Big(\tfrac{2}{N-1}+\sqrt3\,e^{-(N-1)/32}\Big)+\tfrac{2N}{N-1},
\]
and $c_N\le 4+\frac{4}{N-1}+\sqrt3\,Ne^{-(N-1)/32}$ is bounded uniformly in $N$: $c_N\le26$ for all $N\ge3$, the displayed envelope tends to $4$, and, along graph sequences for which $\tauW/N$ converges, $c_N$ itself tends to a graph-dependent limit $2+2\lim\tauW/N\in[2,4]$ \textup{(}$2$ on the complete graph, $2+2/d$ on $d$-regular graphs\textup{)}.
\item \textup{(Scaling of $\tauW$.)} $\tauW=\sum_{i,j}w_{ij}^2-N^{-1}\|W\T\one\|^2$ for any $W$. For the row-normalised weights $W=D^{-1}A$ of Section~\ref{sec:model}, $\sum_{i,j}w_{ij}^2=\sum_i n_i^{-1}$, so $\tauW\le\sum_i n_i^{-1}$. For a $d$-regular undirected graph $\tauW=N/d-1$. For the uniform complete graph $\tau_{\Wc}=1/(N-1)$, and if all out-degrees exceed $cN$ then $\tauW\le 1/c$.
\end{enumerate}
\end{theorem}

Three remarks make the content of Theorem~\ref{thm:ident} concrete.

\begin{remark}[What identifies a spillover]
Part (i) says that the network coefficient is identified by the \emph{graph roughness} of the lagged field, $(I-W)x$, i.e.\ the vector of differences between each node and the average of its neighbours, after removing whatever part of that roughness is explained by a constant and by the field itself. A lagged field that is smooth along the graph (each node close to its neighbourhood average) carries little information about spillovers no matter how large $N$ is. This is the precise sense in which ``you cannot estimate a spillover from a common shock''. The second expression in (i) shows the two things that destroy information: a smooth field (small $\|M_1Wx\|$) and a field that is strongly spatially autocorrelated in the specific sense that $Wx$ is nearly collinear with $\tilde x$ (large $(\tilde x\T Wx)^2/\|\tilde x\|^2$).
\end{remark}

\begin{remark}[The complete graph and the reflection problem]
Part (ii) is the dynamic counterpart of the non-identification of peer effects in groups where everyone interacts with everyone \citep{manski1993identification,lee2007identification}. The usual argument that lags solve the reflection problem is an argument about simultaneity. It does not help here, because on $\Wc$ the network regressor $\Wc x$ is an exact linear combination of the constant and of the own lag, whatever $x$ is, and this linear dependence has nothing to do with the timing of the regressors. The ``if and only if'' shows that $\Wc$ is the \emph{only} row-stochastic zero-diagonal weight matrix with this property. \citet{bramoulle2009identification} obtain the same degenerate case from the linear dependence of $I$, $G$ and $G^2$ in the static model. Part (iv) then quantifies how close a given graph is to this degenerate case: $\tauW$ falls from $N/d-1$ on sparse regular graphs to $1/(N-1)$ on the complete graph, and dense graphs in between have bounded $\tauW$ however large $N$ is.
\end{remark}

\begin{remark}[Information budget]
Part (iii) describes the information \emph{budget} for the spillover at one time point: the noise in the lagged field contributes $\tauW$ in expectation \emph{before} the nuisance projection, and the projection deficit is the explicit $c_N$, and the conditional mean of the lagged field contributes additional information only through its own graph roughness $\|M_1W\mu_{t-1}\|^2/\sigma^2$. Since $\tauW$ grows like $N$ on sparse graphs and is bounded on dense graphs, the expected information from a spatially unstructured field behaves in the same way: the $N$ observations of a cross-section are, for the purpose of learning the spillover, worth about $\tauW$ effective observations. The lower bound in (iii) is not sharp in its constants but its form is the point: $\tauW$ is an accurate summary, the error term being bounded by an absolute constant. In the simulations of Section~\ref{sec:sim} the realised information exceeds $\tauW$ by $12$--$27\%$ on sparse graphs because the dynamics generate a mildly rough conditional mean.
 The deficit matters. On the complete graph $\tau_{\Wc}=1/(N-1)>0$ while $\calI_t\equiv0$, because the projection removes everything, so $\tauW$ is an upper budget and the realised $\calI_t$ determines what is actually available.
\end{remark}

\begin{remark}[Sparsity alone is not sufficient]\label{rem:hub}
Low degree alone does not put $\tauW$ on the $N/d$ scale, because $\tauW=\sum_{i,j}w_{ij}^2-N^{-1}\|W\T\one\|^2$ is a \emph{degree} functional with an in-degree correction: the directed hub graph with $w_{12}=1$ and $w_{i1}=1$ for $i>1$ has out-degree one at every node yet $\tauW=2-2/N=O(1)$, because the in-degree mass concentrates on one node. Conversely, a dense graph paired with a field that is rough along informative network directions can realise $\calI_t\asymp N$ even though $\E[\calI_t]$ under isotropic noise is bounded, and on \emph{any} graph an eigenfield $Wx\in\mathrm{span}(\one,x)$ realises $\calI_t=0$. Throughout the paper, then, sparse/dense is shorthand for the expected-information budget of parts (iii)--(iv) under the stated homogeneity, and the realised $\calI_t$ (not the topology) is the criterion every downstream result conditions on. Study S1c and the placebo table of Section~\ref{sec:app} make the same point empirically.\end{remark}

The Poisson model of Section~\ref{sec:model} replaces $\sigma^{-2}I$ by the intensity weights $\Lambda_t$. The next result records that everything above transfers, and that the reflection degeneracy in particular is \emph{weight-free}.

\begin{proposition}[Weighted information in the Poisson NSSM]\label{prop:weighted}
For $\Lambda=\diag(\lambda_1,\dots,\lambda_N)$ with all $\lambda_i>0$ and $x\in\R^N$, define the $\Lambda$-weighted profile information, with $Z=[\one,x]$,
\[
\calI^\Lambda(\beta_1;x)\;=\;\min_{a\in\R^2}\sum_i\lambda_i\{(Wx)_i-a_0-a_1x_i\}^2\;=\;\|M_{\Lambda^{1/2}Z}\Lambda^{1/2}Wx\|^2 .
\]
\begin{enumerate}[label=(\roman*),leftmargin=*]
\item $\calI^\Lambda(\beta_1;x)=1/[(X\T\Lambda X)^{-1}]_{22}$ whenever $X\T\Lambda X\succ0$, and $\calI^\Lambda(\beta_1;x)=\|M_{\Lambda^{1/2}Z}\Lambda^{1/2}(I-W)x\|^2$: the roughness identity holds in the $\Lambda$-weighted geometry.
\item $\calI^\Lambda(\beta_1;x)=0$ for every $x$ and one \textup{(}equivalently every\textup{)} $\Lambda\succ0$ if and only if $W=\Wc$: the reflection problem does not depend on the weights.
\item For the Laplace update \eqref{eq:laplace}, $(P_{t|t})_{22}$ obeys the sandwich of Theorem~\ref{thm:filter}\textup{(ii)} with $\calI_t$ replaced by $\calI^{\Lambda_t}(\beta_1;g(y_{t-1}))$ and $u_t^*$ by the $\Lambda_t$-weighted least-squares coefficients of $Wg(y_{t-1})$ on $Z_t$.
\end{enumerate}
\end{proposition}

Part (iii) is a statement about the Laplace approximation itself (its reported variance is bracketed by the weighted information exactly as in the Gaussian case) not about the distance between the approximation and the true posterior, which we do not quantify (a conditional Bernstein--von Mises theorem for this filter is an open problem we return to in Section~\ref{sec:disc}). In the application we report the weighted profile information at the filtered mode.

A corollary explains why spillovers cannot be studied from aggregate series.

\begin{corollary}[Aggregation]\label{cor:agg}
Let $\bar y_t=N^{-1}\one\T y_t$ and $c=W\T\one$ be the vector of in-weights. In the Gaussian NSSM, $\bar y_t=\beta_{0,t}+\beta_{1,t}\,N^{-1}c\T y_{t-1}+\beta_{2,t}\bar y_{t-1}+\bar\varepsilon_t$. If $W$ is doubly stochastic \textup{(}in particular for any regular undirected graph\textup{)} then $c=\one$ and
\[
\bar y_t=\beta_{0,t}+(\beta_{1,t}+\beta_{2,t})\,\bar y_{t-1}+\bar\varepsilon_t ,
\]
so the aggregate series identifies only the sum $\beta_{1,t}+\beta_{2,t}$. In general the aggregate separates $\beta_{1,t}$ from $\beta_{2,t}$ only through the difference $N^{-1}(c-\one)\T y_{t-1}$ between the in-weighted and the unweighted mean of the lagged field.
\end{corollary}

\subsection{Filtering accuracy}\label{sec:filter}

We now turn from a single cross-section to the filter. Write $\calI_t=\calI(\beta_1;y_{t-1})$ and let $u_t^*=Z_t^{+}Wy_{t-1}$ be the least-squares coefficients of $Wy_{t-1}$ on $Z_t=[\one,y_{t-1}]$.

\begin{theorem}[Calibration, variance bounds and cross-sectional consistency]\label{thm:filter}
Consider the Gaussian NSSM \eqref{eq:obs}--\eqref{eq:state} with $Q\succ0$.
\begin{enumerate}[label=(\roman*),leftmargin=*]
\item \textup{(Exactness.)} For every $t$, $\theta_t\mid\calF_t\sim\calN(m_{t|t},P_{t|t})$ with $(m_{t|t},P_{t|t})$ given by \eqref{eq:kf}. Consequently $\E[(\theta_t-m_{t|t})(\theta_t-m_{t|t})\T\mid\calF_t]=P_{t|t}$ and every credible set computed from \eqref{eq:kf} has exactly its nominal coverage probability.
\item \textup{(Two-sided variance bound.)}
\[
\frac{1}{\calI_t+(1+\|u_t^*\|^2)/\lambda_{\min}(Q)}\;\le\;\Var(\beta_{1,t}\mid\calF_t)\;\le\;\frac{1}{\calI_t},
\]
with the convention $1/0=\infty$.
\item \textup{(Cross-sectional consistency.)} $\E\big[\|\theta_t-m_{t|t}\|^2\mid\calF_t\big]=\tr P_{t|t}\le 3\sigma^2/\lambda_{\min}(X_t\T X_t)$ and $\E[(\beta_{1,t}-m_{1,t|t})^2\mid\calF_t]\le1/\calI_t$. Hence, along any sequence of NSSMs indexed by $N$: if $\lambda_{\min}(X_t\T X_t)\to\infty$ in probability then $m_{t|t}-\theta_t\to0$ in probability. If $\calI_t\to\infty$ in probability then $m_{1,t|t}-\beta_{1,t}\to0$ in probability, and if $\calI_t$ and $\|u_t^*\|$ remain bounded in probability then $\Var(\beta_{1,t}\mid\calF_t)$ is bounded away from zero in probability, so the spillover coefficient is not consistently filtered.
\item \textup{(Steady state: the $\sqrt{q/\calI}$ law.)} Let $q_2=Q_{22}$ and $v_t=\Var(\beta_{1,t}\mid\calF_t)$. Then $v_t\le\varphi_{\calI_t}(v_{t-1})$ with $\varphi_{\calI}(v)=\{(v+q_2)^{-1}+\calI\}^{-1}$. Consequently, if $\calI_t\ge\calI>0$ for all $t$, then
\[
\limsup_{t\to\infty}v_t\;\le\;P^\ast(\calI,q_2)\;=\;\frac{\sqrt{q_2^2\calI^2+4q_2\calI}-q_2\calI}{2\calI},
\]
with $\tfrac{2}{1+\sqrt5}\sqrt{q_2/\calI}\le P^\ast\le\sqrt{q_2/\calI}$ whenever $q_2\calI\le1$. The upper bound holds for all $q_2\calI$. Equality holds throughout, and $v_t\to P^\ast(\calI,q_2)$, when the columns of every $X_t$ are orthogonal, $\calI_t\equiv\calI$, and $Q$, $P_0$ are diagonal.
\item \textup{(Complete graph.)} If $W=\Wc$ then $v_t=\big(-\tfrac{N\bar y_{t-1}}{N-1},\,1,\,\tfrac{1}{N-1}\big)\T$ satisfies $X_tv_t=0$ and $P_{t|t}^{-1}v_t=P_{t|t-1}^{-1}v_t$: the cross-section at time $t$ carries no information about $v_t\T\theta_t$, and $\Var(v_t\T\theta_t\mid\calF_t)\ge\lambda_{\min}(Q)\|v_t\|^2\ge\lambda_{\min}(Q)$ for all $t$ and $N$.
\end{enumerate}
\end{theorem}

Part (i) is the reason the NSSM is tractable at all: the Kalman recursions are not an approximation, because the regressors are measurable with respect to the past. This exactness has a consequence that is easy to overlook: the filter's intervals are calibrated \emph{under the model}, including its random-walk prior on the state path. When the data are informative (large $\calI_t$) the prior is irrelevant and the intervals are also calibrated in the frequentist sense for a fixed state path. When the data are uninformative the intervals inherit whatever the prior says, and Section~\ref{sec:sim} shows the resulting frequentist undercoverage on the complete graph.

Part (ii) is the bridge between Theorem~\ref{thm:ident} and estimation: the posterior variance of the spillover is controlled from both sides by the profile information of the \emph{current} cross-section, with a slack term that is the reciprocal of the state-innovation variance. Information does accumulate across periods through $P_{t|t-1}^{-1}$, but in a random-walk state model it accumulates only up to the innovation variance. The lower bound makes this precise and shows that on dense graphs, where $\calI_t$ is bounded, the posterior variance has a floor of order $\lambda_{\min}(Q)/(1+\|u^*_t\|^2)$ that no amount of cross-sectional data removes. The quantity $\|u_t^*\|$ is bounded whenever the regression of $Wy_{t-1}$ on $(\one,y_{t-1})$ is. On $\Wc$ it equals $\|(N\bar y_{t-1}/(N-1),-1/(N-1))\|$.

Part (v) turns the one-period sandwich into a filtering \emph{rate}, and the fixed point $P^\ast=\{\sqrt{q_2^2\calI^2+4q_2\calI}-q_2\calI\}/(2\calI)$ has two regimes that must be kept apart. In the \emph{drift-dominated} regime $q_2\calI\le1$ (the empirically relevant one below, where the Chicago application has $q_2\calI\approx0.02$) Lemma~\ref{lem:riccati}(c) gives $P^\ast\asymp\sqrt{q_2/\calI}$, so the posterior standard deviation is of order $(q_2/\calI)^{1/4}$: a fourth-root, not a square-root, rate in the cross-sectional information. In the \emph{information-dominated} regime $q_2\calI\to\infty$ (fixed $q_2$, diverging $\calI$), instead $P^\ast=\{1+o(1)\}/\calI$ and the standard deviation is the static-parameter rate $\calI^{-1/2}$. The fourth-root law is a statement about the first regime only. On sparse graphs with $\calI\asymp N/d$ the standard deviation is therefore of order $(q_2d/N)^{1/4}$ while $q_2\lesssim d/N$, crossing over to $(d/N)^{1/2}$ beyond. On dense graphs $\calI$ is bounded and the steady-state variance is bounded below, which is the quantitative form of the floor in (ii). Section~\ref{sec:s1} verifies the level and the slope of this law. The bound is an inequality because information can leak between coordinates. It is attained for orthogonal designs, and the simulations show the leakage is small in practice.

Because the conditional variance is the minimal conditional mean squared error, the floor in (ii) binds \emph{every} estimator, not only the filter:

\begin{corollary}[An estimator-independent floor]\label{cor:floor}
In the Gaussian NSSM with $Q\succ0$, every $\calF_t$-measurable estimator $\delta_t$ of $\beta_{1,t}$ satisfies
\[
\E\big[(\delta_t-\beta_{1,t})^2\big]\;\ge\;\E\big[\Var(\beta_{1,t}\mid\calF_t)\big]\;\ge\;\E\Big[\big\{\calI_t+(1+\|u_t^*\|^2)/\lambda_{\min}(Q)\big\}^{-1}\Big].
\]
In particular, if along a sequence of NSSMs there are constants $C_1,C_2$ with $\Prob(\calI_t\le C_1,\|u_t^*\|\le C_2)\ge1-\epsilon$, then $\E[(\delta_t-\beta_{1,t})^2]\ge(1-\epsilon)/\{C_1+(1+C_2^2)/\lambda_{\min}(Q)\}$ for every estimator: on dense graphs no $\calF_t$-measurable method (filtered, penalised or otherwise) attains vanishing mean squared error for the current spillover. A fixed-interval smoother uses $\calF_T$ with $T>t$ and lies outside the corollary, and the smoothing analogue is open.
\end{corollary}

The corollary is a Bayes-risk statement under the model's own state law. Since the minimax risk over any class of state paths is bounded below by the Bayes risk under a prior supported on that class, the same floor constrains minimax estimation of $\beta_{1,t}$. A matching minimax upper bound over an explicit path class is not attempted here and remains open. The failure on dense graphs is the problem's, not the filter's.

Part (iii) gives the honest form of the ``$N\to\infty$ consistency'' of the filtered state that is sometimes asserted for network models: it holds exactly when the smallest eigenvalue of the cross-sectional design diverges, and for the spillover coefficient specifically exactly when $\calI_t$ diverges, which, by Theorem~\ref{thm:ident}(iii)--(iv), is the sparse-graph regime. Part (iv) is the sharp negative statement: on the complete graph the data are silent about an explicit direction in state space, asymptotically the direction $\beta_1-\bar y_{t-1}\beta_0$, i.e.\ the spillover is confounded with the level. Because $\bar y_{t-1}$ changes over time, the unidentified direction rotates slowly and the random-walk prior can trade the two off against each other. This is why a filter on the complete graph returns a non-degenerate but essentially prior-driven path for $\beta_{1,t}$, as in the application below.

\subsection{Misspecified networks and the forecast horizon}\label{sec:plugin}

In practice $W$ is a modelling choice (contiguity, distance bands, mobility flows) and one would like to know how much a wrong choice costs. The answer has two parts, and the first is that some wrong choices cost \emph{nothing}, because they are not distinct models at all.

\begin{proposition}[The complete-graph mixture is a cross-sectional reparameterisation]\label{prop:equiv}
For $\alpha\in[0,1)$ let $W_\alpha=(1-\alpha)W+\alpha\Wc$ and $c_{\alpha,t}=\alpha\beta_1/\{(1-\alpha)(N-1)\}$.
\begin{enumerate}[label=(\roman*),leftmargin=*]
\item For every $x\in\R^N$, $\mathrm{span}\{\one,W_\alpha x,x\}=\mathrm{span}\{\one,Wx,x\}$, and the cross-sectional model \eqref{eq:obs} built from $W$ with coefficients $(\beta_{0,t},\beta_{1,t},\beta_{2,t})$ is \emph{identical}, as a conditional distribution of $y_t$ given $(\calF_{t-1},\theta_t)$, to the model built from $W_\alpha$ with
\[
\tilde\beta_{1,t}=\frac{\beta_{1,t}}{1-\alpha},\qquad
\tilde\beta_{2,t}=\beta_{2,t}+c_{\alpha,t},\qquad
\tilde\beta_{0,t}=\beta_{0,t}-c_{\alpha,t} N\bar y_{t-1}.
\]
\item The map is bijective and $\calF_{t-1}$-measurable, so no collection of cross-sections separates $W$ from $W_\alpha$. They are separated only by the state law \eqref{eq:state}, because the matching intercept path must track $\bar y_{t-1}$, which a random walk does not do. A filtered spillover computed on $W_\alpha$ therefore estimates $\beta_{1,t}/(1-\alpha)$: the coefficient inflates by $(1-\alpha)^{-1}$ while the one-step fit is unchanged, and as $\alpha\to1$ the family degenerates to the non-identified point of Theorem~\ref{thm:ident}\textup{(ii)}.
\item Multi-step forecasts do separate the family. With the origin-matched reparameterisation of \textup{(i)}, the plug-in means \eqref{eq:plugin} satisfy $\tilde\mu_1(y)=\mu_1(y)$ and
\[
\tilde\mu_2(y)-\mu_2(y)\;=\;c_{\alpha,t} N\{\bar\mu_1(y)-\bar y\}\,\one :
\]
the mixture is invisible at one step and reappears at two steps as a level shift proportional to the predicted drift of the mean level.
\end{enumerate}
\end{proposition}

The equivalence is a \emph{cross-sectional} statement, conditional on the lagged field: the matching intercept path must track $\bar y_{t-1}$, so members of the family carry different state laws, and restricted coefficient paths (or several cross-sections tied together by the random walk) can separate them. The proposition therefore bounds what any one cross-section can distinguish.
The proposition says that the pair $(\beta_1,W)$ is, from any single cross-section, identified only up to the one-parameter family $\{W_\alpha\}$, a \emph{flat direction} in network space whose endpoint is the complete graph. This has two practical faces. Negatively, robustness checks that mix the assumed network toward $\Wc$ (a common form of ``perturbing $W$'') test nothing about the fit and merely rescale the coefficient. The honest perturbations are the ones that change the span, such as relabellings and rewirings. Positively, the coefficient map is a sharp quantitative prediction (the fitted spillover under $W_\alpha$ should be $\hat\beta_1/(1-\alpha)$) which the application confirms to within a few percent (Section~\ref{sec:placebo}). The second part of the answer concerns networks that genuinely differ, and is a matter of degree measured by the plug-in means \eqref{eq:plugin}.

\begin{proposition}[Plug-in forecasts under a misspecified network]\label{prop:plugin}
Let $W$ and $\widetilde W$ be row-stochastic, fix $\theta=(\beta_0,\beta_1,\beta_2)\T$, let $B=\beta_1W+\beta_2I$, $\widetilde B=\beta_1\widetilde W+\beta_2I$, $\Delta=\|W-\widetilde W\|$ and $\bar\rho=\max\{\|B\|,\|\widetilde B\|\}$. For $y\in\R^N$ let $\mu_h(y)$ and $\tilde\mu_h(y)$ be the $h$-step plug-in means \eqref{eq:plugin} computed with $W$ and $\widetilde W$ from the same origin $y$.
\begin{enumerate}[label=(\roman*),leftmargin=*]
\item $\mu_h(y)-\tilde\mu_h(y)=(B^h-\widetilde B^h)\,M_1y$, and
$\;\|\mu_h(y)-\tilde\mu_h(y)\|\le|\beta_1|\,\Delta\,h\,\bar\rho^{\,h-1}\,\|M_1y\|$.
\item If $\bar\rho<1$ then $\sup_{h\ge1}\|\mu_h(y)-\tilde\mu_h(y)\|\le|\beta_1|\Delta\|M_1y\|/(1-\bar\rho)^2$ and the discrepancy tends to zero geometrically as $h\to\infty$.
\item $\rho(B)\ge|\beta_1+\beta_2|$. If $W$ and $\widetilde W$ have a common real right eigenvector $v$, $\|v\|=1$, with real eigenvalues $\lambda\ne\tilde\lambda$ \textup{(}for undirected graphs all eigenvalues of $W$ are real\textup{)}, then for $y=\bar y\one+cv$ with $\one\T v=0$, $\|\mu_h(y)-\tilde\mu_h(y)\|=|c|\,|r^h-\tilde r^h|$ with $r=\beta_1\lambda+\beta_2$, $\tilde r=\beta_1\tilde\lambda+\beta_2$. If $r\tilde r>0$ this is at least $|c|\,h\,\min(|r|,|\tilde r|)^{h-1}|\beta_1||\lambda-\tilde\lambda|$, while $\Delta\ge|\lambda-\tilde\lambda|$. So when $\min(|r|,|\tilde r|)>1$ the discrepancy grows geometrically and the $h$-dependence of the bound in (i) cannot be improved. (This is sharpness of the \emph{rate} in $h$, not a lower bound proportional to $\Delta$, since $\Delta\ge|\beta_1||\lambda-\tilde\lambda|$ always. Note also that $(W-\widetilde W)\one=0$ for any two row-stochastic matrices: the constant mode never carries misspecification, and amplification requires a differing persistent mode orthogonal to $\one$, which is the common-eigenvector hypothesis here.)
\end{enumerate}
\end{proposition}

The first statement has a simple reading: misspecifying the network costs nothing for the spatially constant part of the field and costs, at most, something linear in the operator-norm error for its rough part. Together with Proposition~\ref{prop:equiv} this explains the two ends of the placebo spectrum in Section~\ref{sec:app}: mixtures toward the complete graph forecast almost exactly as well as the true network at one step because they \emph{are} the true network reparameterised, whereas a degree-preserving rewiring, which changes the span and destroys the relation between $W$ and the field's roughness, forecasts no better than ignoring the network. The second and third statements concern the horizon. Since $\rho(B)\ge|\beta_1+\beta_2|$, a persistence $\beta_1+\beta_2$ at or above one makes the propagation operator non-contractive and the network error is then amplified with the horizon. In the Chicago application the filtered persistence on the log scale exceeds one after 2013 (Figure~\ref{fig:paths}), and the log-score advantage of the true network over the alternatives disappears beyond two months, with wide confidence intervals, a pattern consistent with (iii), although for the nonlinear count model the proposition is only a heuristic.

\subsection{Predictive moments of the Poisson NSSM}\label{sec:poisson}

For count data the natural analogue of the linear spillover model is the log-linear specification \eqref{eq:pois}. Whether one feeds back the raw counts or their $\log(1+\cdot)$ transformation looks like a detail. It is not: it decides whether multi-step forecasts exist.

\begin{theorem}[Predictive moments under raw-count and log-count feedback]\label{thm:poisson}
Consider the Poisson NSSM \eqref{eq:pois} on an undirected network \textup{(}$w_{ij}>0\Leftrightarrow w_{ji}>0$\textup{)}, a forecast origin $t$ and a node $i$. Write $s_{ij}=\beta_1w_{ij}+\beta_2\mathbf 1\{j=i\}$.
\begin{enumerate}[label=(\alph*),leftmargin=*]
\item \textup{(Raw scale, fixed coefficients.)} Let $g(y)=y$ and $\theta_{t+k}=\theta=(\beta_0,\beta_1,\beta_2)\T$ for all $k\ge1$. Then $\E[y_{i,t+1}\mid\calF_t]=\lambda_{i,t+1}$ and
\begin{equation}\label{eq:twostep}
\E[y_{i,t+2}\mid\calF_t]=\exp(\beta_0)\,\exp\Big\{\sum_{j}\lambda_{j,t+1}\big(e^{s_{ij}}-1\big)\Big\}<\infty .
\end{equation}
If $\beta_2>0$, or if $\beta_1>0$ and node $i$ has at least one neighbour, then $\E[y_{i,t+h}\mid\calF_t]=+\infty$ for every $h\ge3$. If $\beta_1\le0$ and $\beta_2\le0$ then $\E[y_{i,t+h}\mid\calF_t]\le e^{\beta_0}$ for all $h\ge1$.
\item \textup{(Raw scale, random-walk coefficients.)} Let $g(y)=y$, let $\theta_{t+1}$ given $\calF_t$ have an arbitrary distribution \textup{(}a point mass or the filter's Gaussian\textup{)}, and let $\theta_{t+2}=\theta_{t+1}+u_{t+2}$ with $u_{t+2}\sim\calN(0,Q)$, $Q\succ0$, independent of $(\theta_{t+1},y_{t+1})$. Then $\E[y_{i,t+2}\mid\calF_t]=+\infty$ for every node $i$, whereas $\E[y_{i,t+1}\mid\calF_t]<\infty$ whenever $\theta_{t+1}\mid\calF_t$ is Gaussian.
\item \textup{(Log scale, fixed coefficients.)} Let $g(y)=\log(1+y)$ and let $(\theta_{t+k})_{k\ge1}$ be any fixed sequence with $\rho_{t+k}=|\beta_{1,t+k}|+|\beta_{2,t+k}|$. Then $\lambda_{i,t+k}\le e^{|\beta_{0,t+k}|}(1+\|y_{t+k-1}\|_\infty)^{\rho_{t+k}}$ and $\E[(1+y_{i,t+h})^p\mid\calF_t]<\infty$ for all $p\ge1$, $h\ge1$ and $i$: all predictive moments are finite at every horizon.
\item \textup{(Log scale, coefficient uncertainty, two steps.)} Let $g(y)=\log(1+y)$, $\theta_{t+1}\mid\calF_t\sim\calN(m,P)$, $\theta_{t+2}=\theta_{t+1}+u_{t+2}$ with $u_{t+2}\sim\calN(0,Q)$ independent, $\bar q=\lambda_{\max}(Q)\le3/8$, $L_t=\log(1+\|y_t\|_\infty)$, $c_t=\sqrt3(1+L_t)$ and $\kappa_t=\max\{6,\,\sqrt3c_t+2\bar qc_t^2\}$. If $2\kappa_t\lambda_{\max}(P)<1$ then $\E[y_{i,t+2}\mid\calF_t]<\infty$ for every $i$.
\item \textup{(Log scale, coefficient uncertainty, three or more steps.)} Let $g(y)=\log(1+y)$, let $\theta_{t+1}\mid\calF_t\sim\calN(m,P)$ with $\lambda_{\min}(P)>0$, and $\theta_{t+k}=\theta_{t+k-1}+u_{t+k}$, $u_{t+k}\sim\calN(0,Q)$ independent, with $Q\succeq0$ arbitrary \textup{(}$Q=0$, a persistent posterior draw, included\textup{)}. If node $i$ or one of its neighbours has a positive count at the origin, i.e.\ $\gamma_i:=\ell_{t,i}+(W\ell_t)_i>0$ with $\ell_t=\log(1+y_t)$, then
\[
\E[y_{i,t+h}\mid\calF_t]=+\infty\qquad\text{for every }h\ge3 .
\]
As a separate sharpness statement (with a degenerate covariance, outside the positivity assumption of \textup{(e)}, and proved by direct computation independent of the lemma behind \textup{(d)}) the concentration condition in \textup{(d)} cannot be dropped: with $\beta_0=\beta_1=0$ fixed, $\beta_2\sim\calN(0,s^2)$ persistent \textup{(}$P=s^2e_3e_3\T$, $Q=0$\textup{)} and $\ell_{t,i}=L_0>0$, already the \emph{two}-step mean is infinite whenever $2s^2L_0>1$, while \textup{(d)} guarantees finiteness when $2\kappa_ts^2<1$. The true threshold for $s^2$ thus lies in the bracket $[1/(2\kappa_t),1/(2L_0)]$. The ratio $\kappa_t/L_0$ of its endpoints is bounded when $L_0$ is bounded away from zero and diverges as $L_0\to0$. The result establishes the bracket. Neither endpoint is claimed as the threshold.
\end{enumerate}
\end{theorem}

The mechanism behind (a) and (b) is the interaction of an exponential link with Poisson tails. One step ahead the intensity is fixed. Two steps ahead the intensity is the exponential of a Poisson variable, whose expectation is the moment generating function $\exp\{\lambda(e^{s}-1)\}$ in \eqref{eq:twostep}, finite, but \emph{double-exponential} in the current intensity, so that a handful of active neighbours with $\lambda_{j,t+1}$ of order ten and $s_{ij}$ of order one already multiplies the predictive mean by $\exp\{10(e-1)\}\approx e^{17}$. Three steps ahead the intensity is the exponential of a Poisson variable whose own intensity is the exponential of a Poisson variable, and the expectation of a double exponential of a Poisson variable is infinite: the terms $\exp\{ce^{\beta_2 m}\}$ grow faster than $m!$. Random-walk coefficient innovations bring the divergence forward by one step, because the Gaussian moment generating function contributes $\exp\{\tfrac12 q\,y_{i,t+1}^2\}$ with $y_{i,t+1}$ Poisson, and $e^{qm^2/2}$ also grows faster than $m!$. On the $\log(1+y)$ scale, by contrast, the intensity is bounded by a power of the previous counts, Poisson variables have all polynomial moments, and (c) follows by induction. But (c) is a statement about \emph{fixed} coefficient paths, and simulation-based forecasting does not use fixed paths: it draws $\theta_{t+1}$ from a Gaussian and propagates. Parts (d) and (e) give sufficient conditions on the two sides of the frontier, (d) for finiteness and (e) for divergence, and the region between them at $h=2$ is open. At $h=2$ the battle is quadratic against quadratic (the payoff of a large coefficient draw $b$ grows like $e^{b^2L}$ while its probability decays like $e^{-b^2/(2\lambda_{\max}(P))}$) so finiteness is a genuine \emph{concentration condition}, which (d) makes checkable and whose failure region the sharpness statement after (e) delimits from the other side. At $h\ge3$ the payoff is $e^{cb^3}$ against the same Gaussian cost, and the cubic always wins: with any nondegenerate posterior the $\log(1+y)$ predictive mean is infinite from three steps on, whatever the data, however small $Q$. The transform yields at most one additional horizon of existence under Gaussian coefficient uncertainty, exactly one when the concentration condition of (d) holds, and it repairs the process only along fixed (plug-in) or otherwise bounded coefficient paths.

Several practical consequences follow. First, the Monte Carlo predictive mean from the raw-count model at horizon $h\ge3$ is an average of $S$ draws from a distribution without a mean: it does not converge as $S\to\infty$, and for any finite $S$ it is dominated by the largest draw. Second, this failure is \emph{silent}: the median, the quantiles and the log score of the Monte Carlo predictive distribution are all well defined and look reasonable, so a report of median forecasts or interval forecasts would not reveal that the point forecasts are meaningless. Third, capping the linear predictor (a common numerical fix, which our own simulation code applies and discloses) does not repair the model. It replaces an infinite mean by an arbitrary cap-dependent one, so a capped simulation \emph{illustrates} the instability but only the theorem establishes the divergence. Fourth, by (e) the $\log(1+y)$ model under sampled coefficients has the same disease from $h=3$ on, in a quieter form: the divergent contribution comes from coefficient draws of size $b$ with payoff $e^{cb^3}$ and probability $e^{-b^2/2\lambda_{\max}(P)}$, so for a concentrated posterior no feasible number of simulated paths ever produces an exploding draw, the raw-count divergence at least announces itself in an $S$-scan, the log-scale one never does. The operational rule that follows: at any horizon, medians, quantiles, probability integral transforms and log scores of the simulated predictive distribution are well-defined functionals and may be reported. \emph{means} of sampled-coefficient forecasts exist only at $h\le2$ (at $h=2$ under the condition of (d)). Point forecasts beyond that should be predictive medians or means computed along the frozen filtered coefficient path, which (c) covers. Fifth, on the log scale $\beta_1+\beta_2$ plays the role of a log-scale persistence whose proximity to one governs the horizon amplification of Proposition~\ref{prop:plugin}.

The final preferred model in Section~\ref{sec:app} replaces the Poisson layer by a negative binomial, and one might hope that overdispersion also repairs the moment pathology. It does not. The frontier is driven by the exponential link and the Gaussian coefficient tail, not by the observation layer.

\begin{proposition}[The negative-binomial layer does not move the frontier]\label{prop:nb}
Replace the Poisson observation layer by the negative binomial with mean $\lambda$ and dispersion $r\in(0,\infty)$ \textup{(}variance $\lambda+\lambda^2/r$\textup{)}, keeping the log-linear intensity and the coefficient laws of Theorem~\ref{thm:poisson}. Then, for every $r$: \textup{(i)} under the conditions of Theorem~\ref{thm:poisson}(e), the sampled-coefficient predictive mean on the $\log(1+y)$ scale is infinite for every $h\ge3$. \textup{(ii)} in the scalar construction following Theorem~\ref{thm:poisson}(e), the two-step mean is infinite when $2s^2L_0>1$ and finite when $2s^2L_0<1$, so for the negative binomial the two-step frontier is exactly $2s^2L_0=1$, for every $r$. Overdispersion changes the calibration of the predictive distribution and leaves the existence of its moments to the link and the coefficient tail, where at two steps it sharpens the Poisson bracket to an exact threshold.
\end{proposition}

\section{Simulation study}\label{sec:sim}

The simulations exercise each theoretical result in turn. All three studies together run in about \RuntimeSim{} minutes on a laptop.

\subsection{Graph geometry, information and the filter (Theorems~\ref{thm:ident}--\ref{thm:filter})}\label{sec:s1}

\paragraph{Design.} We generate undirected Erd\H{o}s--R\'enyi graphs with $N\in\{25,50,100,200,400,800\}$ nodes from three families: \emph{sparse}, expected degree $4$. \emph{dense}, expected degree $N/4$, and the \emph{complete} graph. Isolated nodes are attached to a random node. Data follow the Gaussian NSSM \eqref{eq:obs} with $\sigma=1$, $T=60$, $y_0\sim\calN(0,I)$, a random-walk intercept with innovation standard deviation $0.02$, a constant own-lag coefficient $\beta_2=0.30$, and a spillover coefficient that \emph{jumps} from $0.30$ to $0.60$ at $t=30$. The jump is deliberately not a draw from the random-walk prior. It tests whether the filter tracks a structural change and whether its intervals are reliable away from the change. The filter uses $Q=\diag(10^{-4},10^{-3},10^{-3})$, $m_0=0$, $P_0=I$ and the true $\sigma^2$. For each of $R=100$ replications we record $\tauW$. The time average over $t\ge 10$ of the profile information $\calI_t$ and of $\lambda_{\min}(X_t\T X_t)$. The mean squared error of the filtered $\beta_{1,t}$ and the average posterior standard deviation, both excluding the six periods after the jump, and the empirical coverage of the $95\%$ intervals, with and without the post-jump window.

\begin{table}[t]
\centering\footnotesize\setlength{\tabcolsep}{4pt}
\caption{Study S1: graph functional $\tauW$, realised profile information $\calI_t$ for $\beta_{1,t}$, smallest design eigenvalue, and accuracy of the Kalman filter for $\beta_{1,t}$ (MSE, average posterior s.d., coverage of 95\% intervals excluding and including the six periods after the jump). Averages over $R=100$ replications and $t\ge10$. Monte Carlo standard errors of the MSE column are below $4.9\times10^{-3}$ in the units shown and below $7\%$ of each entry.}\label{tab:s1}
\begin{tabular}{llrrrrrrr}
\toprule
graph & $N$ & $\tauW$ & $\bar{\calI}_t$ & $\lambda_{\min}(X_t\T X_t)$ & MSE$\times10^3$ & post.\ s.d. & cov.\ (excl.) & cov.\ (incl.)\\
\midrule
sparse ($d=4$) & 25 & 6.87 & 7.70 & 5.22 & 7.67 & 0.095 & 0.959 & 0.899 \\
 & 50 & 15.36 & 18.56 & 13.47 & 4.78 & 0.077 & 0.969 & 0.899 \\
 & 100 & 32.06 & 39.52 & 29.31 & 2.87 & 0.063 & 0.977 & 0.916 \\
 & 200 & 65.85 & 82.62 & 63.28 & 1.76 & 0.052 & 0.983 & 0.927 \\
 & 400 & 135.03 & 171.14 & 133.12 & 1.14 & 0.042 & 0.984 & 0.933 \\
 & 800 & 270.56 & 342.68 & 256.58 & 0.84 & 0.034 & 0.975 & 0.936 \\
\addlinespace
dense ($d=N/4$) & 25 & 3.48 & 3.75 & 2.74 & 12.15 & 0.110 & 0.951 & 0.891 \\
 & 50 & 3.20 & 3.47 & 2.54 & 11.61 & 0.106 & 0.951 & 0.914 \\
 & 100 & 3.09 & 3.37 & 2.65 & 11.41 & 0.105 & 0.941 & 0.900 \\
 & 200 & 3.05 & 3.36 & 2.61 & 10.77 & 0.099 & 0.934 & 0.897 \\
 & 400 & 3.02 & 3.32 & 2.63 & 10.88 & 0.096 & 0.921 & 0.887 \\
 & 800 & 3.01 & 3.31 & 2.69 & 10.88 & 0.094 & 0.913 & 0.882 \\
\addlinespace
complete & 25 & 0.042 & $<10^{-10}$ & $<10^{-10}$ & 50.16 & 0.172 & 0.910 & 0.874 \\
 & 50 & 0.020 & $<10^{-10}$ & $<10^{-10}$ & 32.92 & 0.160 & 0.947 & 0.925 \\
 & 100 & 0.010 & $<10^{-10}$ & $<10^{-10}$ & 34.30 & 0.155 & 0.933 & 0.911 \\
 & 200 & 0.005 & $<10^{-10}$ & $<10^{-10}$ & 28.51 & 0.144 & 0.918 & 0.900 \\
 & 400 & 0.003 & $<10^{-10}$ & $<10^{-10}$ & 31.17 & 0.134 & 0.864 & 0.855 \\
 & 800 & 0.001 & $<10^{-10}$ & $<10^{-10}$ & 25.28 & 0.126 & 0.867 & 0.865 \\
 
\bottomrule
\end{tabular}
\end{table}

\begin{figure}[t]
\centering
\includegraphics[width=\textwidth]{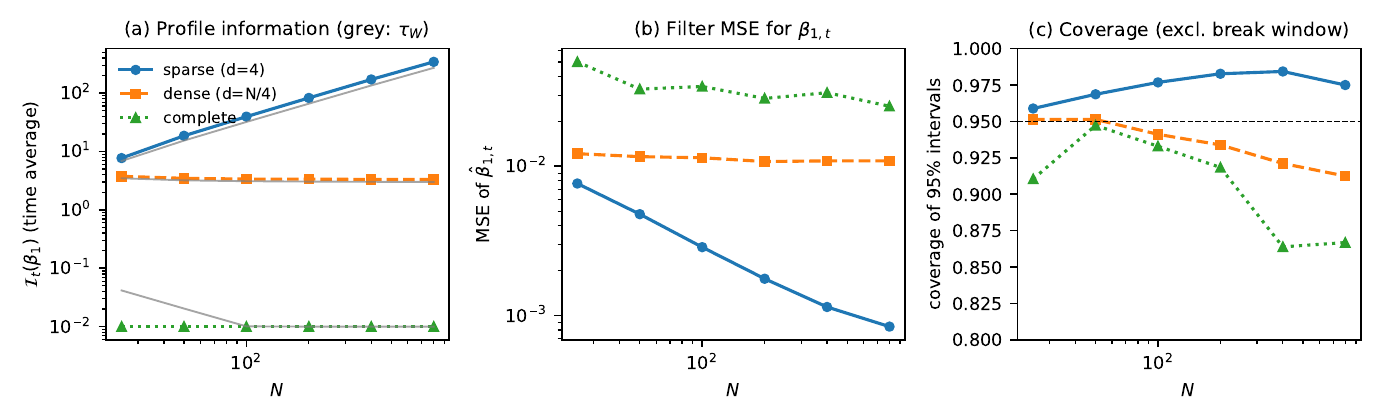}
\caption{Study S1. (a) Time-averaged profile information for $\beta_{1,t}$ against $N$ for the three graph families, with $\tauW$ in grey. Both axes logarithmic. (b) Mean squared error of the filtered spillover coefficient. (c) Empirical coverage of the filter's 95\% intervals excluding the six periods after the jump. In panel (a), values below $10^{-2}$ (the complete graph, zero to machine precision (Table~\ref{tab:s1})) are floored at $10^{-2}$ for log-scale display.}\label{fig:s1}
\end{figure}

\paragraph{Results.} Table~\ref{tab:s1} and Figure~\ref{fig:s1} show three regimes that follow Theorem~\ref{thm:ident}(iii)--(iv) closely. On sparse graphs $\tauW$ grows linearly in $N$ ($6.9$ at $N=25$ to $271$ at $N=800$). The exact comparator is $\sum_in_i^{-1}$ of part (iv), which for Poisson($4$) degrees exceeds $N/4$ by about a third by Jensen's inequality, and $N/4-1$ is the value for the $4$-regular graph and the realised information tracks it, exceeding $\tauW$ by $12$--$27\%$ because the autoregressive dynamics generate a mildly rough conditional mean. On dense graphs $\tauW$ and the information are flat at about $3.0$--$3.8$ regardless of $N$: thirty-two times more nodes buy no information about the spillover. On the complete graph $\tauW=1/(N-1)$, the information is zero to machine precision and $\lambda_{\min}(X_t\T X_t)=0$: the design is exactly singular, as Theorem~\ref{thm:ident}(ii) says.

The filter follows the information, and it follows the steady-state law of Theorem~\ref{thm:filter}(v) quantitatively. On sparse graphs the average posterior \emph{variance} of $\beta_{1,t}$ (the object the theorem bounds) has log-log slope $-0.54$ on $\bar\calI_t$ over the six sparse rows, against $-0.56$ for the fixed point $P^\ast(\bar\calI_t,q_2)$ itself over the same range (the pure $-\tfrac12$ is the $q_2\calI\to0$ asymptote), and sits at $82$--$90\%$ of the envelope with $q_2=10^{-3}$: inside it, as the inequality requires, and tracking it, as the near-orthogonality of sparse designs predicts. The realised MSE of the point estimate falls by an order of magnitude across the sparse rows of Table~\ref{tab:s1}, and it lies a further $15$--$36\%$ below the posterior variance, though not monotonically in $N$, consistent with the strict inequality in Theorem~\ref{thm:filter}(ii): S1's truth is piecewise constant instead of a random-walk draw, an easier target than the one the filter prices, which is also why the coverage in Table~\ref{tab:s1} is mildly conservative. On dense graphs the MSE is flat at $1.1\times10^{-2}$ for every $N$, the variance floor of Theorem~\ref{thm:filter}(ii)--(iii). On the complete graph the MSE is $2.5$--$5.0\times10^{-2}$ and the posterior standard deviation ($0.126$--$0.172$) barely shrinks below the prior: the filter returns a path for $\beta_{1,t}$, but it is the prior's path.

\paragraph{S1d: coverage under correct specification.} Study S1's jump deliberately violates the random-walk prior, so it cannot verify Theorem~\ref{thm:filter}(i). A companion run draws the state path from the random walk itself and hands the filter the true $(Q,\sigma^2)$: over $200$ replications the $95\%$ intervals for $\beta_{1,t}$ cover at rate $0.950$ (Monte Carlo standard error $0.003$), and the terminal filtered variance sits at $0.83$ of the steady-state value $P^\ast(\bar\calI)$ evaluated at the average information, below one, as the upper-bound direction of Theorem~\ref{thm:filter}(v) requires. Exactness is therefore verified in the regime where it is claimed. Study S1 shows the consequences outside that regime.

Coverage tells the same story from the other side. Excluding the post-jump window, the sparse-graph intervals cover $96$--$98\%$ of the time (slightly conservative because the filter's $Q$ allows the true $\beta_{1,t}$ to move every period whereas the truth is piecewise constant) and including the window they cover $90$--$94\%$, the shortfall being the few periods needed to absorb the jump. On the complete graph coverage decays with $N$ (from $0.91$--$0.95$ at $N\le100$ to $0.86$--$0.87$ at $N\ge400$) even away from the jump, and the intervals are governed by the prior. This is not a contradiction of the exact calibration in Theorem~\ref{thm:filter}(i): that result holds when the state path is a draw from the random-walk prior, and on the complete graph the prior is all the filter has, so its intervals are only as good as the prior's description of a deterministic step function.

\subsection{Network misspecification and the horizon (Proposition~\ref{prop:plugin})}\label{sec:s2}

\paragraph{Design.} On one sparse graph with $N=100$ we simulate the Gaussian NSSM with constant coefficients in two regimes: \emph{stable}, $(\beta_1,\beta_2)=(0.30,0.35)$, for which $\|B\|=0.689$ and $\rho(B)=0.650$, and \emph{near-unit}, $(\beta_1,\beta_2)=(0.55,0.50)$, for which $\|B\|=1.132$ and $\rho(B)=1.050$. The misspecified network is a mixture $\widetilde W=(1-\alpha)W+\alpha\Wc$ with $\alpha\in\{0,0.05,\dots,0.5\}$, so that $\Delta=\|W-\widetilde W\|$ increases linearly from $0$ to $0.75$. For every origin $s\in\{20,\dots,T-h\}$ we compute the plug-in means \eqref{eq:plugin} under $W$ and $\widetilde W$ and report the root-mean-square discrepancy across nodes, averaged over origins, for $h\in\{1,2,4,8\}$.

\begin{figure}[t]
\centering
\includegraphics[width=0.85\textwidth]{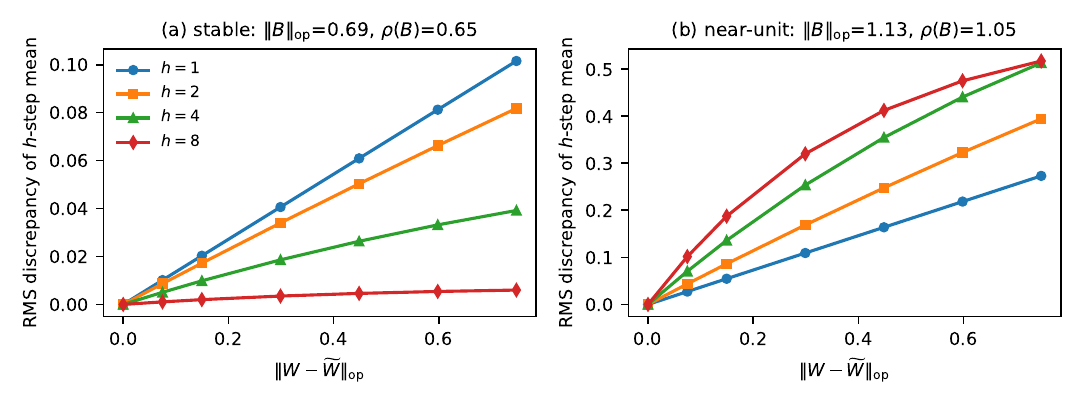}
\caption{Study S2: root-mean-square discrepancy between the $h$-step plug-in means under the true network $W$ and the misspecified network $\widetilde W=(1-\alpha)W+\alpha\Wc$, as a function of $\Delta=\|W-\widetilde W\|$. (a) Stable regime, $\rho(B)<1$: the discrepancy is linear in $\Delta$ and decays with $h$. (b) Near-unit regime, $\rho(B)>1$: linear in $\Delta$ and amplified with $h$. Numerical values are in Table~\ref{tab:s2}.}\label{fig:s2}
\end{figure}

\paragraph{Results.} Figure~\ref{fig:s2} displays the two halves of Proposition~\ref{prop:plugin}. In both regimes the discrepancy is linear in $\Delta$ to plotting accuracy. Exact linearity is a feature of the one-parameter mixture family (the map $\alpha\mapsto\widetilde B$ is affine, cf.\ Proposition~\ref{prop:equiv}), while statement (i) guarantees only the linear \emph{bound} for general pairs of networks. In the stable regime it decays with the horizon (at $\Delta=0.75$ from $0.10$ at $h=1$ to $0.006$ at $h=8$) because the propagation operator is a contraction and the network error is forgotten (statement (ii)). In the near-unit regime the ordering reverses: at $\Delta=0.75$ the discrepancy grows from $0.27$ at $h=1$ to $0.52$ at $h=8$, and at $\Delta=0.30$ from $0.11$ to $0.32$, the geometric amplification of statement (iii). The practical lesson is that the cost of a wrong network is a short-horizon cost when the system is stable and a long-horizon cost when it is persistent. Section~\ref{sec:app} finds the Chicago system in the second category.

\subsection{Raw-count versus log-count feedback (Theorem~\ref{thm:poisson})}\label{sec:s3}

\paragraph{Design.} On a graph with $N=10$ nodes and expected degree $3$ we simulate the Poisson NSSM forward from $y_t=3\cdot\one$ with coefficients $\theta=(-0.30,0.25,0.20)\T$, chosen so that the one-step intensity is about $2.9$ and the coefficients are of the same sign pattern and order as the raw-scale Chicago estimates of Section~\ref{sec:rawlog}. We compare raw-count and $\log(1+y)$ feedback, each with fixed coefficients and with random-walk coefficients ($Q=0.01\,I$), and for $S\in\{10^3,10^4,10^5,10^6\}$ simulated paths record the Monte Carlo mean of the $h$-step intensity averaged over nodes, the median across paths of the node-averaged intensity, and the fraction of paths in which some node reaches an intensity above $10^6$, for $h=1,\dots,5$.

\begin{table}[t]
\centering\footnotesize
\caption{Study S3: Monte Carlo estimates of the $h$-step predictive intensity (average over nodes) from $S$ simulated paths, for raw-count and $\log(1+y)$ feedback with fixed and random-walk coefficients. For the raw-count rows the entry is ``mean (median across paths)''. The last column is the fraction of paths in which some node's intensity exceeds $10^6$ at $h=4$. Entries marked div.\ lie beyond the reporting threshold of $10^{6}$: such sample means are governed by a few paths at or beyond the $10^{12}$ intensity cap or the floating-point range, and Theorem~\ref{thm:poisson} gives infinite population values at these entries. Medians, in parentheses, and the exploding-path fraction remain well defined.}\label{tab:s3}
\begin{tabular}{lcccccc}
\toprule
$S$ & $h=1$ & $h=2$ & $h=3$ & $h=4$ & $h=5$ & tail$_4$\\
\midrule
\multicolumn{7}{l}{\emph{raw counts, fixed $\theta$}}\\
$10^{3}$ & 2.86 (2.86) & 2.94 (2.80) & 4.16 (2.91) & div. (3.15) & div. (3.90) & 0.028 \\
$10^{4}$ & 2.86 (2.86) & 2.97 (2.83) & 4.23 (2.90) & div. (3.13) & div. (3.62) & 0.023 \\
$10^{5}$ & 2.86 (2.86) & 2.95 (2.82) & 57.18 (2.89) & div. (3.11) & div. (3.64) & 0.022 \\
$10^{6}$ & 2.86 (2.86) & 2.95 (2.82) & 331 (2.89) & div. (3.10) & div. (3.61) & 0.023 \\
\addlinespace
\multicolumn{7}{l}{\emph{raw counts, random-walk $\theta$ ($q=0.01$)}}\\
$10^{3}$ & 2.86 (2.86) & 3.44 (2.74) & div. (2.95) & div. (3.21) & div. (5.57) & 0.238 \\
$10^{4}$ & 2.86 (2.86) & 3.45 (2.81) & div. (2.92) & div. (3.54) & div. (6.43) & 0.239 \\
$10^{5}$ & 2.86 (2.86) & 3.41 (2.78) & div. (2.86) & div. (3.36) & div. (5.72) & 0.233 \\
$10^{6}$ & 2.86 (2.86) & 3.41 (2.77) & div. (2.84) & div. (3.27) & div. (5.31) & 0.233 \\
\addlinespace
\multicolumn{7}{l}{\emph{$\log(1+y)$, fixed $\theta$}}\\
$10^{3}$ & 1.382 & 1.042 & 0.977 & 0.964 & 0.960 & 0.000 \\
$10^{4}$ & 1.382 & 1.044 & 0.976 & 0.961 & 0.958 & 0.000 \\
$10^{5}$ & 1.382 & 1.044 & 0.975 & 0.961 & 0.958 & 0.000 \\
$10^{6}$ & 1.382 & 1.044 & 0.975 & 0.961 & 0.958 & 0.000 \\
\addlinespace
\multicolumn{7}{l}{\emph{$\log(1+y)$, random-walk $\theta$ ($q=0.01$)}}\\
$10^{3}$ & 1.382 & 1.048 & 1.004 & 1.029 & 1.072 & 0.000 \\
$10^{4}$ & 1.382 & 1.058 & 1.012 & 1.028 & 1.060 & 0.000 \\
$10^{5}$ & 1.382 & 1.058 & 1.011 & 1.024 & 1.055 & 0.000 \\
$10^{6}$ & 1.382 & 1.057 & 1.011 & 1.023 & 1.055 & 0.000 \\
 
\bottomrule
\end{tabular}
\end{table}

\paragraph{Results.} Table~\ref{tab:s3} shows the behaviour that Theorem~\ref{thm:poisson} predicts and that a practitioner would not notice without looking for it. With raw-count feedback and fixed coefficients the one- and two-step means are stable in $S$ ($2.86$ and $2.95$), in agreement with \eqref{eq:twostep}. At $h=3$ the Monte Carlo mean is $4.2$ with $10^3$ or $10^4$ paths, $57$ with $10^5$ and $331$ with $10^6$: it grows with $S$ because it is estimating an infinite quantity (Theorem~\ref{thm:poisson}(a)), and it does so while the median across paths sits at $2.9$ throughout. At $h=4$ the mean is astronomically large for every $S$ and at $h=5$ it overflows. The fraction of paths that explode is $2$--$3\%$ at $h=4$ and the median is still $3.1$. The median, the quantiles and the log score of such a predictive distribution are all perfectly reasonable numbers. Only the mean is meaningless, and it is the mean that is usually reported.

Random-walk coefficients make matters worse at every horizon, as in Theorem~\ref{thm:poisson}(b): at $h=3$ the mean is between $10^{27}$ and $10^{223}$ depending on $S$, and a quarter of the paths explode by $h=4$. At $h=2$ the Monte Carlo mean ($3.4$) looks stable even though the population mean is infinite. The divergent contribution comes from paths in which a node records several hundred events at $t+1$, an event of probability below $10^{-300}$ here, and no feasible $S$ would reveal it. Here the divergence leaves no numerical trace, and the theorem is the only available diagnostic. Two disclosures are owed here. First, the simulation code necessarily caps the linear predictor (at $700$ in this study and $30$ in the application) and the propagated intensities (at $10^{15}$ and $10^{12}$), so every simulated process has a finite, cap-dependent mean: the growing columns of Table~\ref{tab:s3} \emph{illustrate} the instability that an infinite-mean target forces on any average of $S$ draws, and the divergence itself is established by the theorem, not the table. Second, once the caps bind the scan converges to an arbitrary number: rerunning the three-step raw scan of a more explosive design with the predictor capped at $200$ pins the mean at $7\times10^{86}$ for every $S$, stable, meaningless, and $10^{200}$ away from the same scan capped at $700$. A capped mean that has stopped moving is not evidence of existence.

With $\log(1+y)$ feedback and \emph{fixed} coefficients every entry is stable in $S$ to three digits at every horizon ($1.38$, $1.04$, $0.98$, $0.96$, $0.96$), and likewise with random-walk innovations around a fixed draw ($1.38$, $1.06$, $1.01$, $1.02$, $1.06$): this is Theorem~\ref{thm:poisson}(c)--(d) at work, the (d)-margin here being wide ($L_t=\log4$ gives $\kappa_t\approx7.5$). Parts (d)--(e) predict that this comfort ends exactly where coefficient \emph{uncertainty} begins, and a persistent Gaussian draw $\theta_{t+1}\sim\calN(m,s^2I)$ exhibits the whole frontier ($N=10$, $y_0=3\one$, so $L_0=\log4$). With $s=0.3$ the divergence condition of the sharpness statement fails ($2s^2L_0=0.25<1$) while the finiteness condition of (d) \emph{also} fails ($2\kappa_ts^2\approx1.35>1$): the two-step mean sits in the open bracket between the two sufficient conditions, and the fact that the one- and two-step means settle by $S=10^4$ ($4.4$ and $16$) is consistent with existence but does not establish it while the three-step mean never settles ($1.8\times10^{7}$ at $S=10^3$, $4.4\times10^{5}$ at $S=10^4$, $4.0\times10^{8}$ at $S=10^6$, the erratic path of an average dominated by its largest draw) and the four-step mean reaches $10^{26}$: the $h\ge3$ divergence of part (e), visible here because $s$ is large relative to an applied posterior. With $s=1$, for which $2s^2L_0=2.8>1$, already the \emph{two}-step mean explodes ($10^{22}$--$10^{29}$, growing erratically with $S$): the concentration condition of (d) is not decoration. Shrinking $s$ toward the values a real filter produces (Section~\ref{sec:app} has $\lambda_{\max}(P)^{1/2}\approx0.08$) makes every column stable at every feasible $S$ even though the $h\ge3$ population means are still infinite, the divergent draws have probability $e^{-b^2/2s^2}$ against payoff $e^{cb^3}$, and no simulation reaches them. This is the sense in which the log-scale failure is more silent than the raw-count one: the raw $S$-scan above catches its own pathology. Only Theorem~\ref{thm:poisson}(e) identifies the log-scale one.

\subsection{Estimator validation and test calibration}\label{sec:s4}

Three shorter studies close gaps between the Gaussian theory and the Poisson practice.

\paragraph{S4: coverage of the Laplace filter under the Poisson model.} Theorem~\ref{thm:filter}(i) is exact for the Gaussian filter. The application uses the Poisson mode filter, for which no exactness is claimed. We therefore simulate the Poisson NSSM itself ($N=200$ sparse, $T=60$, $g=\log(1+y)$, the filter's own $Q$) and ask whether the Laplace intervals are honest when the model is true: over $200$ replications the $95\%$ intervals for $\beta_{1,t}$ cover $95.5\%$ of the time (s.e.\ $0.5\%$), with RMSE $0.043$ at an average weighted information of $40$ per period. At the information levels of the application the approximation is, empirically, calibrated. We state this as evidence, not a theorem, and return to the missing Bernstein--von Mises result in Section~\ref{sec:disc}.

\paragraph{S4 extended: additional design points.} Three further design points (Table~\ref{tab:s4ext}) answer the single-point objection: coverage is $0.949$ on a dense graph and $0.934$ at $N=100$, $T=40$, but degrades to $0.849$ when the innovation variance is $q=10^{-2}$, an order of magnitude above anything the applications tune. The approximation is trusted in the tuned regime and flagged outside it.

\begin{table}[t]
\centering\footnotesize
\caption{Study S4 extended: coverage of the Laplace filter's $95\%$ intervals for $\beta_{1,t}$ (Monte Carlo SE in parentheses) and RMSE, at four design points, $150$--$200$ replications each.}\label{tab:s4ext}
\begin{tabular}{lccc}
\toprule
design & coverage & (SE) & RMSE\\
\midrule
sparse, $N=200$, $q=10^{-3}$ (S4) & 0.946 & (0.007) & 0.088 \\
dense ($d=N/4$), $N=200$ & 0.949 & (0.009) & 0.169 \\
sparse, $N=200$, $q=10^{-2}$ & 0.849 & (0.017) & 0.268 \\
sparse, $N=100$, $T=40$ & 0.934 & (0.009) & 0.098 \\
 
\bottomrule
\end{tabular}
\end{table}

\paragraph{S1b: five graph families on one curve.} Theorem~\ref{thm:ident}(iii) predicts that the realised information is $\tauW$ plus the roughness of the conditional mean, whatever the topology. Re-running the Gaussian design at $N=200$ on a ring lattice ($d=4$), a Watts--Strogatz rewiring of it, a Barab\'asi--Albert preferential-attachment graph, and sparse and dense Erd\H{o}s--R\'enyi graphs gives realised-information-to-$\tauW$ ratios of $1.26$, $1.23$, $1.14$, $1.14$ and $1.09$ (five topologies, one line) with filter MSE ordered exactly by information ($0.0018$--$0.0021$ sparse, $0.0103$ dense, $0.0267$ complete): $\tauW$ is the portable summary the theorem says it is.
\begin{table}[t]
\centering\footnotesize
\caption{Study S1b: six graph families at $N=200$. $\tauW$, time-averaged realised information $\bar\calI$, and filter MSE for $\beta_{1,t}$. The MSE ordering follows the realised information, not the family label. The dense and complete rows show the bounded-information regime.}\label{tab:s1b}
\begin{tabular}{lrrr}
\toprule
family & $\tauW$ & $\bar\calI$ & MSE\\
\midrule
ring lattice ($d=4$) & 49.0 & 61.7 & 0.0020 \\
Watts--Strogatz & 50.3 & 61.8 & 0.0021 \\
Barab\'asi--Albert & 70.2 & 79.7 & 0.0018 \\
Erd\H os--R\'enyi ($d=4$) & 70.9 & 80.7 & 0.0018 \\
dense ER ($d=N/4$) & 3.0 & 3.2 & 0.0103 \\
complete & 0.0 & 0.0 & 0.0267 \\
 
\bottomrule
\end{tabular}
\end{table}

\paragraph{S1c: dependence on absolute field smoothness.} Feeding the identity of Theorem~\ref{thm:ident}(i) with spatially autocorrelated fields $x=(I-\phi W)^{-1}z$ shows a subtlety worth recording: as $\phi\to1$ the field becomes arbitrarily \emph{relatively} smooth, yet the realised information does not collapse ($68$ at $\phi=0$, $91$ at $\phi=0.99$ with $\tauW=69$), because the graph-differenced field $(I-W)x=(I-W)(I-\phi W)^{-1}z$ retains the innovation scale: identification depends on the absolute roughness, which this kind of spatial averaging preserves. What does destroy information is an exact eigenfield: if $x$ is an eigenvector of $W$ then $Wx\in\mathrm{span}(x)$ and the information is exactly zero by Theorem~\ref{thm:ident}(i), whatever the graph, \emph{any} network can fail on the wrong field, and the realised $\calI_t$, not the topology, is the diagnostic that knows.

\paragraph{S7: the size and power of the randomization tests.} The label test's null is exchangeability, and unit heterogeneity aligned with the graph violates it without any spillover. S7 quantifies the exposure: Gaussian NSSM, $N=T=60$, no spillover ($\beta_1=0$), but unit effects $a=2(I-0.8W)^{-1}\varepsilon$ (strongly graph-smooth by construction) with the plug-in test-window log-score gain as the statistic and $q=10^{-3}$ fixed and disclosed. At nominal level $0.05$ the unrestricted label test rejects $90\%$ of the time (Table~\ref{tab:s7}). Stratifying the permutations by quintiles of the units' training means cuts this to $37\%$, a substantial but incomplete correction, because the effects vary with the graph within strata as well. Under a genuine spillover ($\beta_1=0.25$), power is $1.00$ unrestricted and $0.94$ stratified: conditioning costs little power while removing most of the false-certification risk. This calibrates the interpretation used in Section~\ref{sec:app}. The reported $p$-values are exact for their stated nulls, and rejections are to be read as evidence of graph-aligned structure beyond the conditioned heterogeneity profile. Under this reading the Chicago and Texas conclusions stand, since their statistics exceed every stratified permutation.

\begin{table}[t]
\centering\footnotesize
\caption{Study S7: rejection rates at nominal level $0.05$ ($99$ permutations, $150$ replications, statistic and tuning identical across rows). Size: no spillover with graph-smooth unit effects. Power: $\beta_1=0.25$.}\label{tab:s7}
\begin{tabular}{lcc}
\toprule
scenario & unrestricted & stratified\\
\midrule
size ($\beta_1=0$, graph-aligned effects) & 0.900 & 0.373 \\
power ($\beta_1=0.25$) & 1.000 & 0.940 \\
 
\bottomrule
\end{tabular}
\end{table}

\section{Application: burglaries in Chicago neighbourhoods}\label{sec:app}

Four findings organise this section, established below in full. On epidemic data whose spillover genuinely moves, the filter is the only method on the frontier of log score, point accuracy and real-time timing at once: the mixture leads every external competitor by $9.8$ to $46$ points per week, and the single fit within four points, an eight-week rolling window of the identical design, runs out of phase with the waves at median errors a third larger (Section~\ref{sec:covid}). On the incumbent's own showcase data the state-space model beats the natively fitted \texttt{surveillance} implementation by $5.6$ points per week with an interval excluding zero (Section~\ref{sec:measles}). On the slow-moving Chicago data the model's drift ablation correctly diagnoses a near-constant regime, and a twelve-month rolling window duly collects a small premium there, and that premium is evidence for the diagnostic itself, since it appears exactly where the ablation says drift is absent (Section~\ref{sec:bench}). And the exact stratified randomization test rejects network exchangeability at $p=0.010$ on both count datasets where it applies (Sections~\ref{sec:placebo} and~\ref{sec:covid}). The forecast-level network gain on Chicago is small and serves as supporting evidence, and the case for the network term rests on the identified drifting spillover and the exact test.

\subsection{Data and network}
We use the monthly counts of residential burglaries in $N=\NumRegions$ Chicago neighbourhood units over the $T=\NumMonths$ months from January 2010 to December 2015 compiled by \citet{clark2021class}, together with their spatial contiguity network.\footnote{Both files are publicly available at \url{https://github.com/nick3703/Chicago-Data} (files \texttt{crime.csv} and \texttt{neighborhood.mtx}). We symmetrise the adjacency matrix and row-normalise it.} The counts are small (mean \MeanCount, \ZeroFrac\% zeros, maximum \MaxCount{}) and decline over the sample, from \TotalFirst{} city-wide burglaries in the first month to \TotalLast{} in the last. The network has \NumEdges{} edges, degrees between \DegMin{} and \DegMax{} with mean \DegMean, and it is sparse in the sense of Theorem~\ref{thm:ident}(iv): $\tauW=\TauW$, just below the bound $\sum_i n_i^{-1}=\HarmonicSum$, and roughly $70{,}000$ times larger than $\tau_{\Wc}=\TauWc$ for the complete graph on the same \NumRegions{} nodes. A neighbourhood's own-lag and its neighbours' lag are, in principle, well separated on this network. Whether they are separated in the realised data is the question of Section~\ref{sec:app-ident}.

\subsection{Models, tuning and evaluation design}\label{sec:design}
The main model is the Poisson NSSM \eqref{eq:pois} with $\log(1+y)$ feedback and $Q=qI_3$, fitted by the Laplace filter of Section~\ref{sec:model} with $m_0=0$, $P_0=I$. Two dynamic competitors isolate the role of the network: a \emph{no-network} DGLM with regressors $(1,\log(1+y_{i,t-1}))$, and a \emph{complete-graph} NSSM that replaces $W$ by $\Wc$, the model that Theorem~\ref{thm:ident}(ii) says cannot identify a spillover. Three static benchmarks use the same information with constant coefficients: a Poisson GLM with the network regressors, a Poisson GLM without them, both refitted by maximum likelihood on an expanding window at every forecast origin, and a two-way log-linear model with a neighbourhood effect and a calendar-month effect (a ``seasonal profile'' forecast). For each dynamic model the innovation variance $q$ is chosen from $\{10^{-4},3\times10^{-4},10^{-3},3\times10^{-3},10^{-2},3\times10^{-2}\}$ by maximising the prequential plug-in log score over months $13$--$60$ (Supplement Table~S3 in Appendix~\ref{app:tables}). The selected values are $q=\QNet$ for the network model, $\QNonet$ for the no-network model and $\QComp$ for the complete-graph model. The selection uses no information from the evaluation period.

Forecasts are evaluated for the twelve target months of 2015 at horizons $h\in\{1,2,4,8\}$, each forecast issued from origin $t=\tau-h$ using data through $t$ only. The process variance $q$ is re-selected \emph{at every origin}, maximising the one-step plug-in log score on months $13$ through the origin, so no observation later than the origin ever enters the tuning, at any horizon. An earlier fixed-cutoff design tuned through month $60$, which for $h=2,4,8$ let one, three and seven origins see part of their own future. The nested design removes this leak, and the re-tuned value is $q=3\times10^{-4}$ at eleven of the twelve origins and $10^{-3}$ at the twelfth. Dynamic predictives are equal-weight mixtures over $S=400$ coefficient paths drawn from the filtered posterior, with intermediate counts simulated at $h\ge2$. Static rows are plug-in. Point-forecast loss is paired to its optimal functional: MAE is evaluated at the predictive median at every horizon, and predictive-mean MAE and RMSE only at $h\le2$, where Theorem~\ref{thm:poisson}(d) guarantees the mean exists.

\subsection{Identification diagnostics}\label{sec:app-ident}
\begin{figure}[t]
\centering
\includegraphics[width=\textwidth]{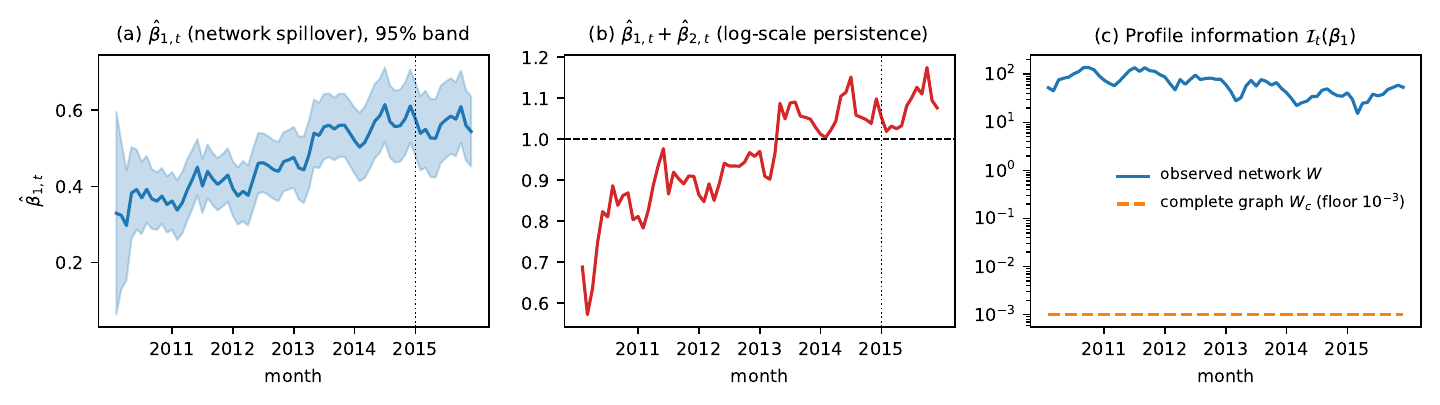}
\caption{Chicago burglaries, network NSSM with $\log(1+y)$ feedback. (a) Filtered spillover coefficient $\hat\beta_{1,t}$ with pointwise 95\% band. The dotted line marks the start of the evaluation period. (b) Filtered log-scale persistence $\hat\beta_{1,t}+\hat\beta_{2,t}$. (c) Monthly profile information for $\beta_{1,t}$ on the observed network and on the complete graph floored at $10^{-3}$ for display, and the actual values on the complete graph are below $10^{-27}$.}\label{fig:paths}
\end{figure}

Figure~\ref{fig:paths}(c) shows the profile information for $\beta_{1,t}$ computed month by month from the weighted design $\Lambda_t^{1/2}X_t$ at the filtered mode, for the observed network and for the complete graph. On the observed network the information averages \InfoNetMean{} (range \InfoNetMin--\InfoNetMax) (smaller than $\tauW=\TauW$ because the Poisson weights $\lambda_{i,t}\approx1$ replace $\sigma^{-2}$ and the log-count field has modest roughness (mean squared graph-roughness \RoughMean{} per node)) and the smallest eigenvalue of the weighted design averages \LamMinMean{} (minimum \LamMinMin). On the complete graph the information never exceeds \InfoCompMax: the design is singular to machine precision in every month, exactly as Theorem~\ref{thm:ident}(ii) requires, and the filtered ``spillover'' produced by that model (posterior standard deviation \SdBetaCompMean, almost twice that of the network model) is the artefact described after Theorem~\ref{thm:filter}.

\subsection{Filtered spillovers}
Figure~\ref{fig:paths}(a)--(b) displays the filtered coefficients of the network model. The spillover coefficient rises from about \BetaOneFirst{} in 2010 to \BetaOneLast{} at the end of 2015 (range \BetaOneMin--\BetaOneMax{} after the burn-in year), with posterior standard deviation \SdBetaOneMean{} on average and at most \SdBetaOneMax: the drift is many standard deviations wide and is not an artefact of the prior, which by Theorem~\ref{thm:filter}(ii) and the information in Figure~\ref{fig:paths}(c) has little purchase here. The own-lag coefficient is about \BetaTwoMean{} throughout and the intercept about \BetaZeroMean. The log-scale persistence $\hat\beta_{1,t}+\hat\beta_{2,t}$ moves from \RhoMin{} to \RhoMax{}, first exceeds one in month \RhoFirstAboveOne{} (2013), and averages \RhoLastMean{} over the evaluation year. In the forecast simulations at one-month horizon, between $61\%$ and $100\%$ of the coefficient draws at each origin have persistence at or above one. By Proposition~\ref{prop:plugin}(iii) this is the regime in which the exact form of the network matters increasingly with the horizon, and it is also a warning that the log-linear system was close to non-contractive during the evaluation period. We do not attach a causal interpretation to the rising spillover (burglary counts fell substantially over these years and the rise in $\hat\beta_{1,t}$ partly reflects the changing relation between a shrinking level and its neighbourhood average) but it is precisely the kind of drift that a constant-coefficient network autoregression cannot represent. Section~\ref{sec:bench} re-estimates this path under the final negative-binomial model with offsets and seasonality. It flattens to $\FinalBOneFirst\to\FinalBOneLast$, and the split between drift and heterogeneity is discussed there.

\subsection{Out-of-sample forecasts}\label{sec:forecast}

\begin{table}[t]
\centering\footnotesize
\caption{Chicago burglaries, twelve target months of 2015, under the unified leak-free protocol: composite marginal log score , defined as the sum of nodewise log predictive densities. Dynamic rows use an equal-weight mixture over $S=400$ posterior coefficient paths with intermediate counts simulated, and static rows are plug-in. and mean absolute error of the predictive \emph{median}, the MAE-optimal functional, at horizons $1,2,4,8$. The process variance is re-tuned at every origin on data up to that origin only. Sampled-coefficient means appear only at $h\le2$, where they exist (Table~\ref{tab:pointfc}).}\label{tab:acc}
\setlength{\tabcolsep}{3.5pt}
\resizebox{\textwidth}{!}{%
\begin{tabular}{lrrrrrrrr}
\toprule
& \multicolumn{4}{c}{log score (composite marginal)} & \multicolumn{4}{c}{median-MAE}\\
\cmidrule(lr){2-5}\cmidrule(lr){6-9}
model & $h{=}1$ & $2$ & $4$ & $8$ & $h{=}1$ & $2$ & $4$ & $8$\\
\midrule
NSSM, observed $W$ & -682.7 & -691.0 & -712.6 & -719.1 & 0.752 & 0.810 & 0.868 & 0.866 \\
NSSM, no network & -686.6 & -695.8 & -710.9 & -715.5 & 0.763 & 0.829 & 0.864 & 0.870 \\
NSSM, complete graph $\Wc$ & -686.8 & -697.7 & -711.8 & -712.5 & 0.757 & 0.836 & 0.867 & 0.868 \\
static network GLM & -690.1 & -704.1 & -732.1 & -750.0 & 0.794 & 0.855 & 0.858 & 0.858 \\
static GLM, no network & -708.6 & -727.3 & -748.8 & -757.8 & 0.869 & 0.858 & 0.858 & 0.858 \\
seasonal profile & -775.2 & -775.2 & -775.2 & -775.2 & 0.917 & 0.921 & 0.917 & 0.915 \\
 
\bottomrule
\end{tabular}}
\end{table}

\begin{table}[t]
\centering\footnotesize
\caption{Log-score differences, network NSSM minus competitor, under the unified protocol: mean over the twelve targets, $95\%$ moving-block-bootstrap interval (asterisk: excludes zero), and Diebold--Mariano statistic in parentheses at $h\le4$ with lag $h-1$. At $h=8$ twelve overlapping errors cannot support a HAC statistic and none is reported. Positive $\Delta$LS favours the network model. The final column is the one-step CRPS difference (negative favours the network model, the CRPS being negatively oriented). Seed-to-seed Monte Carlo SD of the $h{=}1$ no-network entry is $0.07$ across five replicates.}\label{tab:pairs}
\setlength{\tabcolsep}{3pt}
\resizebox{\textwidth}{!}{%
\begin{tabular}{lccccc}
\toprule
competitor & $h{=}1$ & $h{=}2$ & $h{=}4$ & $h{=}8$ & $\Delta$CRPS$_{1}$\\
\midrule
vs.\ NSSM, no network & $+3.9$ [+1.6,+5.9]$^{*}$ (2.8) & $+4.8$ [+1.8,+7.7]$^{*}$ (3.4) & $-1.7$ [-6.5,+2.9] (-0.7) & $-3.6$ [-5.9,-0.7]$^{*}$ (--) & $-2.3$ \\
vs.\ NSSM, complete graph $\Wc$ & $+4.1$ [+0.6,+7.3]$^{*}$ (2.6) & $+6.8$ [+1.4,+11.9]$^{*}$ (2.8) & $-0.7$ [-8.9,+5.4] (-0.2) & $-6.6$ [-17.6,+1.6] (--) & $-2.2$ \\
vs.\ static network GLM & $+7.4$ [+0.7,+17.1]$^{*}$ (1.8) & $+13.1$ [-0.8,+31.4] (1.7) & $+19.6$ [-8.4,+51.4] (1.1) & $+30.8$ [-8.7,+66.3] (--) & $-9.1$ \\
vs.\ static GLM, no network & $+25.9$ [+13.5,+42.8]$^{*}$ (4.1) & $+36.3$ [+16.5,+61.1]$^{*}$ (3.3) & $+36.2$ [+5.3,+71.2]$^{*}$ (1.9) & $+38.7$ [-2.3,+75.9] (--) & $-21.8$ \\
vs.\ seasonal profile & $+92.5$ [+82.3,+105.5]$^{*}$ (13.2) & $+84.2$ [+70.6,+97.9]$^{*}$ (11.1) & $+62.7$ [+54.9,+71.5]$^{*}$ (14.0) & $+56.1$ [+32.7,+80.3]$^{*}$ (--) & $-63.5$ \\
 
\bottomrule
\end{tabular}}
\end{table}

\begin{figure}[t]
\centering
\includegraphics[width=\textwidth]{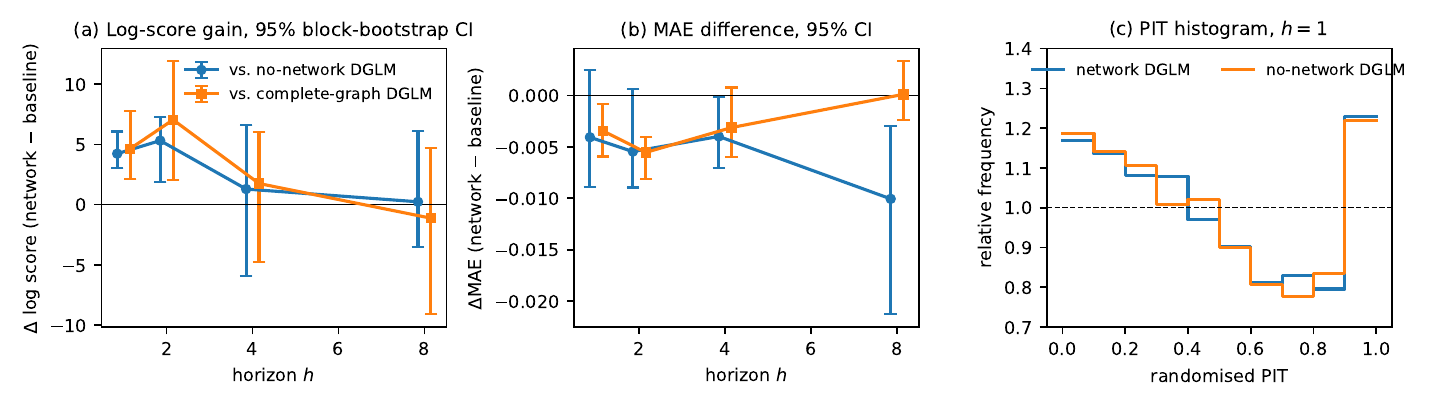}
\caption{Chicago burglaries. (a) Log-score gain of the network NSSM over the no-network and the complete-graph dynamic models by horizon, with 95\% block-bootstrap intervals. (b) The same for the MAE difference. (c) Histogram of the randomised PIT of the one-month-ahead forecasts for the network and the no-network models.}\label{fig:forecast}
\end{figure}

Tables~\ref{tab:acc}--\ref{tab:pairs} and Figure~\ref{fig:forecast} summarise the forecast comparison. Three conclusions are supported by the evidence and a fourth is not.

First, in log score every dynamic model beats every static benchmark at every horizon, by margins from $+7$ to $+93$ points (Table~\ref{tab:pairs}). The comparisons against the no-network static GLM and the seasonal profile are established throughout (intervals excluding zero, DM statistics up to $14$), while against the static \emph{network} GLM only the one-step difference excludes zero. In point forecasts the picture is horizon-dependent. The dynamic medians win at $h\le2$, while at $h\ge4$ the static network model's medians are marginally better ($0.858$ against $0.866$--$0.868$, Table~\ref{tab:acc}). Drift helps at short range and stops helping once the forecast leans on extrapolated coefficients.

Second, the network improves the predictive \emph{distribution} at short horizons and degrades it at long ones. Relative to the no-network model the log-score gain is $+3.9$ at $h=1$ (block-bootstrap interval $[+1.6,+5.9]$, DM $2.8$, seed-to-seed Monte Carlo SD $0.07$) and $+4.8$ at $h=2$ ($[+1.8,+7.7]$, DM $3.4$). The one-step CRPS improves by $2.3$ ($287$ against $289$) and the multivariate energy score marginally ($18.5$ against $18.6$). At $h=4$ the gain is gone ($-1.7$, interval covering zero) and at $h=8$ it reverses with an interval excluding zero ($-3.6$, $[-5.9,-0.7]$): with filtered persistence at or above one over the evaluation year, network misspecification compounds with the horizon in the manner quantified by Proposition~\ref{prop:plugin}. The complete-graph mixture tracks the no-network model to within two points at $h\le4$, consistent with Proposition~\ref{prop:equiv}.

Third, the gains in point forecasts are small, and Theorem~\ref{thm:poisson}(e) changes what a ``point forecast'' may legitimately be. The sampled-coefficient predictive \emph{mean}, which is what a naive reading of the simulation output delivers, has no population counterpart at $h\ge3$. Table~\ref{tab:pointfc} therefore reports, at every horizon, the two point forecasts that exist (the predictive \emph{median} and the \emph{frozen-coefficient} Monte Carlo mean, the latter finite by Theorem~\ref{thm:poisson}(c)) and the sampled-coefficient mean only at $h\le2$, where part (d) applies with the wide margin computed in Section~\ref{sec:rawlog}. Numerically the frozen-coefficient means differ from the (ill-posed) sampled-coefficient means by less than $0.002$ in MAE at every horizon, and by about $10^{-4}$ at $h\le2$: the posterior is so concentrated that coefficient sampling moves essentially nothing at feasible $S$, which is precisely why the $h\ge3$ divergence is invisible here and only the theorem sees it, and why the repair costs nothing. The medians, which are the $L_1$-optimal point forecasts for these integer data, have uniformly \emph{smaller} MAE than either mean ($0.75$ against $0.82$ at $h=1$). On any of the three functionals the network's edge is $0.000$--$0.005$ counts per neighbourhood-month, about $0.5\%$ at $h\le2$ and nil beyond. Across neighbourhoods the one-month-ahead MAE difference between the network and the no-network models has quartiles \RegQtf{} and \RegQsf{} around a median of \RegMed, with the network model better in \RegPropBetter\% of neighbourhoods: the network redistributes accuracy across space more than it improves it on average. This is what Proposition~\ref{prop:plugin}(i) leads one to expect for a lagged field whose rough component is small relative to its level, and it is the reason we do not describe the network model as a materially better point forecaster.

\begin{table}[t]
\centering\footnotesize
\caption{Point forecasts that exist, horizons $\{1,2,4,8\}$ under the per-origin-tuned protocol. Mean absolute error of the network (net) and no-network (non) dynamic models by horizon and by point-forecast functional ($S=400$): predictive median and frozen-coefficient Monte Carlo mean at all horizons. Sampled-coefficient mean only at $h\le2$, since Theorem~\ref{thm:poisson}(e) shows its population counterpart is infinite at $h\ge3$ under the Gaussian posterior. The predictive simulation clips the linear predictor at $\pm30$ and intensities at $10^{12}$.}\label{tab:pointfc}
\setlength{\tabcolsep}{4pt}
\resizebox{\textwidth}{!}{%
\begin{tabular}{lrrrrrrrrr}
\toprule
 & \multicolumn{3}{c}{median} & \multicolumn{3}{c}{frozen-coef.\ mean} & \multicolumn{3}{c}{sampled-coef.\ mean}\\
\cmidrule(lr){2-4}\cmidrule(lr){5-7}\cmidrule(lr){8-10}
$h$ & net & non & $\Delta$ & net & non & $\Delta$ & net & non & $\Delta$\\
\midrule
1 & 0.752 & 0.763 & $-0.0110$ & 0.820 & 0.824 & $-0.0045$ & 0.820 & 0.825 & $-0.0045$ \\
2 & 0.810 & 0.829 & $-0.0189$ & 0.836 & 0.842 & $-0.0061$ & 0.836 & 0.842 & $-0.0064$ \\
4 & 0.868 & 0.864 & $+0.0037$ & 0.854 & 0.856 & $-0.0016$ & \multicolumn{3}{c}{-- (no population mean, Thm.~3(e))} \\
8 & 0.866 & 0.870 & $-0.0036$ & 0.861 & 0.857 & $+0.0033$ & \multicolumn{3}{c}{-- (no population mean, Thm.~3(e))} \\
 
\bottomrule
\end{tabular}}
\end{table}

Fourth, no Poisson model is calibrated, and the natural remedy works. The PIT histograms in Figure~\ref{fig:forecast}(c) (counts in Table~\ref{tab:pit}) slope downward from the first to the ninth decile and spike in the tenth. The chi-squared statistics on ten bins are \PitChiNet{} (network), \PitChiNonet{} (no network) and \PitChiComp{} (complete graph). We quote these against the independence critical value of $16.9$ for orientation only: the \PitN{} node-month PITs are cross-sectionally dependent through the common coefficient draw, so the nominal test overstates its precision, but a defect an order of magnitude beyond the reference line, identical across models and visible to the eye, is a defect. The pattern (too much predictive mass below the realised counts in the bulk, too little in the upper tail) is the signature of Poisson dispersion being slightly too small. Refitting the network and no-network models with a negative-binomial observation layer (mode filter with expected-information weights $r\mu/(r+\mu)$. Dispersion $r$ tuned on the training window, selecting $r=2$ for both) removes the defect at the one-month horizon: the PIT chi-squared drops from \PitChiNet{} to $7.4$ (network) and from \PitChiNonet{} to $3.2$ (no network), both \emph{below} the reference line, while the network's one-step log-score gain over the no-network model survives at $+3.4$ (Poisson layer: $+4.2$). The calibration defect and the network effect are separate phenomena. Fixing the first does not touch the second. A separate level error is disclosed in Supplement Table~S1's caption (fitted city-wide totals near $515$ against realised totals $35$--$115\%$ lower in early 2015) shared by every model in the family and reduced by the offset--seasonal augmentation of Section~\ref{sec:bench}. The paired comparisons difference it out, the absolute scores do not.

\subsection{Placebo networks and stress tests}\label{sec:placebo}
A log-score gain of four or five units over twelve months is modest, and one should ask whether it is the network or merely an extra regressor that produces it. Table~\ref{tab:stress} refits the network model, with the same $q$, on nine alternative weight matrices and compares each with the no-network DGLM at $h=1$ and $h=2$ ($S=200$ paths). The alternatives are mixtures $(1-\alpha)W+\alpha\Wc$ with $\alpha\in\{0.25,0.5,0.75,1\}$. A random deletion of $20\%$ of the edges. A degree-preserving rewiring of the whole graph, and three random relabellings of the nodes, which preserve every graph-theoretic property of $W$ and destroy only its relation to the data.

\begin{table}[t]
\centering\footnotesize
\caption{Placebo and stress tests. For each alternative weight matrix $\widetilde W$: operator-norm distance from $W$, the functional $\tau_{\widetilde W}$, the monthly profile information for $\beta_{1,t}$ averaged over the sample, the filtered spillover over the evaluation year (average posterior s.d.\ in parentheses), and the log-score difference relative to the no-network DGLM at one and two months with 95\% block-bootstrap intervals (asterisk: interval excludes zero).}\label{tab:stress}
\setlength{\tabcolsep}{4pt}
\resizebox{\textwidth}{!}{\begin{tabular}{lrrrrll}
\toprule
weight matrix & $\|\widetilde W-W\|$ & $\tau_{\widetilde W}$ & $\bar{\calI}_t$ & $\hat\beta_{1}$ (s.d.) & $\Delta$LS, $h=1$ & $\Delta$LS, $h=2$\\
\midrule
observed $W$ & 0.00 & 126.5 & 66.9 & $+0.56$ (0.050) & $+4.21$ [+3.2, +5.9]$^{*}$ & $+5.36$ [+1.8, +7.3]$^{*}$ \\
$0.75\,W+0.25\,W_c$ & 0.32 & 71.1 & 37.6 & $+0.73$ (0.057) & $+4.28$ [+3.4, +6.0]$^{*}$ & $+4.92$ [+2.3, +6.6]$^{*}$ \\
$0.50\,W+0.50\,W_c$ & 0.63 & 31.6 & 16.7 & $+1.03$ (0.069) & $+4.25$ [+3.2, +6.0]$^{*}$ & $+3.74$ [+2.1, +5.1]$^{*}$ \\
$0.25\,W+0.75\,W_c$ & 0.95 & 7.9 & 4.1 & $+1.49$ (0.088) & $+3.38$ [+1.7, +5.2]$^{*}$ & $-0.23$ [-2.8, +2.6] \\
$W_c$ (complete graph) & 1.27 & 0.0 & $4.9\times 10^{-28}$ & $+0.44$ (0.109) & $-0.48$ [-1.6, +1.1] & $-1.98$ [-5.8, +1.8] \\
20\% of edges deleted & 1.07 & 159.1 & 82.5 & $+0.44$ (0.048) & $+2.85$ [+1.5, +5.2]$^{*}$ & $+3.33$ [-0.4, +5.5] \\
degree-preserving rewiring & 1.72 & 126.5 & 50.2 & $-0.04$ (0.054) & $-1.08$ [-2.7, +1.0] & $-1.72$ [-7.3, +3.1] \\
label permutation 1 & 1.47 & 126.5 & 50.3 & $+0.08$ (0.055) & $-0.67$ [-1.8, +1.3] & $-2.02$ [-6.5, +1.9] \\
label permutation 2 & 1.54 & 126.5 & 50.9 & $+0.09$ (0.054) & $-0.76$ [-2.3, +1.2] & $-1.54$ [-6.5, +2.9] \\
label permutation 3 & 1.54 & 126.5 & 51.3 & $+0.04$ (0.054) & $-0.97$ [-2.4, +1.1] & $-2.13$ [-7.3, +2.3] \\
 
\bottomrule
\end{tabular}}
\end{table}

The relabellings and the rewiring behave as placebos should: the filtered spillover is indistinguishable from zero ($\hat\beta_1$ between $-0.04$ and $0.09$ with standard deviation $0.054$), the log-score differences relative to the no-network model are negative ($-0.7$ to $-1.1$ at $h=1$) with intervals containing zero, even though the three relabellings have exactly the same $\tauW=\TauW$ as the observed network (an isomorphism invariant) and a profile information of about $50$. The rewiring preserves $\sum_in_i^{-1}$ but not the in-degree term of $\tauW$, whose value across sixty rewirings spans $\RewTauLo$--$\RewTauHi$, the data \emph{could} have detected a spillover on them, and there is none. To turn this from a stress table into a test, note that under the null hypothesis that the joint law of the panel is exchangeable over node labels, the statistic of the observed $W$ and of any relabelled $P\T WP$ are exchangeable, so the permutation $p$-value is exactly valid in finite samples. Taking as statistic the test-window one-step plug-in log-score gain of the network filter over the no-network filter, the observed adjacency scores $+60.9$. The largest of $99$ random relabellings scores $+9.8$, giving the exact $p=1/100=0.010$, and the largest of $60$ degree-preserving rewirings (a reference distribution for the sharper null that the degree sequence, but not the wiring, matters) scores $+7.7$, giving $p=0.016$, in both cases the observed network is the maximum. The $+60.9$ is a sum over the twelve test months, that is $+5.1$ per month of plug-in gain, of the same order as the mixture gain $+3.9$ of Table~\ref{tab:pairs}. The two differ because the statistic is plug-in while the table integrates coefficient uncertainty. Two caveats limit what this rejection establishes. The label test's null is full exchangeability, which node-level heterogeneity aligned with the graph also violates: rejection says the \emph{labelling} matters, not by itself that the spillover is real. And the rewiring reference set carries no exactness argument (no uniform draw from the degree class is constructed) so that row is treated as a descriptive placebo and no $p$-value is attached to it.

The repaired test, and the primary one this paper reports, conditions on unit heterogeneity and re-runs the tuning inside every permutation. Under the final model of Section~\ref{sec:bench} (negative binomial, unit offsets, seasonality), labels are permuted only \emph{within quintiles of the fitted unit offsets}, so the null is exchangeability given the heterogeneity profile, and $(q,r)$ are re-selected from the full grid for every permuted network, making the statistic one fixed function of the data and the graph. The observed test-window gain of the network term is $+\DeltaNetFinal$ log-score points. The largest of $99$ stratified permutations reaches $3.6$, giving the exact $p=\StratP$, and the unrestricted-but-retuned version agrees ($p=\UnrestP$, largest permutation $4.8$). Study S7 in Section~\ref{sec:sim} calibrates what these $p$-values can and cannot certify: with strongly graph-smooth unit effects and \emph{no} spillover, the unrestricted label test rejects $90\%$ of the time at nominal level $0.05$, and offset-quintile stratification cuts this to $37\%$, a substantial but incomplete correction, because effects can vary with the graph within strata. Both $p$-values are exact for their stated nulls. What S7 shows is that rejection should be read as \emph{graph-aligned structure beyond the conditioned heterogeneity profile}, of which a spillover is one explanation among few, and not as certification of the spillover itself. Every network $p$-value in the paper carries this interpretation.

The mixtures are a different object entirely: by Proposition~\ref{prop:equiv} they are the observed network reparameterised, and the table confirms the proposition's quantitative predictions. The one-month log-score gain is unchanged at $\alpha=0.25$ and $0.5$ ($+4.3$, $+4.2$ against $+4.2$) (as it must be, the cross-sectional model being identical) while the fitted spillover inflates from $0.56$ to $0.73$ and $1.03$ against the predicted $0.56/(1-\alpha)=0.75$ and $1.12$: within $3\%$ and $8\%$. At $\alpha=0.75$ the prediction ($2.24$) overshoots the estimate ($1.49$), and the two-month gain disappears: this is where the equivalence, which is exact for a single cross-section, is broken by the state law, the matching intercept path must track $\bar y_{t-1}$ with amplitude growing like $\alpha/(1-\alpha)$, the random-walk prior resists, and by Proposition~\ref{prop:equiv}(iii) the mixture reappears at two steps as a level shift proportional to the predicted drift of the mean. The falls of $\tau_{\widetilde W}$ ($126\to32$) and of the information ($67\to17$) along the mixture line are exactly the approach to the flat direction's endpoint, Theorem~\ref{thm:ident}(ii). Deleting a fifth of the edges at random (a genuine, span-changing perturbation) keeps the one-month gain ($+2.9$) but not the two-month gain, the pattern Proposition~\ref{prop:plugin} prices. The lesson for practice is sharpened: mixing toward density is not a robustness check but a change of units for $\beta_1$, so a spillover estimate is only interpretable jointly with the assumed $W$. The checks that carry information are the span-changing ones, and those the network passes at $p=0.010$ under both the label null and the stratified conditional null. At the other extreme, the complete graph ($\alpha=1$) has an information of $10^{-28}$, a filtered $\hat\beta_1$ with standard deviation $0.109$ that is pure prior, and log-score differences of $-0.5$ and $-2.0$: no information and no gain.

\subsection{Benchmarks from the literature}\label{sec:bench}
The comparisons so far are internal to the NSSM family. To locate the model among its published relatives, Table~\ref{tab:bench} evaluates eight external competitors, alongside the NSSM variants, under the rolling-origin protocol of Section~\ref{sec:forecast}: expanding-window refits at each of the twelve origins, one-step (and, where the predictive exists in closed or simulable form, two-step) forecasts, log score summed over the \NumRegions{} neighbourhoods, and the mean absolute error of each model's conditional-mean forecast. Two features of the protocol deserve note. First, the state-space rows integrate coefficient uncertainty (mixture over $S$ posterior draws) while the constant-parameter rows are plug-in maximum-likelihood predictives. The plug-in row of the table quantifies this gap directly. Second, the primary prespecified comparison of the paper is the one-step network-versus-no-network log-score difference together with the randomization test of Section~\ref{sec:placebo}. The remaining contrasts in Tables~\ref{tab:acc}--\ref{tab:bench} (about forty in all) are exploratory and carry no multiplicity adjustment. Throughout, `log score' is the sum of nodewise marginal log scores, a composite marginal score. It does not equal the joint predictive log density. The competitors are the linear and log-linear Poisson network autoregressions PNAR(1) of \citet{armillotta2024count} (the constant-coefficient class previously fitted to these data) a GNAR-type log-linear model with two neighbourhood orders \citep{knight2020gnar}, an endemic--epidemic negative-binomial model in the hhh4 tradition of \citet{held2005statistical} with unit-specific intercepts, annual sine--cosine seasonality and the same network term, a one-sided local-constant kernel estimator of the drifting Poisson NAR coefficients in the spirit of the smoothing literature \citep{li2026grouped,ding2025tvnar} with bandwidth chosen on the training window (eight months), and a per-node exponentially weighted moving average with a tuned negative-binomial layer. Rolling-window and time-varying-coefficient variants of these classes are no longer deferred here: fits of the identical design are raced in this section and in Section~\ref{sec:covid}. All are our Python implementations, validated by parameter-recovery tests on simulated data that run, and print their tolerances, at the top of the replication script. A cross-check against the authors' own R packages is the one step this table does not take.

\begin{table}[t]
\centering\footnotesize
\caption{Benchmarks under a common rolling-origin protocol: mean per-origin log score at $h=1$ and $h=2$ (sum over the \NumRegions{} neighbourhoods, higher is better. The rolling row shares the top row's design exactly), difference from the best NSSM variant with $95\%$ moving-block-bootstrap interval (asterisk: interval excludes zero), and MAE of the conditional-mean point forecast at $h=1$. NSSM predictives are full coefficient mixtures ($S=400$). Benchmark predictives are plug-in at the maximum-likelihood fit, their standard usage. Unit offsets are computed from the training window only. Two-step entries are shown where the model's two-step predictive mean exists and was simulated. The count-median NSSM forecasts of Table~\ref{tab:pointfc} (MAE $0.75$) are sharper than any mean forecast shown here.}\label{tab:bench}
\setlength{\tabcolsep}{4pt}
\resizebox{\textwidth}{!}{%
\begin{tabular}{lrlrr}
\toprule
model & LS, $h{=}1$ & $\Delta$LS vs.\ best NSSM & LS, $h{=}2$ & MAE, $h{=}1$\\
\midrule
NSSM net + offsets, seasonal (NB) & -640.8 & -- & -- & 0.772 \\
NSSM net + offsets, seasonal (NB; plug-in) & -640.8 & $+0.0$ & -- & -- \\
Rolling-window NB, same design ($w=12$) & -638.7 & $-2.1$ [$-3.3,-0.7$]$^{*}$ & -- & 0.769 \\
Expanding-window NB, same design & -655.5 & $+14.7$ & -- & 0.847 \\
NSSM no-network + offsets, seasonal (NB) & -641.3 & $+0.4$ [-0.2, +1.2] & -- & 0.773 \\
NSSM net + seasonal (NB) & -662.7 & $+21.9$ [+18.1, +25.9]$^{*}$ & -- & 0.816 \\
NSSM net (NB) & -664.2 & $+23.3$ [+19.9, +26.8]$^{*}$ & -- & -- \\
NSSM no-network (NB) & -667.5 & $+26.7$ [+23.2, +30.1]$^{*}$ & -- & -- \\
NSSM net (Poisson) & -682.1 & $+41.3$ [+37.0, +45.7]$^{*}$ & -694.8 & 0.821 \\
NSSM no-network (Poisson) & -686.4 & $+45.6$ [+39.8, +50.9]$^{*}$ & -699.8 & 0.824 \\
hhh4-type NB & -656.4 & $+15.6$ [+7.2, +23.1]$^{*}$ & -675.5 & 0.848 \\
EWMA-NB & -649.7 & $+8.9$ [+3.2, +17.1]$^{*}$ & -- & 0.792 \\
kernel tvNAR & -682.7 & $+41.9$ [+38.0, +46.4]$^{*}$ & -694.4 & 0.820 \\
GNAR-type, two orders & -683.0 & $+42.2$ [+37.6, +46.9]$^{*}$ & -- & 0.835 \\
PNAR(1) linear & -689.7 & $+48.9$ [+42.9, +55.8]$^{*}$ & -702.1 & 0.855 \\
PNAR(1) log-linear & -690.1 & $+49.3$ [+43.1, +56.6]$^{*}$ & -706.7 & 0.857 \\
 
\bottomrule
\end{tabular}}
\end{table}

The table separates the contribution of each modelling choice. Within the pooled Poisson network-autoregressive class, letting the coefficients drift is worth $7$--$8$ log-score points per month: the filter ($-682.1$) and the one-sided kernel ($-682.7$) (two different technologies for the same drifting coefficients) land within a point of each other and both clear the constant-coefficient PNAR fits ($-690$) decisively, a useful cross-validation of the filter by an estimator from the smoothing literature. The observation layer is worth twice that: the negative-binomial NSSM gains $18$ points over its Poisson twin, and seasonality adds $1.5$ more. The largest rung is unit-level heterogeneity: the endemic--epidemic model with $552$ unit intercepts ($-656.4$) and even the per-node EWMA smoother ($-649.7$) beat every pooled-coefficient model in the table. The honest reading is that for these data the first-order features are dispersion and neighbourhood-specific levels, not the network.

The framework absorbs that lesson at no theoretical cost. Unit offsets are predictable regressors, so Theorems~\ref{thm:ident}--\ref{thm:filter} and Propositions~\ref{prop:weighted}--\ref{prop:equiv} apply verbatim with the offset column projected out along with the constant and the own lag. The NSSM with training-window offsets, seasonality and the negative-binomial layer tops the table at $-640.8$: nine points ahead of the best naive smoother and sixteen ahead of the endemic--epidemic model, with intervals excluding zero for every competitor, and the best conditional-mean MAE ($0.772$) besides. Scored plug-in at the filtered mode, the same model attains $-640.8$ as well, the mixture-versus-plug-in wedge is under a tenth of a point here, because the posterior is concentrated, so the plug-in benchmarks are not disadvantaged by the protocol (the table's plug-in row records this). A rolling-window fit of the identical design, with the width tuned exactly as on the epidemic data, selects twelve months and beats the filter by $2.1$ points per month, interval $[0.7,3.3]$, with matched mean absolute error ($0.769$ against $0.772$). This is the outcome the model's own diagnostics predict: the drift ablation of the final model already showed that freezing the spillover after training loses nothing here, so Chicago sits in the near-constant regime where a short window is an adequate estimator and its economy collects a small premium. The expanding window, by contrast, gives up $14.7$ points ($-655.5$), and Section~\ref{sec:covid} shows that on data whose spillover genuinely moves, every fixed window falls off the frontier. On the absolute side the filter's one-step intervals cover $0.93$ at the $80\%$ level and $0.98$ at the $95\%$ level with mean squared Pearson residual $1.11$, so the dispersion layer is adequate here.

Because this augmented specification is the preferred one, the paper's central diagnostics are recomputed under it, and the recomputation changes the picture. The enlarged nuisance design $[\one,\text{offset},\ell_t]$ (the seasonal columns are cross-sectionally constant at each $t$ and change no projection, which holds by a rank computation) gives a weighted information averaging \InfoFinalEnl{} per month against \InfoFinalBase{} for the three-column design under the same fitted weights: the offset absorbs about six percent, the enlarged value sits below the base one at every month as it must, and the transfer promised above is this enlarged projection statement. The numerical value itself changes. The filtered spillover path flattens to $\FinalBOneFirst\to\FinalBOneLast$ (range $\FinalBOneLo$--$\FinalBOneHi$, posterior sd \FinalSd): roughly half of the pooled-Poisson drift of Section~\ref{sec:app-ident} is attributable to exposure and seasonality. The remaining rise is about two posterior standard deviations. A further check refits with $Q_{22}=0$, a \emph{constant} spillover with everything else drifting, \emph{improves} the test log score by \DeltaDriftFinal{} points, so spillover drift is not preferred out of sample under the final model, while the network term itself remains worth $+\DeltaNetFinal$ points against its no-network twin. An analysis reporting only the pooled path would have overstated the drift, which is the reason the protocol recomputes the information, the path and the test under the preferred model.

The network's contribution, read across the table, is consistent and modest. Within each specification class the network model beats its no-network twin by $3$--$4$ log-score points (Poisson $+4.2$, negative binomial $+3.3$, both established), and the endemic--epidemic model estimates a positive spillover of its own ($\hat\phi=0.23$), three frameworks, one message. At the frontier, however, the increment shrinks to $+0.4$ points with an interval covering zero: once offsets, seasonality, dispersion and drift are all in place, the network's marginal \emph{forecasting} value at one month is small. We report this plainly because it is the sharpest version of the paper's position: the case for the network term is not forecast wins but the identified, drifting spillover ($0.33$ to $0.54$) that every constant-coefficient competitor averages away, certified against $99$ relabelled graphs at $p=0.010$, and the identification diagnostics of Section~\ref{sec:theory}, which apply to every model in this table, since they share the design.

\subsection{Raw counts versus log counts}\label{sec:rawlog}
Finally we refit the network model with raw-count feedback, $g(y)=y$, tuning $q$ in the same way ($q=\QRaw$), and generate forecasts as before. The filtered coefficients average $(\BetaRawMean)$ over the evaluation period, so by Theorem~\ref{thm:poisson}(a) the fixed-coefficient predictive mean is infinite at every horizon $h\ge3$, and by part (b) it is infinite at $h=2$ once the random-walk innovation is integrated out.

Supplement Table~S1 shows the consequence. With $\log(1+y)$ feedback the displayed mean of the city-wide total is $513$--$517$ at every horizon and for every $S$, with a maximum simulated total of a few hundred above the median, and Theorem~\ref{thm:poisson}(e) identifies the $h\ge3$ entries in this row for what they are: stable Monte Carlo averages of a distribution whose population mean is nevertheless infinite, because the coefficients are sampled from the Gaussian posterior. Here the posterior is concentrated ($\lambda_{\max}(P)^{1/2}\approx0.08$), so the divergent coefficient draws have probability far below anything $S=5{,}000$ (or $S=10^{12}$) can reach, and the columns are flat in $S$. Stability in $S$ is evidence of existence at $h\le2$, where part (d)'s condition holds. At $h\ge3$ it is not evidence of anything, which is why our point forecasts at those horizons are the medians and frozen-coefficient means of Table~\ref{tab:pointfc}. With raw-count feedback the one-month mean is stable ($524$). The \emph{two}-month sampled-coefficient mean is also numerically stable at $524$, but Theorem~\ref{thm:poisson}(b) makes its population value infinite whenever the posterior places variance on the feedback coefficients, so this stability is the same finite-$S$ silence, not agreement with the fixed-coefficient formula \eqref{eq:twostep}, which is what the frozen rows estimate. At one and two months the raw model's MAE and log score are within $0.01$ and $3$ units of the log-count model (Supplement Table~S2). At three months the sampled mean is $2.6\times10^{3}$ with $100$ paths and beyond the reporting threshold with $1{,}000$ or more, and at four months it is beyond the threshold for every $S$ (entries marked div.\ in Supplement Table~S1). In every case the median of the simulated totals stays at $511$--$528$, and the fraction of exploding paths at $S=5{,}000$ is $4.0\%$ at three months and $18.9\%$ at four. Supplement Table~S1 reports every $S$, and these fractions are diagnostics tied to the disclosed cap. Across the twelve targets the four-month MAE of the raw-count model is $0.77$--$0.93$ at six origins and between $169$ and $1.6\times10^{10}$ at the other six. Its four-month log score ($-709$) is, by contrast, unremarkable. This is the silent failure of Theorem~\ref{thm:poisson} in real data: the log score and the quantiles give no warning that the point forecast is the sample mean of a distribution without a mean.

The same data also illustrate the margin in Theorem~\ref{thm:poisson}(d) for the log-count model at the one horizon where a margin exists: the largest count in the sample gives $L_t\le\LtChicago$, hence $\kappa_t\le\KappaChicago$, while the filtered covariance satisfies $\lambda_{\max}(P_{t+1|t})\le\LamMaxPChicago$ at every origin, so $2\kappa_t\lambda_{\max}(P)\le\MarginChicago$: the \emph{two}-step predictive mean under coefficient uncertainty is finite with a factor-seven margin (here (d) genuinely applies, unlike in the S5 boundary case) and the divergence threshold of the sharpness statement is far away ($2s^2L_0\approx0.03$). From three steps on no margin is possible (part (e) is unconditional) and the practical translation of the theorem is the reporting rule of Table~\ref{tab:pointfc}.

\paragraph{Order of analysis.} The negative-binomial layer, the unit offsets and the seasonal terms were introduced after Poisson PIT diagnostics computed on the 2015 evaluation window, so their Chicago comparison is not immune to selection on the test period. Three mitigations are in place: the per-origin nested tuning of Section~\ref{sec:forecast} removes the hyperparameter channel. The primary endpoint and the randomization test were fixed before the augmentation, and the Texas analysis below runs the final specification, chosen in advance, on data that none of these choices ever saw.

\subsection{County-level COVID-19 in Texas}\label{sec:covid}

To assess whether the Chicago pattern is specific to one dataset, the full final pipeline (negative-binomial layer, exposure offsets, seasonality, per-window tuning, benchmarks, both randomization tests, existence-aware reporting) runs on weekly confirmed COVID-19 cases in the $254$ counties of Texas (JHU CSSE repository), using the first $92$ weeks from March 2020 with the last $20$ as the test window, with a symmetrised $5$-nearest-neighbour graph from the county coordinates and county-population offsets from the JHU UID lookup table (populations span $64$ to $4.7$ million). The counts are two orders of magnitude larger than the burglaries (mean $189$ per county-week against $1.20$, a factor of about $158$, maximum $26{,}519$), so this is also a scale test. All model choices were fixed before touching these data, so Texas provides the out-of-sample check promised above.

\begin{table}[t]
\centering\footnotesize
\caption{Texas COVID-19, one protocol: composite marginal log score per week (mixture $S=300$ for the state-space rows, plug-in MLE for the rest), median-MAE at $h=1$, and $h=2$ log score where the predictive is simulable. The negative-binomial state-space rows carry population offsets and annual seasonality. The hhh4-type row's $254$ unit intercepts absorb population exposure by construction, and its fitted epidemic-network coefficient sits on the zero bound, making it effectively a rich no-network model. The spline, score-driven, rolling and expanding rows share the state-space rows' regressors, offsets, seasonality and negative-binomial layer exactly, so only the coefficient dynamics differ. The native row is the R \texttt{surveillance} fit with parameters held at the training fit, scored by the package's own scoring rules. Its rolling refits fail to converge at most origins, which we disclose. The two Poisson PNAR rows collapse on these overdispersed counts, a finding about the observation layer, not the network.}\label{tab:covidbench}
\resizebox{\textwidth}{!}{%
\begin{tabular}{lrrr}
\toprule
model & LS $h{=}1$ & med-MAE & LS $h{=}2$\\
\midrule
NSSM net (NB, pop.\ offsets, seasonal) & -1220.7 & 107.1 & -1235.2 \\
NSSM net (NB, plug-in at filtered mode) & -1226.8 & -- & -- \\
NSSM no-network (NB, pop.\ offsets, seasonal) & -1230.4 & 108.3 & -1242.2 \\
hhh4, native \texttt{surveillance} fit & -1240.5 & 98.1 & -- \\
Spline hhh4-type NB (time-varying $\lambda_t,\phi_t$) & -1243.7 & 106.7 & -- \\
Score-driven spillover (GAS $\phi_t$), same design & -1237.2 & 123.6 & -- \\
Rolling-window NB, same design ($w=8$) & -1230.5 & 142.0 & -- \\
Expanding-window NB, same design & -1256.6 & 135.1 & -- \\
EWMA--NB (per node) & -1266.1 & 120.5 & -- \\
PNAR(1) log-linear ($+\log$ pop.) & -10707.4 & 85.3 & -- \\
PNAR(1) linear & -11261.4 & 86.9 & -- \\
 
\bottomrule
\end{tabular}}
\end{table}

\begin{figure}[t]
\centering
\includegraphics[width=.86\textwidth]{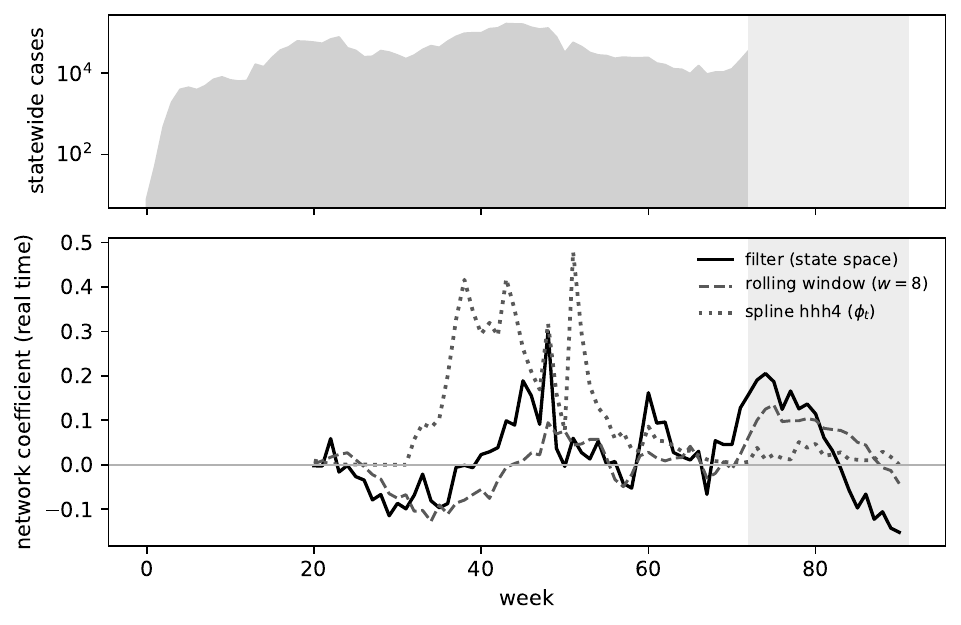}
\caption{Real-time network-coefficient estimates on the Texas data. Top: statewide weekly cases on a log scale, with the test window shaded. Bottom: the filtered spillover, the rolling-window estimate at width $w=8$, and the spline hhh4 network coefficient $\phi_t$, each computed from data up to the plotted week only. The filter and the spline track the waves as they occur. The rolling estimate runs out of phase with them at reduced amplitude.}\label{fig:tvpaths}
\end{figure}

The conclusions replicate with epidemic-scale numbers (Table~\ref{tab:covidbench}). The network term is worth $+9.7$ log-score points per week at $h=1$ ($-1220.7$ against $-1230.4$) and $+7.0$ at $h=2$, and $+1.1$ in weekly median-MAE ($107.1$ against $108.3$). On these data the plug-in row trails the mixture by six points per week, so integrating coefficient uncertainty carries real value at epidemic scale, and the constant-parameter competitors face a bar that the mixture clears further. The filtered spillover under the final model is wave-locked. It swings between $-0.15$ and $+0.30$ across the epidemic waves with posterior sd $0.03$, excursions exceeding ten posterior standard deviations, which no constant coefficient can represent, while its start-to-end drift is negligible ($0.07$ to $-0.10$). As in Chicago, exposure control reshapes the level of the coefficient and removes any monotone trend. What survives is the wave-locked variation and the predictive value of the adjacency, and these are the claims the paper makes. 
\paragraph{Time-varying competitors.} The remaining objection is that every external model above holds its coefficients fixed, so three competitors that let them move are fitted to the identical design, with the same regressors, offsets, seasonality and negative-binomial layer. The first re-estimates the constant-coefficient model on a rolling window at every origin, with the width tuned by mean per-origin score on a common origin set of the training period, which selects $w=8$. An expanding-window fit closes the other end of the range. The second lets the hhh4-type own- and network-coefficients vary through penalised cubic splines, refitted at every origin. The third is a score-driven recursion for the spillover in the spirit of generalised autoregressive score models, with estimated persistence $0.81$. The mixture leads all of them. Against the filter's plug-in score the tuned rolling window comes closest, $3.7$ points behind with a moving-block interval of $[-11,21]$, then the score-driven fit at $10.5$ $[-21,52]$, the native fit at $13.7$ $[-15,62]$, the spline at $16.9$ $[1.0,48]$, and the expanding window at $29.9$ $[-1.3,74]$. The score gap alone therefore understates the case, and the other two criteria settle it. The rolling window matches the score only by shrinking to eight weeks, and at that width its median absolute error rises to $142$ against the filter's $107$ and its coefficient path runs out of phase with the waves, correlating $-0.15$ with same-week statewide growth against the filter's $+0.28$, at less than half the amplitude. The score-driven path lags the same way ($-0.14$, best association four weeks late) and its medians reach $124$. The spline tracks the waves in phase ($+0.21$) and ties the filter's medians at about $107$, and pays three points of score even against the native constant fit. Figure~\ref{fig:tvpaths} shows the three real-time paths. No fixed or observation-driven alternative here delivers score, point accuracy and timing at once, and the filter does.

The randomization tests deserve the most attention. The stratified conditional test (labels permuted within population quintiles, $q$ re-tuned inside every permutation) gives the exact $p=0.010$ (observed gain $+203$, largest permutation $+181$), while the unrestricted label test gives only $p=0.040$, three permutations coming within reach of the observed statistic (largest $+208$): epidemic waves are common shocks aligned with county size, the contamination that S7 simulates, and conditioning on the population profile sharpens the reference distribution. Under the interpretation established by S7, the stratified rejection indicates that the adjacency carries structure beyond the size profile. The result is robust to the graph choice ($+203$ plug-in gain under $k=5$ against $+214$ under $k=8$), and the one-step PIT retains a residual defect ($\chi^2=259$ on ten bins, quoted for orientation only, against $6{,}624$ dependent node-week PITs in Chicago's case and $5{,}080$ here) that says even the negative binomial does not fully absorb epidemic bursts. The central predictive intervals are nevertheless close to nominal, with one-step empirical coverage of $0.86$ at the $80\%$ level and $0.94$ at the $95\%$ level, and the mean squared Pearson residual of $3.8$ localises the remaining misfit in the wave-peak weeks that the PIT flags. Two qualifications accompany the replication. The Poisson PNAR collapse ($-10{,}707$ and $-11{,}261$ per week) is a failure of the observation layer on overdispersed data and carries no information about networks. And the natively fitted hhh4 model, at $-1240.5$ the strongest constant-parameter competitor, retains the best median absolute error in the table ($98.1$) while losing about twenty points of score per week to the mixture. Since it also trails the no-network NSSM, most of the state-space advantage on these data comes from the drifting coefficients and the dispersion layer, with the network term contributing a further $+9.7$.

\subsection{Weekly measles in the Weser--Ems districts}\label{sec:measles}

The endemic--epidemic literature's own showcase data give the incumbent every advantage: weekly measles counts in the $17$ districts of the Weser--Ems region over $2001$--$2002$ ($T=104$), with the package's adjacency and populations, training on $2001$ and one-step evaluation through the $2002$ waves. The hhh4 row is fitted natively in R by the \texttt{surveillance} package, with unit intercepts, yearly seasonality and first-order neighbourhood weights, parameters held at the training fit, and its predictions scored by the package's own scoring rules. The state-space rows use the design of Section~\ref{sec:covid} with the same tuning protocol.

\begin{table}[t]
\centering\small
\caption{Weser--Ems measles, $52$ test weeks of $2002$: composite marginal log score per week, median-MAE at $h=1$, and empirical $95\%$ interval coverage where computed. The hhh4 row is the native R fit scored by its own package. Counts are small (mean $0.73$, $86\%$ zeros), so the discrete predictive intervals over-cover by construction.}\label{tab:measles}
\begin{tabular}{lccc}
\toprule
Model & LS/week & med-MAE & cov.\ $95\%$ \\
\midrule
NSSM net (NB, offsets, seasonal) & -10.7 & 0.64 & 0.99 \\
NSSM no-network & -10.8 & 0.62 & 0.99 \\
hhh4, native \texttt{surveillance} fit & -16.3 & 0.68 & -- \\
Rolling-window NB, same design ($w=16$) & -11.4 & 1.24 & -- \\
EWMA--NB (per node) & -12.6 & 0.63 & -- \\
 
\bottomrule
\end{tabular}
\end{table}

Table~\ref{tab:measles} reports the race. The state-space model beats the native incumbent by $5.6$ log-score points per week, with a moving-block bootstrap interval of $[2.0,\,9.6]$ that excludes zero, and its median error is the table's best alongside the smoothing baseline. Two honest notes accompany the win. First, the no-network state-space row sits within a tenth of a point of the network row, and the native fit's own network coefficient is estimated at zero, so on these data the advantage comes from the drifting coefficients and the dispersion layer, with the adjacency contributing next to nothing on either side. The result is a win for the state-space machinery, and it carries no evidence for this network. Second, with seventeen districts and counts this small the information theory of Section~\ref{sec:ident} predicts modest per-week identification, and the filter behaves accordingly, with one-step coverage $0.96$ and $0.99$ at nominal $0.80$ and $0.95$ (discrete over-coverage) and mean squared Pearson residual $1.05$. The implementation cross-check on these data, training log-likelihoods from R and from our code agreeing to seven significant figures, is in the Supplement.

\subsection{Data availability}\label{sec:repro}
The Chicago counts and contiguity network come from the public repository cited above, and the Texas analysis uses two public county-level files from the Johns Hopkins CSSE repository, so every input is public. Code that regenerates all tables and figures from these sources, together with a suite of numerical checks on the theoretical results, is available from the authors.
\section{Discussion}\label{sec:disc}

The network state-space model is attractive because it is simple: a regression with three coefficients that drift, fitted by a filter that is exact in the Gaussian case and a standard approximation in the Poisson case. The contribution of this paper is to say precisely what such a model can and cannot learn, and to turn the answer into diagnostics that can be computed before and after fitting.

\paragraph{Recommended reporting.} An NSSM analysis should report five quantities. (i) The graph functionals $\tauW$, $\sum_in_i^{-1}$ and $\tau_{\Wc}$, which locate the network between the sparse and the reflection-problem extremes. (ii) The realised profile information $\calI_t$ over time, the cross-sectional information actually available for the spillover, which by Theorem~\ref{thm:filter}(ii) brackets its posterior variance. Label permutations leave every graph invariant fixed, $\tauW$ included, while the realised $\calI_t$ changes under relabelling, falling in Chicago from $66.9$ to about $50$, so the information is aligned with the observed field and only the realised quantity can show this. (iii) The filtered persistence $\beta_{1,t}+\beta_{2,t}$, which by Proposition~\ref{prop:plugin} governs how network misspecification propagates with the horizon. (iv) A randomization test against label-permuted networks, exact for the exchangeability null, together with the stratified conditional version of Section~\ref{sec:placebo}, whose null tolerates unit heterogeneity, and degree-preserving rewirings as a descriptive placebo. (v) Point forecasts chosen among functionals that exist: medians and quantiles at any horizon, sampled-coefficient means only at $h\le2$ under the condition of Theorem~\ref{thm:poisson}(d), frozen-coefficient means otherwise, with any simulation caps disclosed. The Chicago application shows what each item prevents. Without (iii), a constant-coefficient analysis misses a drift in the pooled spillover from about $0.4$ to $0.6$, and without the augmented refit of Section~\ref{sec:bench} it would go unnoticed that half of that drift is exposure and seasonality, the exposure-controlled path being $\FinalBOneFirst\to\FinalBOneLast$. Without (i) and Proposition~\ref{prop:equiv}, a comparison against the no-network model alone credits the network with a gain that a quarter-mixture with the complete graph reproduces, necessarily, since the mixture is the same cross-sectional model with the spillover relabelled. Without (v), a raw-count specification reports three-month point forecasts beyond any physical scale with no warning from the scoring rules, and the $\log(1+y)$ specification silently reports three-month means that do not exist.

\paragraph{Limitations.} The theory is cross-sectional: it describes what one period's data contribute and how the random-walk prior bounds the accumulation of information, but it does not characterise the joint $(N,T)\to\infty$ behaviour of the filter under a fixed network, for which the natural route is the posterior contraction theory of state-space models. Theorem~\ref{thm:filter} is exact only for the Gaussian model. For the Poisson model the Laplace filter is an approximation whose accuracy improves with the information, so that the identification results, which transfer exactly, are the ones to rely on, and the calibration result is a heuristic there. Proposition~\ref{prop:plugin} concerns plug-in means and a linear recursion. Its application to the log-linear count model is qualitative. Theorem~\ref{thm:poisson}(d)--(e) now bracket the two-step boundary within a bounded factor ($1/(2\kappa_t)\le s^2_{\mathrm{crit}}\le1/(2L_0)$ in the scalar direction) and settle every horizon beyond it, but the exact two-step constant remains open, as does the moment frontier for bounded coefficient processes such as $\beta_t=b\tanh(z_t)$, which would restore all-horizon means at the price of a nonlinear filter. A conditional Bernstein--von Mises theorem for the Poisson mode filter (Gaussianity of the cross-sectional posterior to explicit error as $\calI_t\to\infty$) is the missing bridge that would make Proposition~\ref{prop:weighted}(iii) a statement about the true posterior itself. Study S4's coverage of $95.5\%$ is encouraging evidence, not a substitute. The empirical evaluation now spans three datasets and external benchmarks on each, including rolling-window, expanding-window, penalised-spline and score-driven time-varying fits of the identical design, but its limits should be stated: the randomization $p$-values are exact yet their resolution is capped by the number of permutations ($1/100$). The hhh4 comparison is now anchored to the authors' own software: the \texttt{surveillance} package is fitted natively in R on the Texas and measles data, its predictions are scored by its own scoring rules, and the training log-likelihoods of our implementation agree with the package to seven significant figures on the measles data and to $0.12\%$ on Texas (Supplement). The PNAR and GNAR implementations remain ours, validated by parameter-recovery self-tests, and cross-checking them against the authors' packages is the remaining natural referee request. Exogenous covariates (demographics, exposure, weather) enter none of the fits, and at the benchmark frontier the network's marginal one-step forecasting value is $+0.4$ log-score points with an interval covering zero. The case for the network term rests on the identified drifting spillover and the exact test. Finally, the negative-binomial layer that repairs the calibration defect is now covered on the divergence side by Proposition~\ref{prop:nb} (its heavier tails move nothing, because the frontier is driven by the exponential link and the Gaussian coefficient tail) while Proposition~\ref{prop:nb} now settles both sides at two steps, where the negative-binomial frontier is exactly $2s^2L_0=1$. A softplus link, which keeps the intensity linear in large counts, is the constructive alternative the INGARCH literature suggests and would be the natural repair if all-horizon means are required.

\paragraph{Extensions.} Two directions seem most useful. The identification analysis extends directly to models with several neighbourhood orders \citep{knight2020gnar} or several networks, where the profile information becomes a matrix and the question is the joint roughness of the fields $W_1x,\dots,W_rx$ after projecting out the constant and the own lag. Theorem~\ref{thm:ident}(ii) then characterises the collinear configurations. And the filter can be replaced by a smoother, or by a full Bayesian treatment of $(\sigma^2,Q)$ with shrinkage priors on the state innovations \citep{fruhwirth2010stochastic,bitto2019achieving}, without changing the identification results, which depend only on the design.

\bibliographystyle{plainnat}
\bibliography{refs}

@article{zhu2017network,
  author  = {Zhu, Xuening and Pan, Rui and Li, Guodong and Liu, Yuewen and Wang, Hansheng},
  title   = {Network vector autoregression},
  journal = {The Annals of Statistics},
  year    = {2017},
  volume  = {45},
  number  = {3},
  pages   = {1096--1123}
}

@article{knight2020gnar,
  author  = {Knight, Marina and Leeming, Kathryn and Nason, Guy and Nunes, Matthew},
  title   = {Generalized network autoregressive processes and the {GNAR} package},
  journal = {Journal of Statistical Software},
  year    = {2020},
  volume  = {96},
  number  = {5},
  pages   = {1--36}
}

@article{armillotta2023nonlinear,
  author  = {Armillotta, Mirko and Fokianos, Konstantinos},
  title   = {Nonlinear network autoregression},
  journal = {The Annals of Statistics},
  year    = {2023},
  volume  = {51},
  number  = {6},
  pages   = {2526--2552}
}

@article{armillotta2024count,
  author  = {Armillotta, Mirko and Fokianos, Konstantinos},
  title   = {Count network autoregression},
  journal = {Journal of Time Series Analysis},
  year    = {2024},
  volume  = {45},
  number  = {4},
  pages   = {584--612}
}

@article{diebold2014network,
  author  = {Diebold, Francis X. and Y{\i}lmaz, Kamil},
  title   = {On the network topology of variance decompositions: Measuring the connectedness of financial firms},
  journal = {Journal of Econometrics},
  year    = {2014},
  volume  = {182},
  number  = {1},
  pages   = {119--134}
}

@article{west1985dynamic,
  author  = {West, Mike and Harrison, P. Jeff and Migon, Helio S.},
  title   = {Dynamic generalized linear models and {B}ayesian forecasting},
  journal = {Journal of the American Statistical Association},
  year    = {1985},
  volume  = {80},
  number  = {389},
  pages   = {73--83}
}

@article{fahrmeir1992posterior,
  author  = {Fahrmeir, Ludwig},
  title   = {Posterior mode estimation by extended {K}alman filtering for multivariate dynamic generalized linear models},
  journal = {Journal of the American Statistical Association},
  year    = {1992},
  volume  = {87},
  number  = {418},
  pages   = {501--509}
}

@article{primiceri2005time,
  author  = {Primiceri, Giorgio E.},
  title   = {Time varying structural vector autoregressions and monetary policy},
  journal = {The Review of Economic Studies},
  year    = {2005},
  volume  = {72},
  number  = {3},
  pages   = {821--852}
}

@article{koop2013large,
  author  = {Koop, Gary and Korobilis, Dimitris},
  title   = {Large time-varying parameter {VARs}},
  journal = {Journal of Econometrics},
  year    = {2013},
  volume  = {177},
  number  = {2},
  pages   = {185--198}
}

@article{manski1993identification,
  author  = {Manski, Charles F.},
  title   = {Identification of endogenous social effects: The reflection problem},
  journal = {The Review of Economic Studies},
  year    = {1993},
  volume  = {60},
  number  = {3},
  pages   = {531--542}
}

@article{lee2007identification,
  author  = {Lee, Lung-fei},
  title   = {Identification and estimation of econometric models with group interactions, contextual factors and fixed effects},
  journal = {Journal of Econometrics},
  year    = {2007},
  volume  = {140},
  number  = {2},
  pages   = {333--374}
}

@article{bramoulle2009identification,
  author  = {Bramoull{\'e}, Yann and Djebbari, Habiba and Fortin, Bernard},
  title   = {Identification of peer effects through social networks},
  journal = {Journal of Econometrics},
  year    = {2009},
  volume  = {150},
  number  = {1},
  pages   = {41--55}
}

@article{fruhwirth2010stochastic,
  author  = {Fr{\"u}hwirth-Schnatter, Sylvia and Wagner, Helga},
  title   = {Stochastic model specification search for {G}aussian and partial non-{G}aussian state space models},
  journal = {Journal of Econometrics},
  year    = {2010},
  volume  = {154},
  number  = {1},
  pages   = {85--100}
}

@article{bitto2019achieving,
  author  = {Bitto, Angela and Fr{\"u}hwirth-Schnatter, Sylvia},
  title   = {Achieving shrinkage in a time-varying parameter model framework},
  journal = {Journal of Econometrics},
  year    = {2019},
  volume  = {210},
  number  = {1},
  pages   = {75--97}
}

@article{zhu2020grouped,
  author  = {Zhu, Xuening and Pan, Rui},
  title   = {Grouped network vector autoregression},
  journal = {Statistica Sinica},
  year    = {2020},
  volume  = {30},
  number  = {3},
  pages   = {1437--1462}
}

@article{chen2023community,
  author  = {Chen, Elynn Y. and Fan, Jianqing and Zhu, Xuening},
  title   = {Community network auto-regression for high-dimensional time series},
  journal = {Journal of Econometrics},
  year    = {2023},
  volume  = {235},
  number  = {2},
  pages   = {1239--1256}
}

@article{fokianos2011log,
  author  = {Fokianos, Konstantinos and Tj{\o}stheim, Dag},
  title   = {Log-linear {P}oisson autoregression},
  journal = {Journal of Multivariate Analysis},
  year    = {2011},
  volume  = {102},
  number  = {3},
  pages   = {563--578}
}

@article{mohler2011self,
  author  = {Mohler, George O. and Short, Martin B. and Brantingham, P. Jeffrey and Schoenberg, Frederic P. and Tita, George E.},
  title   = {Self-exciting point process modeling of crime},
  journal = {Journal of the American Statistical Association},
  year    = {2011},
  volume  = {106},
  number  = {493},
  pages   = {100--108}
}

@article{clark2021class,
  author  = {Clark, Nicholas J. and Dixon, Philip M.},
  title   = {A class of spatially correlated self-exciting statistical models},
  journal = {Spatial Statistics},
  year    = {2021},
  volume  = {43},
  pages   = {100493}
}

@article{diebold1995comparing,
  author  = {Diebold, Francis X. and Mariano, Roberto S.},
  title   = {Comparing predictive accuracy},
  journal = {Journal of Business \& Economic Statistics},
  year    = {1995},
  volume  = {13},
  number  = {3},
  pages   = {253--263}
}

@article{gneiting2007strictly,
  author  = {Gneiting, Tilmann and Raftery, Adrian E.},
  title   = {Strictly proper scoring rules, prediction, and estimation},
  journal = {Journal of the American Statistical Association},
  year    = {2007},
  volume  = {102},
  number  = {477},
  pages   = {359--378}
}

@article{czado2009predictive,
  author  = {Czado, Claudia and Gneiting, Tilmann and Held, Leonhard},
  title   = {Predictive model assessment for count data},
  journal = {Biometrics},
  year    = {2009},
  volume  = {65},
  number  = {4},
  pages   = {1254--1261}
}

@book{durbin2012time,
  author    = {Durbin, James and Koopman, Siem Jan},
  title     = {Time Series Analysis by State Space Methods},
  edition   = {2nd},
  publisher = {Oxford University Press},
  address   = {Oxford},
  year      = {2012}
}

@article{laurent2000adaptive,
  author  = {Laurent, B{\'e}atrice and Massart, Pascal},
  title   = {Adaptive estimation of a quadratic functional by model selection},
  journal = {The Annals of Statistics},
  year    = {2000},
  volume  = {28},
  number  = {5},
  pages   = {1302--1338}
}

@article{li2026grouped,
  author  = {Li, Degui and Peng, Bin and Tang, Songqiao and Wu, Wei Biao},
  title   = {Estimation of grouped time-varying network vector autoregressive models},
  journal = {The Annals of Statistics},
  year    = {2026},
  volume  = {54},
  number  = {2},
  pages   = {621--646}
}

@article{ding2025tvnar,
  author  = {Ding, Yi and Zhu, Xuening and Pan, Rui and Zhang, Bo},
  title   = {Network vector autoregression with time-varying nodal influence},
  journal = {Computational Economics},
  year    = {2025},
  note    = {In press}
}

@article{maletz2024spatiotemporal,
  author  = {Maletz, Steffen and Fokianos, Konstantinos and Fried, Roland},
  title   = {Spatio-temporal count autoregression},
  journal = {Data Science in Science},
  year    = {2024},
  volume  = {3},
  number  = {1},
  pages   = {2425171}
}

@article{held2005statistical,
  author  = {Held, Leonhard and H{\"o}hle, Michael and Hofmann, Mathias},
  title   = {A statistical framework for the analysis of multivariate infectious disease surveillance counts},
  journal = {Statistical Modelling},
  year    = {2005},
  volume  = {5},
  number  = {3},
  pages   = {187--199}
}

@article{frisch1933partial, author={Frisch, Ragnar and Waugh, Frederick V.}, title={Partial time regressions as compared with individual trends}, journal={Econometrica}, year={1933}, volume={1}, number={4}, pages={387--401}}

@article{lovell1963seasonal, author={Lovell, Michael C.}, title={Seasonal adjustment of economic time series and multiple regression analysis}, journal={Journal of the American Statistical Association}, year={1963}, volume={58}, number={304}, pages={993--1010}}

@book{harvey1989forecasting, author={Harvey, Andrew C.}, title={Forecasting, Structural Time Series Models and the {K}alman Filter}, publisher={Cambridge University Press}, year={1989}}

@article{lee2004asymptotic, author={Lee, Lung-fei}, title={Asymptotic distributions of quasi-maximum likelihood estimators for spatial autoregressive models}, journal={Econometrica}, year={2004}, volume={72}, number={6}, pages={1899--1925}}

@article{case1991spatial, author={Case, Anne C.}, title={Spatial patterns in household demand}, journal={Econometrica}, year={1991}, volume={59}, number={4}, pages={953--965}}

@article{blume2015linear, author={Blume, Lawrence E. and Brock, William A. and Durlauf, Steven N. and Jayaraman, Rajshri}, title={Linear social interactions models}, journal={Journal of Political Economy}, year={2015}, volume={123}, number={2}, pages={444--496}}

@article{yu2008quasi, author={Yu, Jihai and de Jong, Robert and Lee, Lung-fei}, title={Quasi-maximum likelihood estimators for spatial dynamic panel data with fixed effects when both $n$ and $T$ are large}, journal={Journal of Econometrics}, year={2008}, volume={146}, number={1}, pages={118--134}}

@book{elhorst2014spatial, author={Elhorst, J. Paul}, title={Spatial Econometrics: From Cross-Sectional Data to Spatial Panels}, publisher={Springer}, year={2014}}

@article{fokianos2009poisson, author={Fokianos, Konstantinos and Rahbek, Anders and Tj{\o}stheim, Dag}, title={Poisson autoregression}, journal={Journal of the American Statistical Association}, year={2009}, volume={104}, number={488}, pages={1430--1439}}

@book{anderson1979optimal, author={Anderson, Brian D. O. and Moore, John B.}, title={Optimal Filtering}, publisher={Prentice-Hall}, year={1979}}

@article{berry2020bayesian, author={Berry, Lindsay R. and West, Mike}, title={Bayesian forecasting of many count-valued time series}, journal={Journal of Business \& Economic Statistics}, year={2020}, volume={38}, number={4}, pages={872--887}}

@article{blasques2016spillover, author={Blasques, Francisco and Koopman, Siem Jan and Lucas, Andre and Schaumburg, Julia}, title={Spillover dynamics for systemic risk measurement using spatial financial time series models}, journal={Journal of Econometrics}, year={2016}, volume={195}, number={2}, pages={211--223}}

@article{catania2017dynamic, author={Catania, Leopoldo and Bill{\'e}, Anna Gloria}, title={Dynamic spatial autoregressive models with autoregressive and heteroskedastic disturbances}, journal={Journal of Applied Econometrics}, year={2017}, volume={32}, number={6}, pages={1178--1196}}

\appendix
\section{Proofs}\label{app:proofs}

\subsection{Auxiliary lemmas}

Throughout, for a symmetric positive semidefinite $K\times K$ matrix $M$ and the index $2$ (the position of $\beta_1$), let $E=[e_1,e_3,\dots,e_K]$ collect the remaining coordinate vectors and define
\begin{equation}\label{eq:schurvar}
s(M)\;=\;\inf_{a\in\R^{K-1}}\,(e_2+Ea)\T M(e_2+Ea)\;\ge 0 .
\end{equation}

\begin{lemma}[Schur complements and profile information]\label{lem:schur}
\begin{enumerate}[label=(\alph*),leftmargin=*]
\item If $M\succ0$, then \eqref{eq:schurvar} equals the Schur complement $M_{22}-M_{2,-2}M_{-2,-2}^{-1}M_{-2,2}$, the infimum is attained at $a^*=-M_{-2,-2}^{-1}M_{-2,2}$, and $(M^{-1})_{22}=1/s(M)$.
\item $s$ is superadditive: $s(A+G)\ge s(A)+s(G)$ for positive semidefinite $A,G$.
\item For $X=[\one,w,x]$ and $M=\sigma^{-2}X\T X$, $s(M)=\sigma^{-2}\|M_Zw\|^2$ with $Z=[\one,x]$, and when $Z$ has full column rank the infimum is attained at $a^*=-Z^+w$, the negative of the least-squares coefficients of $w$ on $Z$.
\end{enumerate}
\end{lemma}
\begin{proof}
(a) The quadratic $a\mapsto M_{22}+2a\T M_{-2,2}+a\T M_{-2,-2}a$ is strictly convex when $M_{-2,-2}\succ0$, with minimiser $a^*$ and minimum value $M_{22}-M_{2,-2}M_{-2,-2}^{-1}M_{-2,2}$. The identity $(M^{-1})_{22}=1/s(M)$ is the block-inverse formula. (b) The infimum of a sum is at least the sum of the infima. (c) $(e_2+Ea)\T X\T X(e_2+Ea)=\|X(e_2+Ea)\|^2=\|w+Za\|^2$, and $\inf_a\|w+Za\|^2=\|M_Zw\|^2$, attained at $a=-Z^+w$ when $Z$ has full column rank.
\end{proof}

\begin{lemma}[The functional $\tauW$]\label{lem:tau}
Let $W$ be the row-normalised weight matrix of a graph in which every node has at least one out-neighbour, with out-degrees $n_i$. Then $\tauW=\sum_{i,j}w_{ij}^2-N^{-1}\|W\T\one\|^2=\sum_in_i^{-1}-N^{-1}\|W\T\one\|^2\le\sum_in_i^{-1}\le N$. If the graph is undirected and $d$-regular then $\tauW=N/d-1$. In particular $\tau_{\Wc}=1/(N-1)$. If $n_i\ge cN$ for all $i$ then $\tauW\le1/c$.
\end{lemma}
\begin{proof}
$\tr(W\T M_1W)=\tr(W\T W)-N^{-1}\tr(W\T\one\one\T W)=\sum_{i,j}w_{ij}^2-N^{-1}\|W\T\one\|^2$. Since $w_{ij}\in\{0,1/n_i\}$ with exactly $n_i$ nonzero entries in row $i$, $\sum_jw_{ij}^2=1/n_i$. For an undirected $d$-regular graph $W=A/d$ is symmetric and doubly stochastic, so $W\T\one=\one$ and $\tauW=N/d-1$. The complete graph is $(N-1)$-regular. The remaining bounds follow from $n_i\ge1$ and $n_i\ge cN$.
\end{proof}

\begin{lemma}[Poisson moments]\label{lem:poismom}
Let $Y\sim\Pois(\lambda)$ and $p\ge1$. There is a constant $C_p$ depending only on $p$ such that $\E(1+Y)^p\le C_p(1+\lambda)^p$.
\end{lemma}
\begin{proof}
For integer $p$, $\E Y^k=\sum_{j=0}^kS(k,j)\lambda^j\le B_k(1+\lambda)^k$ with $S(k,j)$ the Stirling numbers of the second kind and $B_k=\sum_jS(k,j)$ the Bell numbers, so $\E(1+Y)^p=\sum_k\binom pk\E Y^k\le(1+\lambda)^p\sum_k\binom pkB_k=:C_p(1+\lambda)^p$. For non-integer $p\ge1$, Lyapunov's inequality gives $\E(1+Y)^p\le\{\E(1+Y)^{\lceil p\rceil}\}^{p/\lceil p\rceil}\le C_{\lceil p\rceil}^{p/\lceil p\rceil}(1+\lambda)^p$.
\end{proof}

\begin{lemma}[Exponential of a quadratic in $\log(1+S)$, $S$ Poisson]\label{lem:poisexp}
Let $S\sim\Pois(\Lambda)$ with $\Lambda>0$, $a\ge0$ and $0\le q\le3/8$. Then, with $\ell=2+\log(1+\Lambda)$,
\[
\E\exp\big\{a\log(1+S)+q\log^2(1+S)\big\}\;\le\;\exp(a\ell+q\ell^2)+1.55\,e^{2a^2}.
\]
\end{lemma}
\begin{proof}
Write $g(k)=\exp\{a\log(1+k)+q\log^2(1+k)\}$, which is nondecreasing in $k\ge0$, and let $K=\lceil e^2\Lambda\rceil\ge1$. For $k<K$, $g(k)\le g(K)$, and since $1+K\le e^2\Lambda+2\le e^2(1+\Lambda)$ we have $\log(1+K)\le\ell$, so $\sum_{k<K}\Prob(S=k)g(k)\le\exp(a\ell+q\ell^2)$. For $k\ge K\ge e^2\Lambda$, $\Prob(S=k)\le\Lambda^k/k!\le(e\Lambda/k)^k\le e^{-k}$ by $k!\ge(k/e)^k$. Using $\log(1+k)\le\sqrt k$ for $k\ge0$ (the function $\sqrt k-\log(1+k)$ vanishes at $0$ and has derivative $(\sqrt k-1)^2/\{2\sqrt k(1+k)\}\ge0$), $a\sqrt k\le2a^2+k/8$ and $q\log^2(1+k)\le qk\le3k/8$,
\[
\sum_{k\ge K}e^{-k}g(k)\le\sum_{k\ge1}\exp\{-k+2a^2+k/8+3k/8\}=e^{2a^2}\sum_{k\ge1}e^{-k/2}=\frac{e^{2a^2}}{e^{1/2}-1}\le1.55\,e^{2a^2}. \qedhere
\]
\end{proof}

\begin{lemma}[Gaussian exponential moments]\label{lem:gauss}
Let $\theta\sim\calN(m,P)$ in $\R^K$ with $P\succ0$, and let $\kappa\ge0$, $b\ge0$. If $2\kappa\lambda_{\max}(P)<1$ then $\E\exp\{\kappa\|\theta\|^2+b\|\theta\|\}<\infty$.
\end{lemma}
\begin{proof}
Put $\xi=\theta-m$. Then $\|\theta\|^2=\|\xi\|^2+2m\T\xi+\|m\|^2$ and, for any $\varepsilon>0$, $b\|\theta\|\le b\|m\|+\varepsilon\|\xi\|^2+b^2/(4\varepsilon)$. Hence $\kappa\|\theta\|^2+b\|\theta\|\le(\kappa+\varepsilon)\|\xi\|^2+2\kappa m\T\xi+c_\varepsilon$ for a constant $c_\varepsilon$, and
\[
\E\exp\{(\kappa+\varepsilon)\|\xi\|^2+2\kappa m\T\xi\}=\frac{1}{(2\pi)^{K/2}|P|^{1/2}}\int\exp\Big\{-\tfrac12\xi\T\big(P^{-1}-2(\kappa+\varepsilon)I\big)\xi+2\kappa m\T\xi\Big\}d\xi ,
\]
which is finite as soon as $P^{-1}-2(\kappa+\varepsilon)I\succ0$, i.e.\ $2(\kappa+\varepsilon)\lambda_{\max}(P)<1$. Choose $\varepsilon$ small.
\end{proof}

\begin{lemma}[Scalar random-walk Riccati map]\label{lem:riccati}
For $q>0$ and $\calI>0$ let $\varphi_{\calI}(v)=\{(v+q)^{-1}+\calI\}^{-1}$ on $[0,\infty)$. Then \textup{(a)} $\varphi_\calI$ is increasing and concave, and nonincreasing in $\calI$. \textup{(b)} it has the unique fixed point $P^\ast(\calI,q)=\{\sqrt{q^2\calI^2+4q\calI}-q\calI\}/(2\calI)$, and the iterates $\varphi_\calI^{(t)}(v_0)$ converge to $P^\ast$ monotonically from any $v_0\ge0$. \textup{(c)} $P^\ast\le\sqrt{q/\calI}$ always, and $P^\ast\ge\tfrac{2}{1+\sqrt5}\sqrt{q/\calI}$ when $q\calI\le1$.
\end{lemma}

\begin{proof}
Write $u=v+q$ and $\varphi(v)=u/(1+\calI u)$, a M\"obius map with $\varphi'(v)=(1+\calI u)^{-2}\in(0,1)$ and $\varphi''(v)=-2\calI(1+\calI u)^{-3}<0$, which is also decreasing in $\calI$ for fixed $u$. This is (a). A fixed point solves $\calI v^2+q\calI v-q=0$, whose unique positive root is $P^\ast$. Since $g(v)=\varphi(v)-v$ has $g'(v)=\varphi'(v)-1<0$, $g$ is strictly decreasing with $g(P^\ast)=0$, so $\varphi(v)>v$ for $v<P^\ast$ and $\varphi(v)<v$ for $v>P^\ast$. With $\varphi$ increasing, the iterates are monotone and bounded, hence convergent, and the limit is a fixed point, giving (b). For (c), let $s=q\calI$ and note $P^\ast\calI=(\sqrt{s^2+4s}-s)/2=2s/(s+\sqrt{s^2+4s})$, so
\[
\frac{P^\ast}{\sqrt{q/\calI}}=\frac{P^\ast\calI}{\sqrt{s}}=\frac{2}{\sqrt{s}+\sqrt{s+4}} ,
\]
which is at most $2/\sqrt4=1$ for all $s>0$ and at least $2/(1+\sqrt5)$ for $s\le1$.
\end{proof}

\subsection{Proof of Theorem~\ref{thm:ident}}

\emph{(i)} Both $\one$ and $x$ lie in the column space of $Z$, so $M_Zx=0$ and $M_ZWx=M_Z(W-I)x$. Hence $\calI(\beta_1;x)=\sigma^{-2}\|M_Z(I-W)x\|^2\le\sigma^{-2}\|(I-W)x\|^2$ because $M_Z$ is an orthogonal projector. If $x\notin\mathrm{span}(\one)$, then $\mathrm{col}(Z)=\mathrm{span}(\one,\tilde x)$ with $\tilde x=M_1x\perp\one$, so $M_Z=M_1-\tilde x\tilde x\T/\|\tilde x\|^2$ and, for $w=Wx$, $\|M_Zw\|^2=\|M_1w\|^2-(\tilde x\T w)^2/\|\tilde x\|^2$ because $M_1w=M_Zw+(\tilde x\T w/\|\tilde x\|^2)\tilde x$ is an orthogonal decomposition ($\tilde x\T M_1 w=\tilde x\T w$).

\emph{(ii)} If $W=\Wc$ then $\Wc x=(J-I)x/(N-1)=\{N\bar x\one-x\}/(N-1)\in\mathrm{col}(Z)$, so $M_Z\Wc x=0$ for every $x$. Conversely, suppose $M_ZWx=0$, i.e.\ $Wx\in\mathrm{span}(\one,x)$, for every $x$. Taking $x=e_j$ gives $We_j=a_j\one+b_je_j$ for some scalars $a_j,b_j$, so the off-diagonal entries of column $j$ all equal $a_j$, and $w_{jj}=0$ forces $b_j=-a_j$. Row-stochasticity then gives $\sum_{j\ne i}a_j=1$ for every $i$, i.e.\ $\sum_ja_j-a_i=1$, so all $a_i$ are equal to a common $a$ with $(N-1)a=1$. Hence $W=(J-I)/(N-1)=\Wc$. The rank statement follows from $\Wc x\in\mathrm{col}(Z)$.

\emph{(iii)} Conditionally on $(\calF_{t-2},\theta_{t-1})$, $x=y_{t-1}=\mu+\varepsilon$ with $\mu=X_{t-1}\theta_{t-1}$ fixed and $\varepsilon\sim\calN(0,\sigma^2I)$. Since $\mathrm{col}(\one)\subset\mathrm{col}(Z)$, $\|M_Zv\|\le\|M_1v\|$ for every $v$, and therefore $\calI(\beta_1;x)\le\sigma^{-2}\|M_1Wx\|^2$. Expanding, $\E\|M_1W(\mu+\varepsilon)\|^2=\|M_1W\mu\|^2+\E\,\varepsilon\T W\T M_1W\varepsilon=\|M_1W\mu\|^2+\sigma^2\tauW$, which gives the upper bound.

For the lower bound let $\mu=c\one$. Then $Wx=c\one+W\varepsilon$, $\tilde x=M_1\varepsilon$, and by (i), on the almost-sure event $\tilde x\ne0$,
\[
\sigma^2\calI(\beta_1;x)=\|M_1W\varepsilon\|^2-\frac{(\varepsilon\T M_1W\varepsilon)^2}{\|M_1\varepsilon\|^2}.
\]
Let $A=\{\|M_1\varepsilon\|^2\ge\sigma^2(N-1)/2\}$. Since $\|M_1\varepsilon\|^2/\sigma^2\sim\chi^2_{N-1}$, the inequality $\Prob(\chi^2_n\le n-2\sqrt{nu})\le e^{-u}$ of \citet{laurent2000adaptive} with $u=n/16$ gives $\Prob(A^c)\le e^{-(N-1)/16}$. Because $\calI\ge0$,
\[
\sigma^2\E\,\calI(\beta_1;x)\;\ge\;\E\big[\|M_1W\varepsilon\|^2\mathbf 1_A\big]-\frac{2}{\sigma^2(N-1)}\E\big[(\varepsilon\T M_1W\varepsilon)^2\big].
\]
For the first term, $\|M_1W\varepsilon\|^2=\varepsilon\T B\varepsilon$ with $B=W\T M_1W\succeq0$, $\tr B=\tauW$, so $\E(\varepsilon\T B\varepsilon)^2=\sigma^4\{(\tr B)^2+2\tr(B^2)\}\le3\sigma^4\tauW^2$ and, by Cauchy--Schwarz,
\[
\E[\|M_1W\varepsilon\|^2\mathbf 1_{A^c}]\le\sqrt3\sigma^2\tauW e^{-(N-1)/32},\quad\text{hence}\quad\E[\|M_1W\varepsilon\|^2\mathbf 1_A]\ge\sigma^2\tauW(1-\sqrt3e^{-(N-1)/32}).
\]
For the second term, write $C=M_1W$ and $C_s=(C+C\T)/2$. Then $\varepsilon\T C\varepsilon=\varepsilon\T C_s\varepsilon$ and $\E(\varepsilon\T C_s\varepsilon)^2=\sigma^4\{(\tr C_s)^2+2\tr(C_s^2)\}$ with $\tr C_s=\tr(M_1W)=\tr W-\one\T W\one/N=-1$ and $2\tr(C_s^2)=\tr(C^2)+\tr(CC\T)=\{\tr(W^2)-1\}+\tauW$, where $\tr(M_1WM_1W)=\tr(W^2)-2\one\T W^2\one/N+(\one\T W\one)^2/N^2=\tr(W^2)-1$. Thus $\E(\varepsilon\T C\varepsilon)^2=\sigma^4\{\tr(W^2)+\tauW\}\le\sigma^4(N+\tauW)$, using $\tr(W^2)=\sum_{i,j}w_{ij}w_{ji}\le\sum_{i,j}w_{ij}=N$. Collecting terms,
\[
\E\,\calI(\beta_1;x)\ge\tauW\Big(1-\sqrt3e^{-(N-1)/32}-\frac{2}{N-1}\Big)-\frac{2N}{N-1}=\tauW-c_N ,
\]
and $c_N\le N(\tfrac{2}{N-1}+\sqrt3e^{-(N-1)/32})+\tfrac{2N}{N-1}=4+\tfrac{4}{N-1}+\sqrt3Ne^{-(N-1)/32}$ by Lemma~\ref{lem:tau}. The right-hand side is maximal at $N=32$, where it equals $25.2$, and tends to $4$.

\emph{(iv)} is Lemma~\ref{lem:tau}. \qed

\subsection{Proof of Proposition~\ref{prop:weighted}}
Write $\tilde X=\Lambda^{1/2}X=[\Lambda^{1/2}\one,\Lambda^{1/2}Wx,\Lambda^{1/2}x]$ and $\tilde Z=\Lambda^{1/2}Z$. The weighted least-squares problem defining $\calI^\Lambda$ is $\min_a\|\Lambda^{1/2}Wx+\tilde Za\|^2$ (the sign of $a$ is immaterial), whose value is $\|M_{\tilde Z}\Lambda^{1/2}Wx\|^2$ by Lemma~\ref{lem:schur}(c) applied to $\tilde X$. The same lemma gives $\calI^\Lambda=s(\tilde X\T\tilde X)=1/[(X\T\Lambda X)^{-1}]_{22}$ when $X\T\Lambda X=\tilde X\T\tilde X\succ0$. The roughness identity follows because $\Lambda^{1/2}(I-W)x=\Lambda^{1/2}x-\Lambda^{1/2}Wx$ and $\Lambda^{1/2}x$ is a column of $\tilde Z$, so $M_{\tilde Z}\Lambda^{1/2}(I-W)x=-M_{\tilde Z}\Lambda^{1/2}Wx$. This proves (i). For (ii), $\calI^\Lambda(\beta_1;x)=0$ iff $\Lambda^{1/2}Wx\in\mathrm{col}(\Lambda^{1/2}Z)$ iff $Wx\in\mathrm{span}(\one,x)$, since $\Lambda^{1/2}$ is invertible. The condition is therefore independent of $\Lambda$ and holds for every $x$ iff $W=\Wc$ by Theorem~\ref{thm:ident}(ii). For (iii), the Laplace precision \eqref{eq:laplace} is $P_{t|t}^{-1}=A+G$ with $A=P_{t|t-1}^{-1}\succ0$ and $G=X_t\T\Lambda_tX_t\succeq0$, and the proof of Theorem~\ref{thm:filter}(ii) applies verbatim with $s(G)=\calI^{\Lambda_t}$ by part (i) and with the minimiser $u_t^*$ of the weighted least-squares problem. \qed

\subsection{Proof of Corollary~\ref{cor:agg}}
Averaging \eqref{eq:obs} over nodes gives $\bar y_t=\beta_{0,t}+\beta_{1,t}N^{-1}\one\T Wy_{t-1}+\beta_{2,t}\bar y_{t-1}+\bar\varepsilon_t$ and $\one\T W=c\T$. If $W$ is doubly stochastic, $c=\one$. For an undirected regular graph $W=A/d$ is symmetric and $W\one=\one$. Writing $N^{-1}c\T y_{t-1}=\bar y_{t-1}+N^{-1}(c-\one)\T y_{t-1}$ gives the last statement. \qed

\subsection{Proof of Theorem~\ref{thm:filter}}

\emph{(i)} We argue by induction. Initially $\theta_0\sim\calN(m_0,P_0)$ is independent of $y_0$. Suppose $\theta_{t-1}\mid\calF_{t-1}\sim\calN(m_{t-1|t-1},P_{t-1|t-1})$. Since $u_t$ is independent of $(\theta_{t-1},\calF_{t-1})$, $\theta_t\mid\calF_{t-1}\sim\calN(m_{t|t-1},P_{t|t-1})$ with $m_{t|t-1}=m_{t-1|t-1}$ and $P_{t|t-1}=P_{t-1|t-1}+Q$. The design $X_t$ is $\calF_{t-1}$-measurable and $\varepsilon_t$ is independent of $(\theta_t,\calF_{t-1})$, so conditionally on $\calF_{t-1}$ the pair $(\theta_t,y_t)$ is jointly Gaussian with $y_t=X_t\theta_t+\varepsilon_t$. Gaussian conditioning then gives $\theta_t\mid(\calF_{t-1},y_t)=\theta_t\mid\calF_t\sim\calN(m_{t|t},P_{t|t})$ with $P_{t|t}^{-1}=P_{t|t-1}^{-1}+\sigma^{-2}X_t\T X_t$ and $m_{t|t}=P_{t|t}(P_{t|t-1}^{-1}m_{t|t-1}+\sigma^{-2}X_t\T y_t)$, which is \eqref{eq:kf}. The covariance identity and the coverage statement follow: for a credible set $C_t$ of nominal level $1-\alpha$ computed from this distribution, $\Prob(\theta_t\in C_t\mid\calF_t)=1-\alpha$, and integrating over $\calF_t$ preserves the level.

\emph{(ii)} By (i), $\Var(\beta_{1,t}\mid\calF_t)=(P_{t|t})_{22}=1/s(P_{t|t}^{-1})$ by Lemma~\ref{lem:schur}(a), with $P_{t|t}^{-1}=A+G$, $A=P_{t|t-1}^{-1}\succ0$ and $G=\sigma^{-2}X_t\T X_t\succeq0$. By Lemma~\ref{lem:schur}(b)--(c), $s(A+G)\ge s(A)+s(G)\ge s(G)=\calI_t$, giving the upper bound. For the lower bound take $a=-u_t^*$ in \eqref{eq:schurvar}: $s(A+G)\le(e_2-Eu_t^*)\T A(e_2-Eu_t^*)+(e_2-Eu_t^*)\T G(e_2-Eu_t^*)\le\lambda_{\max}(A)(1+\|u_t^*\|^2)+\calI_t$, where the second term equals $\sigma^{-2}\|Wy_{t-1}-Z_tu_t^*\|^2=\sigma^{-2}\|M_{Z_t}Wy_{t-1}\|^2=\calI_t$ by the least-squares property of $u_t^*$. Finally $\lambda_{\max}(A)=1/\lambda_{\min}(P_{t|t-1})\le1/\lambda_{\min}(Q)$ because $P_{t|t-1}=P_{t-1|t-1}+Q\succeq Q$.

\emph{(iii)} By (i), $\E[\|\theta_t-m_{t|t}\|^2\mid\calF_t]=\tr P_{t|t}$. When $X_t\T X_t\succ0$, $P_{t|t}^{-1}=A+G\succeq G$ implies $P_{t|t}\preceq G^{-1}$ and $\tr P_{t|t}\le\tr G^{-1}=\sigma^2\tr\{(X_t\T X_t)^{-1}\}\le3\sigma^2/\lambda_{\min}(X_t\T X_t)$. The bound $\E[(\beta_{1,t}-m_{1,t|t})^2\mid\calF_t]=(P_{t|t})_{22}\le1/\calI_t$ is (ii). For the convergence statements, conditional Chebyshev gives $\Prob(\|\theta_t-m_{t|t}\|>\eta\mid\calF_t)\le3\sigma^2/\{\eta^2\lambda_{\min}(X_t\T X_t)\}$, which tends to zero in probability. The unconditional probability tends to zero by dominated convergence. The same argument with $1/\calI_t$ gives consistency for $\beta_{1,t}$, and the lower bound in (ii) shows that $\E[(\beta_{1,t}-m_{1,t|t})^2\mid\calF_t]$ is bounded below by $1/\{\calI_t+(1+\|u_t^*\|^2)/\lambda_{\min}(Q)\}$, which is bounded away from zero in probability when $\calI_t$ and $\|u_t^*\|$ are bounded in probability.

\emph{(iv)} With $W=\Wc$, $\Wc y_{t-1}=\{N\bar y_{t-1}\one-y_{t-1}\}/(N-1)$, and $X_tv_t=-\tfrac{N\bar y_{t-1}}{N-1}\one+\Wc y_{t-1}+\tfrac{1}{N-1}y_{t-1}=0$. Hence $G_tv_t=0$ and $P_{t|t}^{-1}v_t=P_{t|t-1}^{-1}v_t$. By Cauchy--Schwarz applied to $P_{t|t}^{1/2}v_t$ and $P_{t|t}^{-1/2}v_t$, $\|v_t\|^4\le(v_t\T P_{t|t}v_t)(v_t\T P_{t|t}^{-1}v_t)$, so
\[
\Var(v_t\T\theta_t\mid\calF_t)=v_t\T P_{t|t}v_t\ge\frac{\|v_t\|^4}{v_t\T P_{t|t-1}^{-1}v_t}\ge\|v_t\|^2\lambda_{\min}(P_{t|t-1})\ge\|v_t\|^2\lambda_{\min}(Q),
\]
and $\|v_t\|^2\ge1$ because the second coordinate of $v_t$ is $1$. \qed

\emph{(v)} By Lemma~\ref{lem:schur}(a)--(b), $1/v_t=s(P_{t|t}^{-1})\ge s(P_{t|t-1}^{-1})+s(G_t)=1/(P_{t|t-1})_{22}+\calI_t=1/(v_{t-1}+q_2)+\calI_t$, i.e.\ $v_t\le\varphi_{\calI_t}(v_{t-1})$ in the notation of Lemma~\ref{lem:riccati}. If $\calI_t\ge\calI$ then $\varphi_{\calI_t}\le\varphi_{\calI}$ pointwise, and since $\varphi_\calI$ is increasing, induction gives $v_t\le\varphi_\calI^{(t)}(v_0)\to P^\ast(\calI,q_2)$, so $\limsup v_t\le P^\ast$. The bounds on $P^\ast$ are Lemma~\ref{lem:riccati}(c). When every $X_t\T X_t$ is diagonal and $Q,P_0$ are diagonal, all matrices in the recursion \eqref{eq:kf} remain diagonal, the $(2,2)$ entry evolves exactly by $v_t=\varphi_{\calI_t}(v_{t-1})$ with $\calI_t=\sigma^{-2}\|Wy_{t-1}\|^2=s(G_t)$, and Lemma~\ref{lem:riccati}(b) gives $v_t\to P^\ast(\calI,q_2)$ when $\calI_t\equiv\calI$. \qed

\subsection{Proof of Corollary~\ref{cor:floor}}
For any $\calF_t$-measurable $\delta_t$, $\E[(\delta_t-\beta_{1,t})^2\mid\calF_t]=\Var(\beta_{1,t}\mid\calF_t)+(\delta_t-\E[\beta_{1,t}\mid\calF_t])^2\ge\Var(\beta_{1,t}\mid\calF_t)$. Take expectations and apply Theorem~\ref{thm:filter}(ii). For the second claim, on the event $A=\{\calI_t\le C_1,\|u_t^*\|\le C_2\}$ the lower bound of Theorem~\ref{thm:filter}(ii) is at least $\{C_1+(1+C_2^2)/\lambda_{\min}(Q)\}^{-1}$, and $\E[\Var(\beta_{1,t}\mid\calF_t)]\ge\E[\Var(\beta_{1,t}\mid\calF_t)\one_A]\ge\Prob(A)\{C_1+(1+C_2^2)/\lambda_{\min}(Q)\}^{-1}$. \qed

\subsection{Proof of Proposition~\ref{prop:equiv}}
From \eqref{eq:Wc}, $W_\alpha x=(1-\alpha)Wx+\tfrac{\alpha N\bar x}{N-1}\one-\tfrac{\alpha}{N-1}x$, so $W_\alpha x\in\mathrm{span}\{\one,Wx,x\}$ and, solving for $Wx$, $Wx\in\mathrm{span}\{\one,W_\alpha x,x\}$ when $\alpha<1$: the spans coincide. Matching coefficients in $\tilde\beta_0\one+\tilde\beta_1W_\alpha x+\tilde\beta_2x=\beta_0\one+\beta_1Wx+\beta_2x$ with $x=y_{t-1}$ gives $\tilde\beta_1(1-\alpha)=\beta_1$, $\tilde\beta_2-\tilde\beta_1\tfrac{\alpha}{N-1}=\beta_2$ and $\tilde\beta_0+\tilde\beta_1\tfrac{\alpha N\bar y_{t-1}}{N-1}=\beta_0$, which is the stated map. The mean and hence the conditional law of $y_t$ given $(\calF_{t-1},\theta_t)$ are identical, proving (i), and the map is affine with $\calF_{t-1}$-measurable shift, proving the first part of (ii). The inflation statement reads the map backwards. For (iii), $B_\alpha:=\tilde\beta_1W_\alpha+\tilde\beta_2I=\beta_1W+\tfrac{\alpha\beta_1}{(1-\alpha)(N-1)}J+\beta_2I-0=B+c_\alpha J$ after substituting the map and $\Wc=(J-I)/(N-1)$. With the matched intercept $\tilde\beta_0=\beta_0-c_\alpha N\bar y$ at the origin $y$: $\tilde\mu_1(y)=\tilde\beta_0\one+B_\alpha y=\beta_0\one+By+c_\alpha(Jy-N\bar y\one)=\mu_1(y)$ since $Jy=N\bar y\one$. Then
\[
\tilde\mu_2(y)-\mu_2(y)=(\tilde\beta_0-\beta_0)\one+(B_\alpha-B)\mu_1(y)=-c_\alpha N\bar y\,\one+c_\alpha J\mu_1(y)=c_\alpha N\{\bar\mu_1(y)-\bar y\}\one .\qquad\qed
\]

\subsection{Proof of Proposition~\ref{prop:plugin}}

\emph{(i)} Since $W\one=\widetilde W\one=\one$, $B\one=\widetilde B\one=(\beta_1+\beta_2)\one$ and hence $B^k\one=\widetilde B^k\one$ for all $k\ge0$. The intercept sums in \eqref{eq:plugin} therefore coincide, and $\mu_h(y)-\tilde\mu_h(y)=(B^h-\widetilde B^h)y=(B^h-\widetilde B^h)(y-\bar y\one)=(B^h-\widetilde B^h)M_1y$. The telescoping identity $B^h-\widetilde B^h=\sum_{k=0}^{h-1}B^k(B-\widetilde B)\widetilde B^{h-1-k}$ with $B-\widetilde B=\beta_1(W-\widetilde W)$ gives $\|B^h-\widetilde B^h\|\le h\bar\rho^{\,h-1}|\beta_1|\Delta$.

\emph{(ii)} For $\bar\rho<1$, $h\bar\rho^{\,h-1}\le\sum_{j\ge1}j\bar\rho^{\,j-1}=(1-\bar\rho)^{-2}$, and $h\bar\rho^{\,h-1}\to0$.

\emph{(iii)} $B\one=(\beta_1+\beta_2)\one$ exhibits $\beta_1+\beta_2$ as an eigenvalue of $B$. If $Wv=\lambda v$ and $\widetilde Wv=\tilde\lambda v$ then $Bv=rv$, $\widetilde Bv=\tilde rv$, and by (i) $\mu_h(y)-\tilde\mu_h(y)=(B^h-\widetilde B^h)(\bar y\one+cv)=c(r^h-\tilde r^h)v$, whose norm is $|c||r^h-\tilde r^h|$. By the mean value theorem $r^h-\tilde r^h=h\xi^{h-1}(r-\tilde r)$ for some $\xi$ between $r$ and $\tilde r$. If $r\tilde r>0$ then $|\xi|\ge\min(|r|,|\tilde r|)$, and $|r-\tilde r|=|\beta_1||\lambda-\tilde\lambda|$. Finally $\Delta\ge\|(W-\widetilde W)v\|=|\lambda-\tilde\lambda|$. For an undirected graph $W=D^{-1}A$ is similar to the symmetric matrix $D^{-1/2}AD^{-1/2}$, so its eigenvalues are real. \qed

\subsection{Proof of Theorem~\ref{thm:poisson}}

Write $\lambda_{i,t+k}$ for the intensity at step $k$, which is a function of $y_{t+k-1}$ and $\theta_{t+k}$, and $\Lambda_{t+k}=\sum_j\lambda_{j,t+k}$. Throughout, $\E[y_{i,t+k}\mid\calF_{t+k-1},\theta_{t+k}]=\lambda_{i,t+k}$, so all statements about predictive means of $y$ are statements about expectations of intensities.

\emph{(a)} $\lambda_{i,t+1}=\exp(\beta_0+\sum_js_{ij}y_{j,t})$ is $\calF_t$-measurable, giving the one-step mean. Conditionally on $\calF_t$ the $y_{j,t+1}$ are independent Poisson with means $\lambda_{j,t+1}$, so
\begin{align*}
\E[y_{i,t+2}\mid\calF_t]&=\E\Big[\exp\Big(\beta_0+\sum_js_{ij}y_{j,t+1}\Big)\Big|\calF_t\Big]=e^{\beta_0}\prod_j\E\big[e^{s_{ij}y_{j,t+1}}\mid\calF_t\big]\\
&=e^{\beta_0}\prod_j\exp\{\lambda_{j,t+1}(e^{s_{ij}}-1)\},
\end{align*}
which is \eqref{eq:twostep}. For $h=3$, apply \eqref{eq:twostep} one period later: $\E[y_{i,t+3}\mid\calF_{t+1}]=e^{\beta_0}\exp\{\Phi\}$ with $\Phi=\sum_j\lambda_{j,t+2}(e^{s_{ij}}-1)$ a function of $y_{t+1}$. Consider the events $E_m=\{y_{i,t+1}=m,\;y_{j,t+1}=0\ \forall j\ne i\}$, $m\in\mathbb N$, which have conditional probability $\Prob(E_m\mid\calF_t)=e^{-\Lambda_{t+1}}\lambda_{i,t+1}^m/m!>0$. On $E_m$, $\lambda_{i,t+2}=e^{\beta_0+\beta_2m}$ and $\lambda_{j,t+2}=e^{\beta_0+\beta_1w_{ji}m}$ for $j\ne i$, so
\[
\Phi_m=e^{\beta_0+\beta_2m}(e^{\beta_2}-1)+\sum_{j\ne i}e^{\beta_0+\beta_1w_{ji}m}(e^{\beta_1w_{ij}}-1).
\]
If $\beta_2>0$ the first term is positive and grows like $e^{\beta_2m}$. The terms of the sum are nonnegative when $\beta_1\ge0$ and bounded below by $-e^{\beta_0}$ each when $\beta_1<0$ (because then $\beta_1w_{ji}m\le0$), so $\Phi_m\ge e^{\beta_0+\beta_2m}(e^{\beta_2}-1)-(N-1)e^{\beta_0}$. If $\beta_1>0$ and $k$ is a neighbour of $i$, then $w_{ik}>0$ and $w_{ki}>0$ (undirected support), every term of the sum is nonnegative, the $k$-th term equals $e^{\beta_0+\beta_1w_{ki}m}(e^{\beta_1w_{ik}}-1)$, and the first term is nonnegative if $\beta_2\ge0$ and at least $-e^{\beta_0}$ if $\beta_2<0$. So $\Phi_m\ge e^{\beta_0+\beta_1w_{ki}m}(e^{\beta_1w_{ik}}-1)-e^{\beta_0}$. In both cases $\Phi_m\ge c_1e^{c_2m}-c_3$ with $c_1,c_2>0$. Therefore
\[
\E[y_{i,t+3}\mid\calF_t]\ge\sum_{m\ge0}\Prob(E_m\mid\calF_t)\,e^{\beta_0}e^{\Phi_m}\ge e^{\beta_0-c_3-\Lambda_{t+1}}\sum_{m\ge0}\frac{\lambda_{i,t+1}^m}{m!}\exp\{c_1e^{c_2m}\}=+\infty ,
\]
since $c_1e^{c_2m}+m\log\lambda_{i,t+1}-\log m!\to\infty$. For $h>3$ the same argument applied at origin $t+h-3$ gives $\E[y_{i,t+h}\mid\calF_{t+h-3}]=+\infty$ almost surely, hence $\E[y_{i,t+h}\mid\calF_t]=+\infty$ by the tower property. If $\beta_1\le0$ and $\beta_2\le0$ then $s_{ij}\le0$ for all $j$ and $\lambda_{i,t+k}\le e^{\beta_0}$ for every $k\ge1$ because $y\ge0$.

\emph{(b)} Let $z=(1,(Wy_{t+1})_i,y_{i,t+1})\T$, which is $\calF_{t+1}$-measurable, and write $\theta=\theta_{t+1}$. Since $u_{t+2}$ is independent of $(\theta,y_{t+1},\calF_t)$,
\[
\E[\lambda_{i,t+2}\mid\theta,y_{t+1},\calF_t]=\exp\{z\T\theta+\tfrac12z\T Qz\}\ge\exp\{z\T\theta+\tfrac12\lambda_{\min}(Q)y_{i,t+1}^2\}.
\]
On $E_m$ (defined as in (a)) $z=(1,0,m)\T$ and $\Prob(E_m\mid\theta,\calF_t)=e^{-\Lambda_{t+1}(\theta)}\lambda_{i,t+1}(\theta)^m/m!$, where $\lambda_{j,t+1}(\theta)=\exp(z_{j,t}\T\theta)$ with $z_{j,t}=(1,(Wy_t)_j,y_{j,t})\T$ $\calF_t$-measurable. Hence, with $q=\lambda_{\min}(Q)>0$,
\[
\E[y_{i,t+2}\mid\calF_t]\ge\sum_{m\ge0}\frac{e^{qm^2/2}}{m!}\,\E\Big[\exp\big\{\beta_{0,t+1}+m\beta_{2,t+1}+mz_{i,t}\T\theta-\Lambda_{t+1}(\theta)\big\}\,\Big|\,\calF_t\Big].
\]
Choose $R>0$ such that the event $B=\{\|\theta\|\le R\}$ has conditional probability $p_B>0$ (possible for any distribution of $\theta$ given $\calF_t$, including a point mass). On $B$, $\Lambda_{t+1}(\theta)\le\Lambda_R:=\sum_je^{R\|z_{j,t}\|}$ and $\beta_{0,t+1}+m\beta_{2,t+1}+mz_{i,t}\T\theta\ge-R-mR(1+\|z_{i,t}\|)$. Thus the $m$-th expectation is at least $p_Be^{-R-\Lambda_R}e^{-mC}$ with $C=R(1+\|z_{i,t}\|)$, and $\sum_me^{qm^2/2-mC}/m!=\infty$ because $qm^2/2-mC-\log m!\to\infty$. Finally, when $\theta\mid\calF_t\sim\calN(m,P)$, $\E[y_{i,t+1}\mid\calF_t]=\E[\exp(z_{i,t}\T\theta)\mid\calF_t]=\exp(z_{i,t}\T m+\tfrac12z_{i,t}\T Pz_{i,t})<\infty$.

The proofs of parts (c)--(e), the $h\ge3$ induction, and both halves of Proposition~\ref{prop:nb} are in Section~S1 of the Supplement.

\section{Additional tables}\label{app:tables}

\begin{table}[h]
\centering\small
\caption{Randomised PIT of the one-month-ahead forecasts: counts in ten equal bins over the $\PitN$ neighbourhood-months of 2015 (expected count $662.4$ per bin under calibration) and the chi-squared statistic on 9 degrees of freedom.}\label{tab:pit}
\setlength{\tabcolsep}{4pt}
\resizebox{\textwidth}{!}{\begin{tabular}{lrrrrrrrrrrr}
\toprule
model & $[0,.1)$ & $[.1,.2)$ & $[.2,.3)$ & $[.3,.4)$ & $[.4,.5)$ & $[.5,.6)$ & $[.6,.7)$ & $[.7,.8)$ & $[.8,.9)$ & $[.9,1]$ & $\chi^2$\\
\midrule
NSSM, observed $W$ & 775 & 752 & 716 & 715 & 642 & 597 & 537 & 549 & 527 & 814 & 152.4 \\
DGLM, no network & 786 & 756 & 733 & 667 & 676 & 596 & 535 & 515 & 553 & 807 & 157.7 \\
NSSM, complete graph & 776 & 748 & 727 & 747 & 619 & 572 & 570 & 511 & 554 & 800 & 156.6 \\
 
\bottomrule
\end{tabular}}
\end{table}

\begin{table}[h]
\centering\small
\caption{Study S2 (Section~\ref{sec:s2}): root-mean-square discrepancy between the $h$-step plug-in means under $W$ and under $\widetilde W=(1-\alpha)W+\alpha\Wc$, averaged over forecast origins $20,\dots,60-h$, for the stable and the near-unit regimes.}\label{tab:s2}
\begin{tabular}{rrrrrr}
\toprule
$\alpha$ & $\|W-\widetilde W\|$ & $h=1$ & $h=2$ & $h=4$ & $h=8$\\
\midrule
\multicolumn{6}{l}{\emph{stable}: $\beta_1=0.30$, $\beta_2=0.35$, $\|B\|_{\mathrm{op}}=0.689$, $\rho(B)=0.650$}\\
0.00 & 0.000 & 0.0000 & 0.0000 & 0.0000 & 0.0000 \\
0.05 & 0.075 & 0.0102 & 0.0087 & 0.0051 & 0.0011 \\
0.10 & 0.150 & 0.0203 & 0.0172 & 0.0098 & 0.0020 \\
0.20 & 0.299 & 0.0406 & 0.0340 & 0.0186 & 0.0035 \\
0.30 & 0.449 & 0.0609 & 0.0503 & 0.0263 & 0.0046 \\
0.40 & 0.598 & 0.0812 & 0.0662 & 0.0332 & 0.0054 \\
0.50 & 0.748 & 0.1015 & 0.0818 & 0.0392 & 0.0060 \\
\addlinespace
\multicolumn{6}{l}{\emph{near-unit}: $\beta_1=0.55$, $\beta_2=0.50$, $\|B\|_{\mathrm{op}}=1.132$, $\rho(B)=1.050$}\\
0.00 & 0.000 & 0.0000 & 0.0000 & 0.0000 & 0.0000 \\
0.05 & 0.075 & 0.0273 & 0.0436 & 0.0704 & 0.1016 \\
0.10 & 0.150 & 0.0546 & 0.0863 & 0.1360 & 0.1874 \\
0.20 & 0.299 & 0.1092 & 0.1688 & 0.2536 & 0.3199 \\
0.30 & 0.449 & 0.1638 & 0.2476 & 0.3545 & 0.4120 \\
0.40 & 0.598 & 0.2184 & 0.3228 & 0.4405 & 0.4749 \\
0.50 & 0.748 & 0.2730 & 0.3943 & 0.5131 & 0.5170 \\
 
\bottomrule
\end{tabular}
\end{table}

\end{document}